\documentclass[11pt,letterpaper]{article}
 \pdfoutput=1

\usepackage{jcappub}

\usepackage{caption}
\usepackage{floatrow}
\usepackage{journals}

\usepackage{subfig}

\usepackage{bm}

\def\be{\begin{eqnarray}}
\def\ee{\end{eqnarray}}

\def\k{\bm{k}}

\def\x{\bm{x}}

\def\q{\bm{q}}
\def\s{\bm{s}}

\def\d{\delta_D}

\newsavebox{\tempbox}

\title{Lagrangian or Eulerian; Real or Fourier? Not All Approaches to Large-Scale Structure Are Created Equal.}
\author{Svetlin Tassev$^{a}$}

\affiliation{ \sl $^{a}$ Department of Astrophysical Sciences, Princeton University, 4 Ivy Lane, Princeton, \\NJ 08544, USA
}

\abstract{We present a pedagogical systematic investigation of the accuracy of Eulerian and Lagrangian perturbation theories of large-scale structure. We show that significant differences exist between them especially when trying to model the Baryon Acoustic Oscillations (BAO). We find that the best available model of the BAO in real space is the Zel'dovich Approximation (ZA), giving an accuracy of $\lesssim3\%$ at redshift of $z=0$ in modelling the matter 2-pt function around the acoustic peak. All corrections to the ZA around the BAO scale are perfectly perturbative in real space. Any attempt to achieve better precision  requires  calibrating the theory to simulations because of the need to renormalize those corrections. In contrast, theories which do not fully preserve the ZA as their solution, receive $\mathcal{O}(1)$  corrections around the acoustic peak in real space at $z=0$, and are thus of suspicious convergence at low redshift around the BAO. As an example, we find that a similar accuracy of $3\%$ for the acoustic peak is achieved by  Eulerian Standard Perturbation Theory (SPT) at linear order  only at $z\approx4$. Thus even when SPT is perturbative, one needs to include loop corrections for $z\lesssim4$ in real space.
In Fourier space, all models perform similarly, and are controlled by the  overdensity amplitude, thus recovering standard results. However, that comes at a price.
Real space cleanly separates the BAO signal from non-linear dynamics. In contrast, Fourier space mixes signal from short mildly non-linear scales with the linear signal from the BAO to the level that non-linear contributions from short scales dominate. Therefore, one has little hope in constructing a systematic theory for the BAO in Fourier space.
}

\usepackage{graphicx}
\begin{document}
\maketitle

\section{Introduction}

Establishing the homogeneity and isotropy of the universe was the first step in the development of the standard cosmological model. Later, confirming the predictions of linear theory for the Cosmic Microwave Background (CMB) and  Large-Scale Structure (LSS) using surveys such as COBE \cite{1992ApJ...396L...1S} and SDSS  (e.g. \cite{2005ApJ...633..560E}) earned the field the title of precision science. 

Cosmology has since moved on to studying higher-order effects. Those provide consistency checks for our standard cosmological model, information about the initial conditions set by inflation, as well as understanding the effective physics of LSS, such as those governing the distribution of halos and galaxies.

However, our understanding of structure formation beyond linear order is still far from perfect. Neglecting baryons, and working at scales much smaller than the horizon, the dynamics is completely described by the dynamics of Cold Dark Matter (CDM). Even then, non-linear (NL) evolution at small scales presents a major challenge due to its non-trivial feedback on large scales. Indeed just recently \cite{2010arXiv1004.2488B} it was realized that CDM on large scales behaves effectively like a fluid with sound speed and bulk viscosity, which arise from backreaction effects of small non-virialized structures onto the large scales. At the same time many seemingly successful attempts at going beyond the linear regime have been shown to be plagued by problems (e.g. \cite{2012JCAP...04..013T,2013ApJ...769..106S,2013JCAP...05..031P}).

Thus, it is not surprising that most questions of structure formation are usually addressed through numerical simulations. Those, however, suffer from either finite volume or finite resolution effects, or both, which prevent an efficient sampling of the cosmological parameter space. Consequently, it is still difficult to get a robust handle on the uncertainties of many astrophysical probes, since that requires running hundreds and even thousands of simulations \cite{2013MNRAS.428.1036M}. 

Yet, we cannot abandon understanding structure formation beyond the linear order, since going from the linear into the mildly non-linear regime gives us access to two orders of magnitude more Fourier modes, equivalent to conducting observational surveys with a hundred times more volume. Thus, devising efficient models to study the non-linear and mildly non-linear regimes of structure formation is crucial for utilizing current and planned cosmological observations (such as galaxy clustering and weak lensing) from both ground-based (e.g. SDSS, LSST), as well as space observatories (e.g. Planck, WFIRST, Euclid). 
%
%

One observable which has been a major  target of theoretical investigations beyond linear order is the Baryon Acoustic Oscillations (BAO) peak in the matter 2-point function. The acoustic peak provides a standard ruler, which has been used extensively for constraining the cosmological parameters (e.g. \cite{2012MNRAS.427.2168M}).  Current and future surveys (e.g. BigBOSS, WFIRST) will measure the BAO scale with sub-percent accuracy, thus putting strong constraints on the equation of state of dark energy. However, the acoustic peak is broadened with time, resulting in degradation in the accuracy of the measured BAO scale to $\sim$3\%, and in shifts in the peak of $\sim$0.3\% \cite{2012MNRAS.427.2146X}. Fully understanding that degradation is still a challenge, and is badly needed for fully utilizing  forthcoming observations. 

In a recent paper (Appendix C of \cite{2013arXiv1311.2168P}) it was shown using a simple toy model that in real space the correlation function around the BAO peak should be receiving $\mathcal{O}(1)$ corrections at $z=0$ if one works following Eulerian Standard Perturbation Theory (SPT; e.g. \cite{2002PhR...367....1B}). This runs contrary to the usual intuition, which says that any corrections should be less than a percent, given that the BAO are deep in the linear regime. In the same toy model, it was shown that the correction should be fully captured if one works with Lagrangian Perturbation Theory (LPT) \cite{1995A&A...296..575B}. Even more interestingly, the correction should completely disappear for all models if one works in Fourier space.
Those results should not have been a surprise. From numerical studies it is well known that the fractional difference between the true matter 2-point function and the linear (i.e. lowest order SPT) result is indeed $\mathcal{O}(1)$ around the acoustic peak.

Such an overlook calls for a better systematic investigation of the sources of errors arising at BAO scales, which we try to do in this paper. Given the plethora of perturbative approaches to large-scale structure proposed in the literature, we will frame our discussion by taking a step back and trying to answer two 
questions:
\begin{itemize}
\item Is there any sense in which working in Fourier space is different from working in real space? 
\item Is there any (dis)advantage in working with Eulerian rather than Lagrangian perturbation theory?
\end{itemize} 
In answering those questions we will find that:
\begin{itemize}
\item The Zel'dovich approximation (the lowest order LPT solution) \cite{zeldovich} has an $\mathcal{O}(3\%)$ accuracy\footnote{That includes the errors in the recovered smooth component. Errors in the peak position and spread need not be that large.} around the Baryon Acoustic Oscillations (BAO) peak in real space at $z=0$. Any attempt to improve that requires renormalizing the theory, i.e.  calibrating the theory to observations.
\item Theories not preserving the Zel'dovich approximation intact (e.g. as a lowest order approximation), such as SPT, suffer from the following:
1) In real space we confirm the toy model result of \cite{2013arXiv1311.2168P} that such theories incur an order 1 expansion parameter at $z=0$ when studying the BAO peak, thus invalidating their predictions for the BAO at low redshift. The parameter is given by the square of the ratio between the rms motions of particles within patches of size $\sim100$Mpc$/h$ (set by the acoustic scale) and the width of the BAO peak; 2) At $z\approx4$, that parameter is  $\sim3\%$ in real space, and therefore even at that redshift one has to include 1-loop corrections to achieve a $3\%$ accuracy in the matter 2-pt correlation function.
\item Fourier space mixes linear BAO scales and mildly non-linear scales to a level such that a systematic perturbative theory of the shape of the BAO peak may not be possible to construct in Fourier space.
\item The only way to produce realizations of the density field with an accuracy in halo positions  of better than $\sim 10$Mpc$/h$ at $z=0$  is if one uses the Zel'dovich approximation (or improvements of it).
\end{itemize}
Note that the first three  results are completely dependent on the particular shape of the power spectrum of our universe.  If the power spectrum was scale invariant for example, then the results  will not depend on the choice between real and redshift space, or Lagrangian and Eulerian space in which to expand the fields. This is precisely the reason why we focus on the BAO feature in this paper when comparing those spaces.

Given the somewhat surprising nature of the above results, we will try to introduce them in a more pedagogical format, moving most calculations to appendixes. In the next section we give an overview of the basic categories of models of LSS. In Section~\ref{ftorreal} we give some introductory remarks on the differences between real and Fourier space. In Section~\ref{sec:truth} we compare Lagrangian and Eulerian models by giving  a comparison between models of LSS and N-body results both on a realization by realization basis, and comparing statistics of the matter density. In Section~\ref{sec:xi} we recast the matter 2-pt function in a way suitable for systematically studying the difference between the true result and Lagrangian and Eulerian predictions. In Section~\ref{sec:exp1} we analytically identify the parameters controlling that difference; and in Section~\ref{sec:numEps} we numerically evaluate those parameters from simulations. We discuss our results in Section~\ref{sec:why}, and then summarize them in Section~\ref{sec:summary}.

\section{Models of large-scale structure}

Models of LSS can be crudely grouped into five categories\footnote{\label{foot:foot}
There are a multitude of modifications to SPT and LPT  that have been proposed claiming good performance for the matter power spectrum or real space 2-pt function, e.g. \cite{2008PhRvD..78j3521B,2008PhRvD..77b3533C,2008PhRvD..77f3530M} -- which may or may not be easy to classify according to Table~\ref{table}. However, many of these schemes suffer from problems such as: breaking translation invariance \cite{2012JCAP...04..013T,2013JCAP...05..031P}; using ad hoc approaches to model small-scales; demonstrably not being better than simple SPT \cite{2013ApJ...769..106S}; not being renormalized and thus of suspicious convergence properties; not clear how to extend beyond the power spectrum to higher order statistics. Another possible issue they face which needs further investigation is purely methodological: what gets published are models which show some success, which is usually measured by how well they fit the matter power spectrum extracted from simulations at several redshifts. However, such success is somewhat misleading as the fits usually work well below the non-linear scale, where the power spectrum can be written as the linear power spectrum plus a $k$-dependent correction with a time dependence fixed by the growth factor. So, the results boil down to fitting \textit{one} relatively smooth curve (the $k$-dependent correction) -- the one which the models were designed to fit from the start. Indeed, when put to the test in fitting other observables, many of these models have been shown not to perform well \cite{2009PhRvD..80d3531C}.  Given all the issues listed above, we choose not to discuss these models further in this paper.}, 
summarized in Table~\ref{table}. The straightforward application of perturbation theory in Eulerian space results in what is known as Standard Perturbation Theory (SPT), which has a really long history of applications (e.g. \cite{2002PhR...367....1B}). However, the non-perturbative small scales are known  in SPT to lead to divergencies at large scales, which have only recently been addressed (at least in Fourier space, see Appendix~\ref{app:cc}) with the development of Eulerian Effective Field Theory (EEFT; e.g. \cite{2010arXiv1004.2488B}). EEFT has demonstrated that the small scales modify the dynamics of the CDM fluid to include an effective non-zero sound speed  and bulk viscosity -- parameters which need to be extracted from simulations, a relatively mild price to pay for making the theory convergent.

\begin{table}[t!]
\centering
\caption{ {Categories of models and their properties of large-scale structure ordered according to increasing realism from left to right. By stating that (modified) ZA is ultimately convergent, we assume one renormalizes the theory as in \cite{2013arXiv1311.2168P}. The price to pay is the introduction of free parameters in the theory beyond the ZA. The EEFT entries are explained in Appendix~\ref{app:cc}. The COLA halo position accuracy is for 10 timesteps. See the text for further discussion.}\label{table}}
\begin{tabular}{p{7cm}||c|c|c|c|c}
 &\footnotesize  SPT &\footnotesize  EEFT &\footnotesize  {(modified) ZA} &\footnotesize  {COLA} &\footnotesize  N-body \\ 
\hline \hline 
\footnotesize  no free parameter & \footnotesize  YES & \footnotesize  no &\footnotesize   {no (beyond ZA)} &\footnotesize   {YES} & \footnotesize  YES \\ 
\hline 
\footnotesize convergent in Fourier (below NL scale)&\footnotesize   no &\footnotesize   YES & \footnotesize  {YES} & \footnotesize  {YES} &\footnotesize   YES \\ 
\hline \footnotesize  
 convergent in real space ($z=0$, BAO) &\footnotesize   {no} &\footnotesize   {no} &\footnotesize   {{YES}} & \footnotesize  {YES} &\footnotesize   YES \\ 
\hline \footnotesize  
analytical (\footnotesize  no averaging over realizations) & \footnotesize  YES & \footnotesize  YES &\footnotesize   {YES} &\footnotesize   {no} &\footnotesize   no \\ 
\hline 
\footnotesize  halo position accuracy at $z=0$ [Mpc$/h$] & \footnotesize  $\sim$ 10 &\footnotesize   $\sim$ 5 & \footnotesize  {$\sim$  1} & \footnotesize  {$\sim$ 0.1} & \footnotesize  ``truth'' \\ 
\hline 
\footnotesize  works for non-linear scales & \footnotesize  no &\footnotesize   no & \footnotesize  {no} & \footnotesize  {YES} & \footnotesize  YES 
\end{tabular}
\end{table}

An alternative to Eulerian models is Lagrangian Perturbation Theory, which has the Zel'dovich Approximation (ZA; \cite{zeldovich}; see Appendix~\ref{sec:ZA} as well) as its lowest order solution. Along the course of this paper we will find the ZA to be crucial to understanding the CDM dynamics in real space around the acoustic scale, with the higher orders being small corrections to it. That is why we separate the ZA and its (consistent) modifications, which leave the full unexpanded ZA as a lowest order solution, in their own column in Table~\ref{table}. Examples of such modified ZA models are  LPT  and CLPT \cite{2013MNRAS.429.1674C} for example, as well as LEFT \cite{2013arXiv1311.2168P} and the model in \cite{2012JCAP...12..011T}.

Until recently the ZA had a very serious drawback -- analytical solutions existed only for the 2-point correlation function (e.g. \cite{2013MNRAS.429.1674C}). Thus, (modified) ZA has mostly been regarded as a cheap way of performing very crude N-body simulations (e.g. \cite{2013MNRAS.428.1036M}), and rarely as an analytical power horse. In an attempt to change that, in a companion paper \cite{ZApaper} we give an analytical closed form expression for the $n$-point functions in the ZA, evaluated some of them numerically, and discuss how those results can be  extended to arbitrary orders (including corrections arising when the theory is renormalized).

Going into the truly non-linear regime requires us to make sacrifices to our models. One option is to introduce ansatzes to capture halo physics, such as the halo model \cite{2002PhR...372....1C}. However, unlike the free parameters in  effective field theories, such as EEFT and LEFT,  there is no  systematic prescription of parametrizing 
those ansatzes, thus introducing an element of art. 

Another option is to abandon the goal of obtaining analytical results and recover statistics of large-scale structure as ensemble averages over realizations done with N-body simulations. However, using full-blown N-body simulations is not always practical as sample variance often times makes those prohibitively expensive \cite{2013MNRAS.428.1036M}. But detailed N-body simulations should not be needed for statistics which do not require one to resolve internal halo profiles, since large scales are well-captured by the (modified) ZA. Therefore, recently several numerical prescriptions have been proposed (e.g. \cite{2013JCAP...06..036T,2013MNRAS.433.2389M}) to speed-up N-body simulations by using a hybrid between the (modified) ZA and some ansatz for modelling halo physics. Of them, the Comoving Lagrangian Acceleration (COLA) method \cite{2013JCAP...06..036T} presents the most transparent interpolation between perturbation theory and a full N-body code since it can behave as an N-body implementation of the ZA (for 1 timestep), or as a full-blown N-body code (timesteps $\gg 10$). And more importantly, it requires no free parameters.

From Table~\ref{table} we can see that the (modified) ZA  pushes the limits of what analytical calculations can do. Therefore, it is not surprising that in this paper we will spend a lot of time analyzing its properties, concentrating on some of the conclusions summarized in the table  along the way.

\section{Prelude I. Fourier or real space?}\label{ftorreal}

As has become a standard practice in the field, the Eulerian schemes discussed above are developed and tested in Fourier space because modes of different wavevectors decouple at linear order, simplifying the analytics. In contrast, Lagrangian methods based on the ZA are usually developed in real space (e.g. \cite{2013MNRAS.429.1674C,2012JCAP...12..011T}). Thus, a discussion of Fourier vs real space is very often coupled to the question of whether one writes down a perturbation theory in Eulerian or Lagrangian space.
 
One sphere where Fourier space is not ideal is trying to address actual observations. Observational surveys live in real space, featuring both non-trivial selection functions, as well as geometries -- complications which are much more easily addressed in real space than Fourier.

Moreover, observations target tracers of the CDM -- often non-linear structures such as halos and galaxies. We know that non-linearities, shot noise and halo exclusion are local processes in real space. Thus, one can be optimistic to build \textit{both} physically motivated \textit{and} successful parametric models of tracers, such as the halo bias framework and halo occupation distributions, in real (possibly Lagrangian), but not Fourier space.   An illustration of that is the recent success in modelling 2-pt statistics of biased tracers in both real and redshift space using a modified ZA model \cite{2013MNRAS.429.1674C}.

Yet, one may not think of working in Fourier space as a real drawback since Fourier and real space are conjugate and one should be able to freely transform between the two. That is, if one can model all of Fourier and all of configuration space. However, that is clearly not the case due to the existence of the non-linear regime. So, for each application, one has to make a careful choice between the two spaces. We will stumble upon an illustration of this problem in Section~\ref{sec:FT}, and we will then discuss it at length in Section \ref{sec:why}.

In what follows,  we will focus  on the ability of perturbation theories to recover the CDM density field and its statistics. We will uncover some interesting results even without introducing complications such as tracers, and we will only briefly comment on redshift space distortions. Therefore, the discussion of those we relegate to the proverbial future work.

\section{Prelude II. Lagrangian and Eulerian models compared to the truth.}\label{sec:truth}

Maybe one of the most striking examples of the non-trivial way in which Lagrangian and Eulerian perturbation theories are related  is in trying to capture the large-scale coherent flows in the universe. Those flows live well in the linear regime (e.g. \cite{2012JCAP...04..013T}), and therefore naively must be recoverable in any perturbative expansion. In what follows, we investigate the effects of those large-scale flows on the accuracy of the ZA and SPT.

Imagine we have modelled a region of the universe, and the fields in our model (such as density and velocity) have the correct  statistics. Then imagine that the region of the universe we modelled is translated by a very large-scale coherent motion, which our model did not take into account. The statistics of the fields between the model and true regions will remain exactly the same due to translation invariance. 

Thus, if one calculates $n$-point statistics of the density, for example, one would find that the flows that affect their calculation are only those which are non-coherent on scales smaller than the largest separation between the $n$ points. Both SPT and the ZA satisfy this criterion (e.g. \cite{1996ApJS..105...37S}).

By being the linear order result in LPT, the Zeldovich approximation follows the advection of short-scale modes by the dominant bulk linear motions in the universe. In contrast,  linear theory, being the lowest order in SPT, focuses only on capturing the 2-pt statistics correctly. Thus, flows which are coherent on scales larger than the separation between the two points are completely neglected\footnote{Linear theory gives a density field which is a simple rescaling (by the growth factor) of the initial density field. Thus, it does not capture advection of the short-scales by the large-scale motions.}. Yet, one can hope that higher-order SPT could recover the advection of the small scales by the large-scale flows as it must model higher $n$-point statistics, which couple small and large scales, which in turn requires correctly handling bulk flows on a wider range of scales. Below we will see when this intuition breaks down.

\subsection{Comparing realization by realization}

\begin{figure}[t!]
\centering
\subfloat{\includegraphics[width=0.32\textwidth]{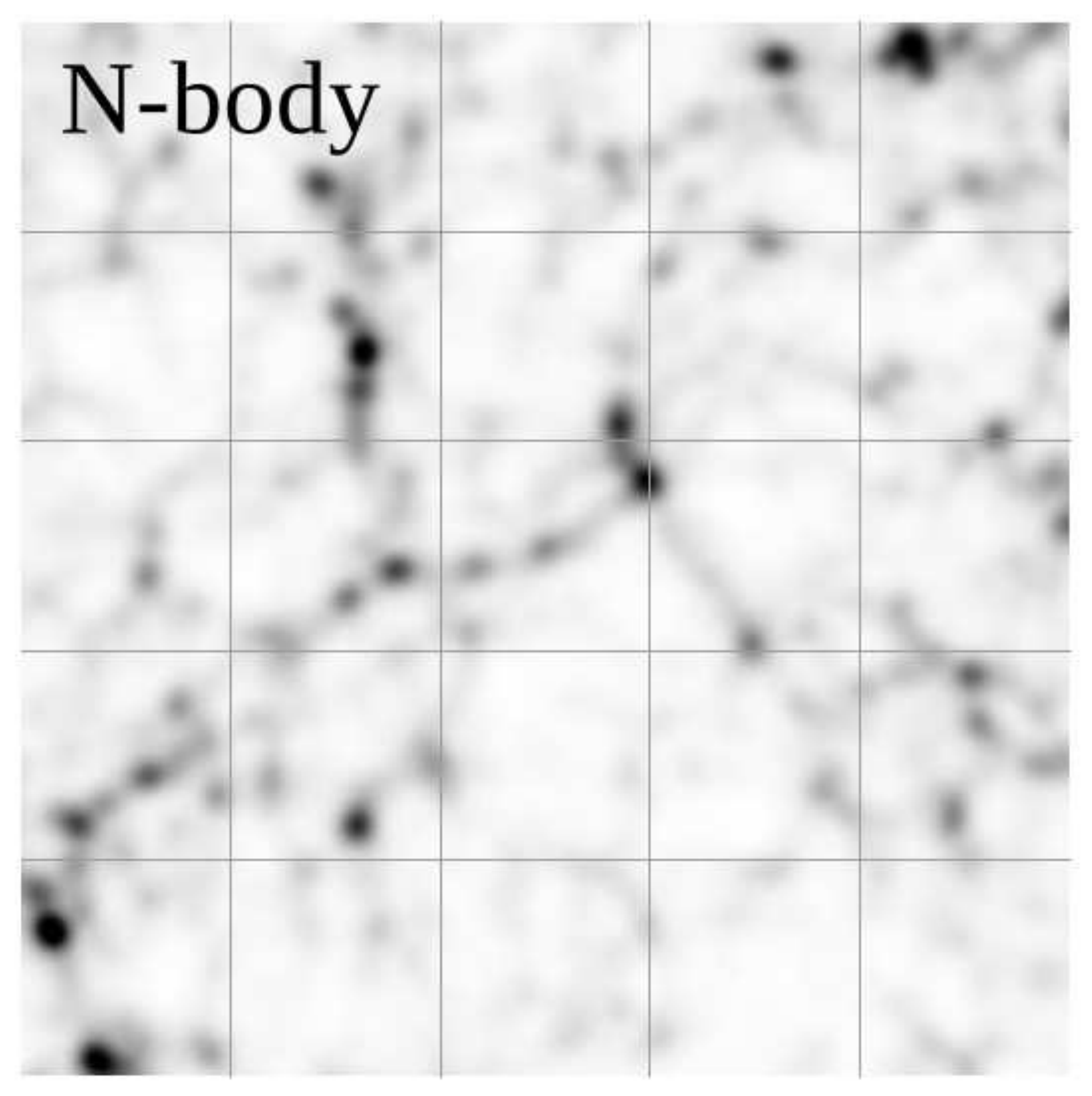}}     
\subfloat{\includegraphics[width=0.32\textwidth]{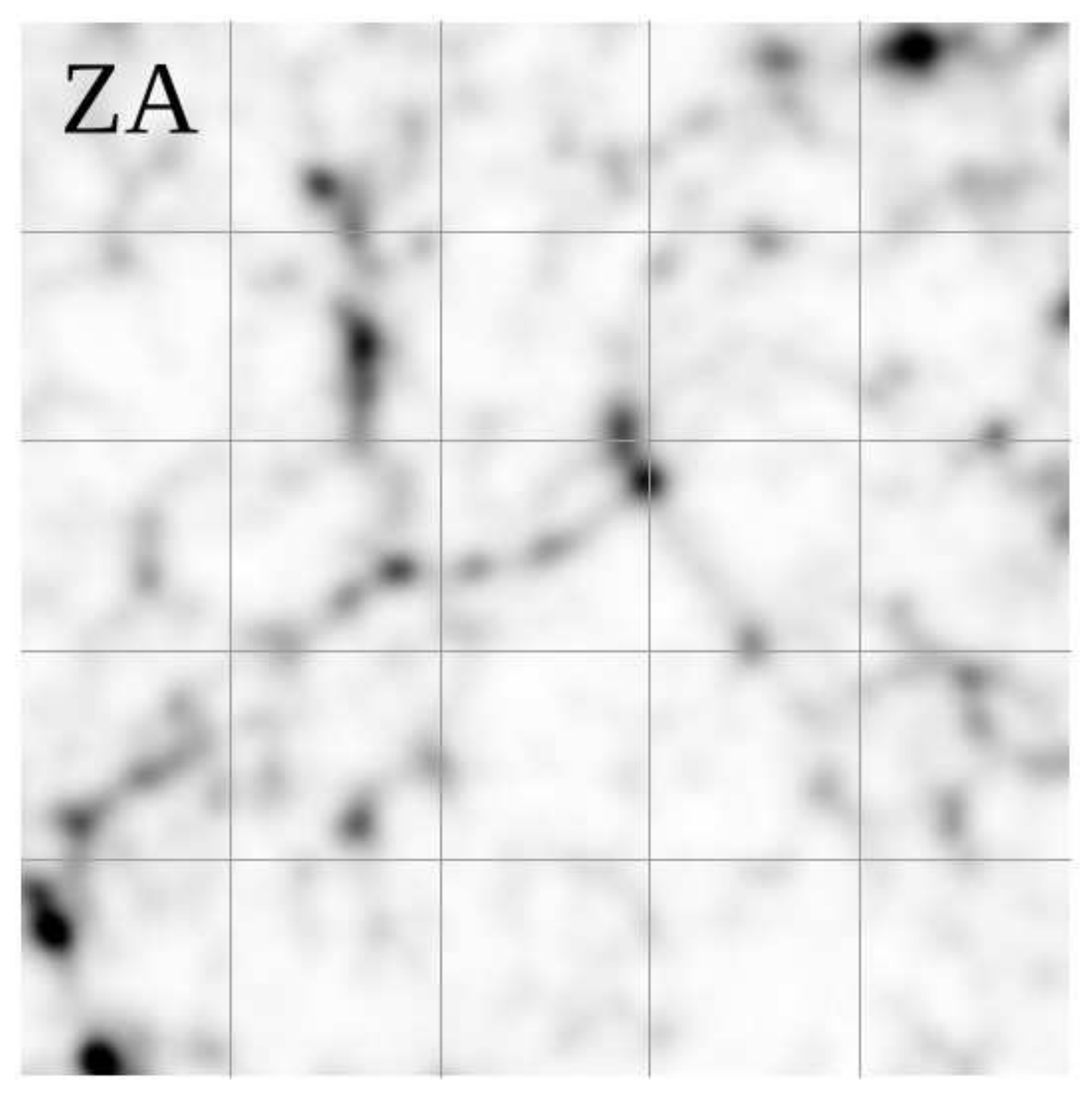}}
\hfill
\\
\subfloat{\includegraphics[width=0.32\textwidth]{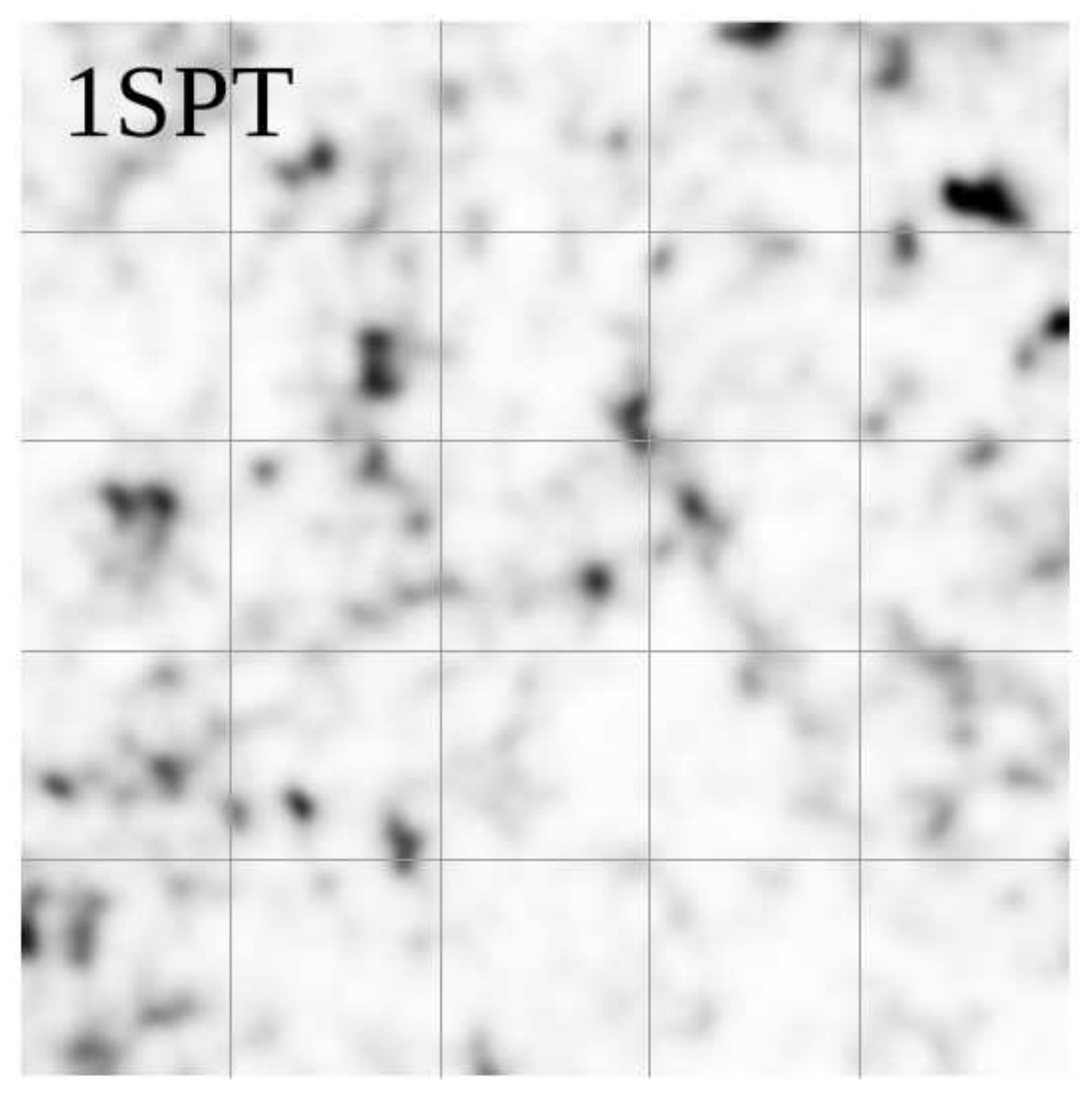}}     
\subfloat{\includegraphics[width=0.32\textwidth]{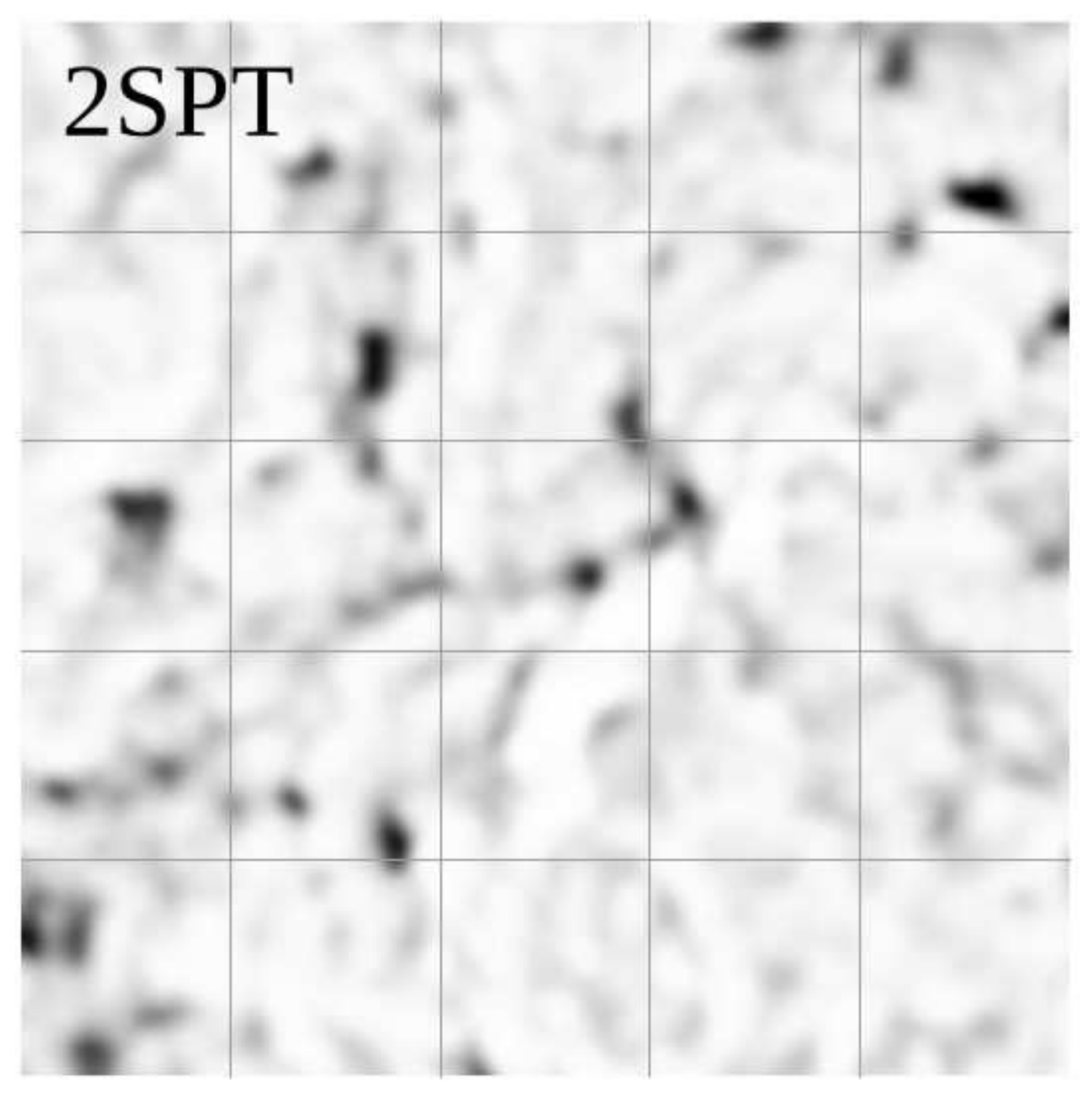}}
\subfloat{\includegraphics[width=0.32\textwidth]{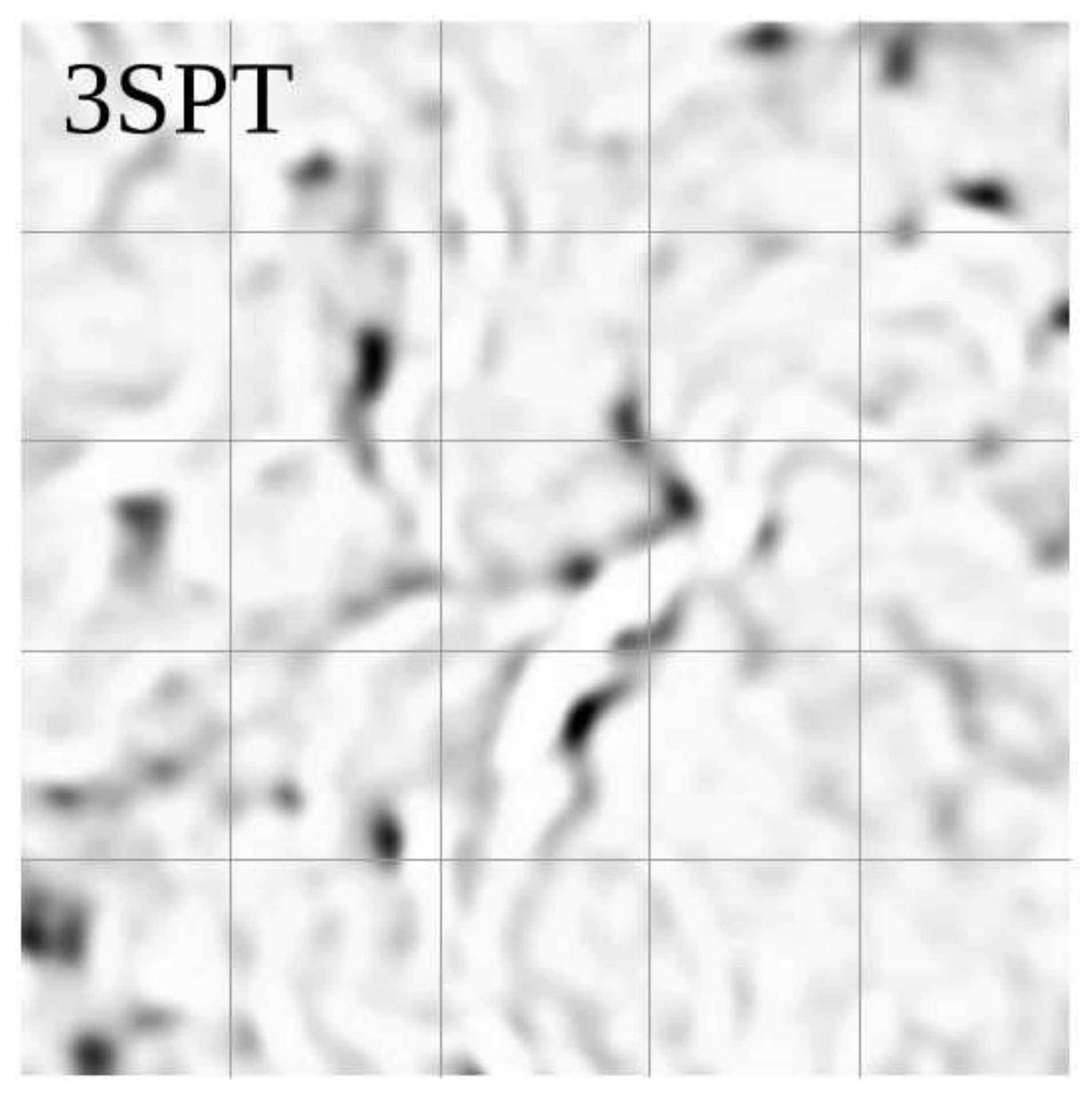}}
\hfill
\caption{Slices (250Mpc$/h$ on the side; $z=0$) through  one and the same  CDM density field as obtained with an N-body and with various models of LSS. One should note that halo positions are well recovered in the ZA, while the errors in SPT are $\sim$10Mpc$/h$. Given the comparable thickness of our slices, some halos which are in the N-body slice are not in the SPT slices, and vice versa.} \label{fig:slices}
\end{figure}

Let us do some numerical experiments.\footnote{For early comparisons between the ZA and linear theory similar to the ones in this section but for power-law spectra, see \cite{1993MNRAS.260..765C}.} In Figure~\ref{fig:slices} we show five slices through one  and the same CDM density field (at $z=0$; 250Mpc$/h$ on the side; 15.6Mpc$/h$ thick; Gaussian smoothed on 2.65Mpc$/h$) as calculated using 5 different methods: using a full N-body, using the ZA, and using the density field as obtained  up to first (linear theory or 1SPT), up to second (2SPT) and up to third order (3SPT) in SPT.\footnote{To the best of our knowledge, the 2SPT and 3SPT slices we show in Figure~\ref{fig:slices} are the first presentations of realizations in real space calculated in SPT beyond  linear order.} To make any visual comparison possible, we transformed the  histograms of the images in bins of intensity to match the N-body result (i.e. approximately matched the 1-pt functions of the density fields).

To obtain the ZA realization one moves actual particles along trajectories calculated using linear theory. Thus, the ZA has particle number conservation explicitly built  into it, and the density is always $\geq 0$, resulting in a nice image whether one does histogram matching or not. 

In contrast, in SPT there is no notion of particles, and the density histogram (i.e. the density 1-pt function) develops strong non-Gaussian wings which go to large both positive and \textit{negative} densities, which results in the strange-looking wisps of near white especially prominent in 3SPT after the histogram matching. Those  artifacts are somewhat dependent on the $k$ cutoffs we impose in calculating the higher order corrections to the density\footnote{This is something which  may or may not be resolved using EEFT, which addresses precisely questions of how one deals with short scale physics when fields evaluated at the same point in real space are multiplied -- an operation required to get the 2SPT and 3SPT realizations.} -- thus, we leave them aside. One thing that does not depend on those cutoffs, however, and still captures the eye is the fact that ``halos'' (dark clumps) are offset from their ``true'' positions as seen in the N-body by $\sim10$Mpc$/h$ in \textit{all} SPT panels. 

That is a striking result, given that halo positions are recovered to $\sim1$Mpc$/h$ accuracy with the ZA (see also \cite{2012JCAP...12..011T}), which employs Lagrangian \textit{linear} theory to solve for the particle motions. As described in the beginning of this section, we already expected that SPT at linear level would be poor in capturing bulk flows. However, as those flows are deep in the linear regime, we expected the picture to be significantly  improved at higher order SPT. Yet looking at the images that does not seem to be the case. Therefore, already at this point one should start wondering whether what is perturbative depends on the choice of space one decides to work in.

\begin{figure}[t!]
\centering
\subfloat{\includegraphics[width=0.48\textwidth]{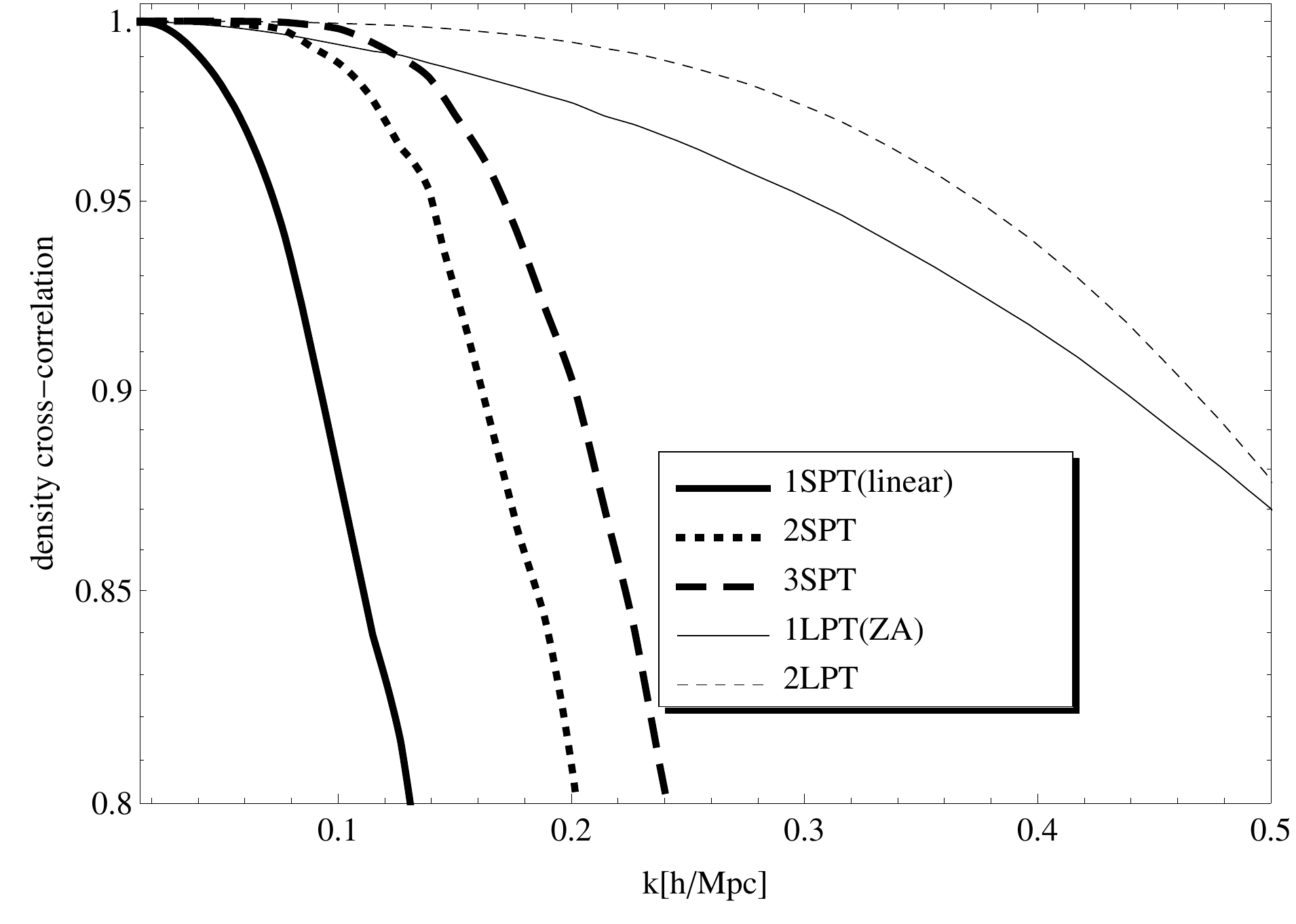}}     
\hfill
\subfloat{\includegraphics[width=0.48\textwidth]{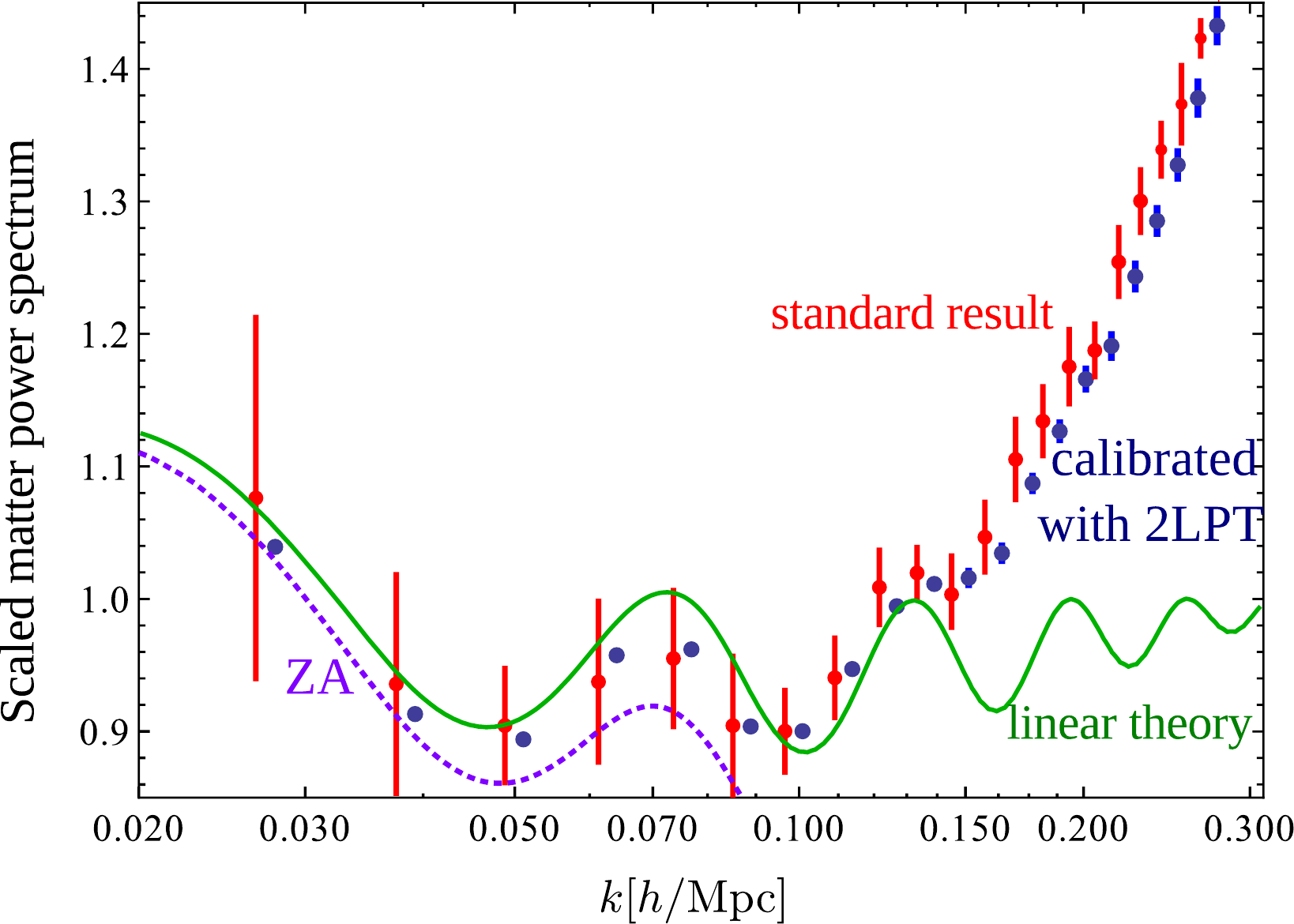}}
\caption{On the left we show the cross-correlation coefficient between the CDM density as obtained from an N-body on the one hand, and various models of LSS on the other. For that, we used the results shown in Figure~\ref{fig:slices}. The cross-correlation is very close to 1 for the ZA (and LPT in general), which implies that LPT captures the phases of short-scale modes much better than SPT. That can be used to calibrate statistics extracted from simulations to reduce their sample variance. An example is shown on the right, adopted from \cite{2012JCAP...04..013T}.} \label{fig:cc}
\end{figure}

Let us be more quantitative, however. In the left panel of Figure~\ref{fig:cc} we show the cross-correlation coefficient\footnote{To remind the reader, a cross-correlation of 1 means that the true and model densities match in phase for a given $\k$, even though their amplitudes may be off. As long as  the phases are captured correctly, the full density field can be easily recovered by fixing the amplitudes with a smooth transfer function dependending on $k$ (but not its direction) as done in \cite{2012JCAP...12..011T,2012JCAP...04..013T} for example.} (at $z=0$) as a function of $k$ between the ``true'' CDM density as calculated from an N-body simulation on the one hand, and the CDM density as calculated at various orders in SPT and LPT  on the other.

One can see that the modified ZA (whether the ZA itself or second order LPT, denoted 2LPT) models perform really well, capturing the phases of the density field well into the non-linear regime ($k\sim0.25h/$Mpc). 
That is a well known result \cite{2012JCAP...04..013T,2012JCAP...12..011T}, used successfully to make cheap mock catalogs \cite{2013MNRAS.428.1036M}, as well as to eliminate sample variance from statistics extracted from simulations \cite{2012JCAP...04..013T} by calibrating with modified ZA models, an illustration of which we show in the right panel of Figure~\ref{fig:cc} \cite{2012JCAP...04..013T}. The same kind of calibration is employed in this paper in Appendix~\ref{app:Npt}.

Going back to the left panel of Figure~\ref{fig:cc}, we see a drastically decaying cross-correlation coefficient for linear theory as already expected. We also confirm that higher-order SPT indeed does a relatively poor job in capturing the effects of the bulk flows when compared with the ZA.

Using a simplified model, in Appendix~\ref{app:cc} we show explicitly that it is indeed the large-scale coherent flows which cause the decaying cross-correlation between the true result and SPT -- a result well known for linear theory (e.g.\cite{2008PhRvD..77b3533C,2012JCAP...04..013T}) but to the best of our knowledge, not for higher order SPT. Moreover, as shown in Appendix~\ref{app:cc}, higher order SPT cannot improve the cross-correlation coefficient between SPT and the truth beyond the rms particle displacement scale $\sim 10$Mpc$/h$ at $z=0$ when the effects of the bulk flows become $\mathcal{O}(1)$ (but see Appendix~\ref{app:cc} where we discuss that the true scale may eventually be the non-linear scale, which unfortunately is still $\sim 10$Mpc$/h$ at $z=0$, albeit slightly smaller). 

How do we reconcile that with the fact that when SPT is consistently expanded to one and the same order in the linear power spectrum, none of its statistics depends on the bulk flows and translation invariance is indeed restored \cite{1996ApJS..105...37S}?

To do that, one has to go back to Figures~\ref{fig:slices} and \ref{fig:cc} and realize that in fact what they tell us is that the ``halos'' are displaced from their correct positions \textit{within} the volume of the simulation box. So, in evaluating Fig.~\ref{fig:slices} we actually calculated by eye higher $n$-point functions coupling the short ``halo'' scales with the large ``box'' scales. In Figure~\ref{fig:cc} we just quantified that coupling. If our simulation box was much smaller, we would not have seen as big of an effect, because rms motions within the volume would have been smaller. 

So, the error in the halo positions\footnote{We use the term ``halo'' very loosely to refer to short-scale modes, while ``halo positions'' refers to the phases of those modes.} in SPT is really given by the rms parte motions within the large-scale patch we simulated in our boxes. And that is a perfectly observable effect even in SPT if one quantified it as the coupling between small- and large-scale modes. In (\ref{SPTfail}) we quantify the error in halo positions in a box of a given size. We confirm that it is given by the  rms motions within the box, unless (for small boxes)  those motions become comparable to the non-linear scale, which is the ultimate lower limit. All of those scales are comparable, and so, we find that the halo position accuracy of SPT is $\sim10$Mpc$/h$ for $z=0$.

The ZA has little problem modelling those rms motions. Moreover, it is able to avoid the non-linear scale limit in predicting halo positions, since non-linear contributions to the displacement are small compared to the linear displacements (see Section~\ref{sec:why}). As linear bulk flows move particles on distances on the order of the non-linear scale, we see that in a sense the ZA separates what is trivially non-linear (the linear bulk motions) from what is truly non-linear (the non-linear density), and captures the trivially non-linear piece. We will see a  manifestation of this when we come to discuss the 2-pt statistics around the BAO peak next.

\subsection{Comparing statistics at the acoustic scale}\label{sec:bao}

From the previous discussion, we can can already anticipate problems for recovering the BAO shape in SPT. If one coherently moves patches of the universe of size roughly given by the acoustic scale ($\sim100$Mpc$/h$), then clearly this will not affect the BAO peak shape. However, motions non-coherent within that scale will definitely distort the peak. So, our ``box'' size, in the terminology of the previous section, is $\sim100$Mpc$/h$, and the accuracy in halo position is bounded by the rms motions inside the box to $\sim 10$Mpc$/h$, comparable with the BAO width of $\sim 20$Mpc$/h$. So, we should definitely start worrying about recovering the BAO shape in SPT.

\begin{figure}[t!]
\centering
    \subfloat{\includegraphics[width=0.48\textwidth]{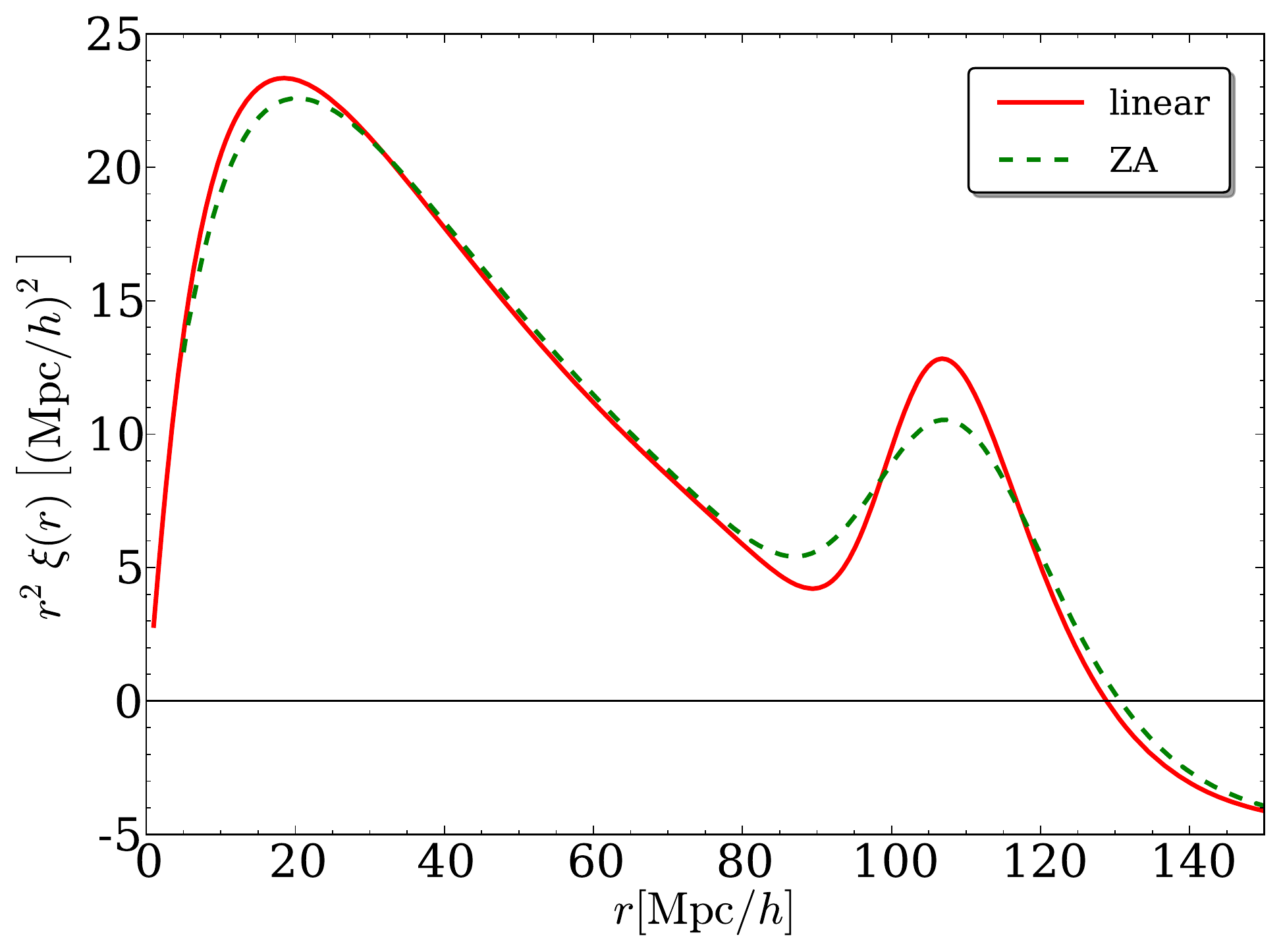}}\hfill
  \subfloat{\includegraphics[width=0.48\textwidth]{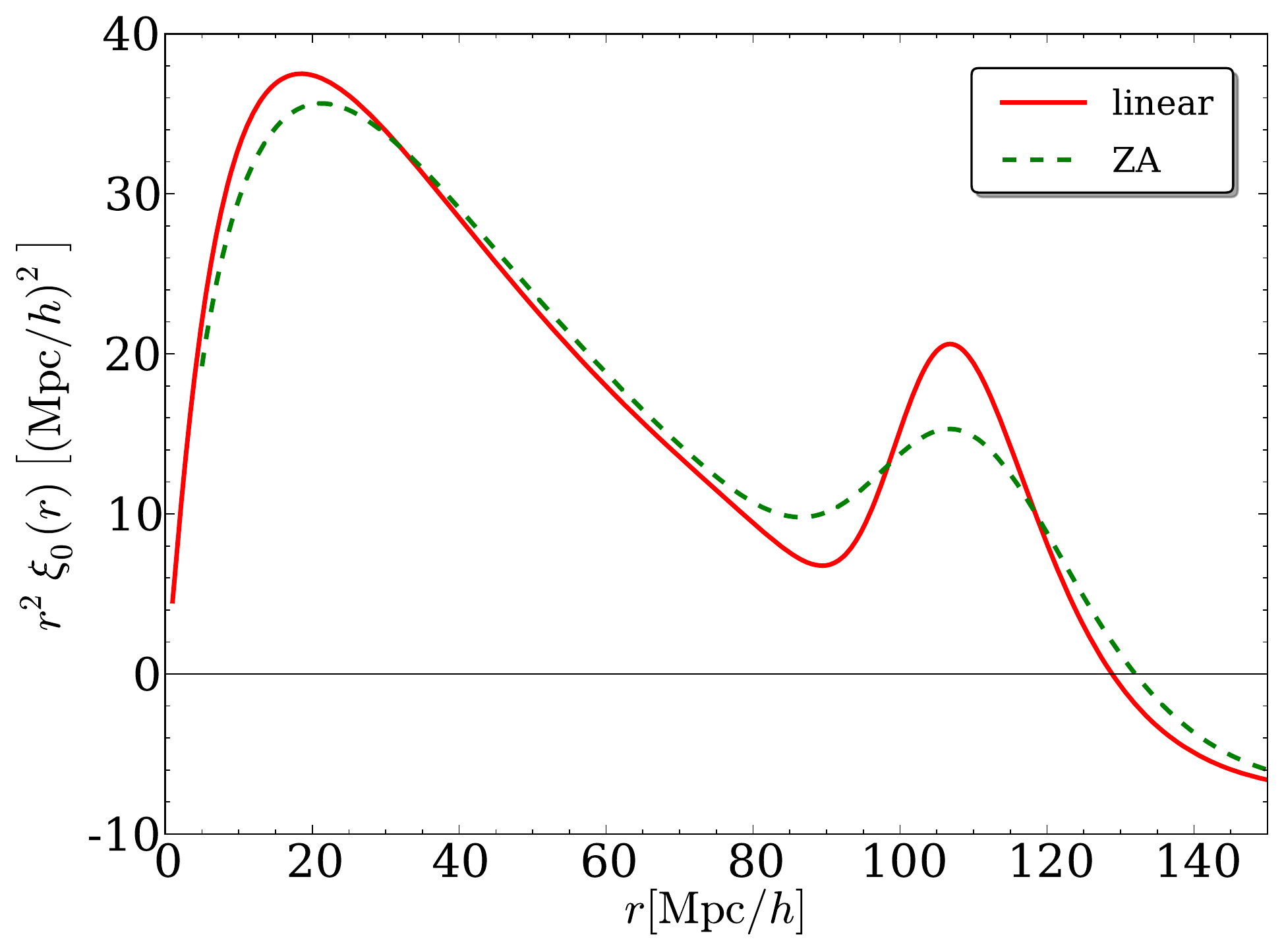}}
\\
    \subfloat{\includegraphics[width=0.48\textwidth]{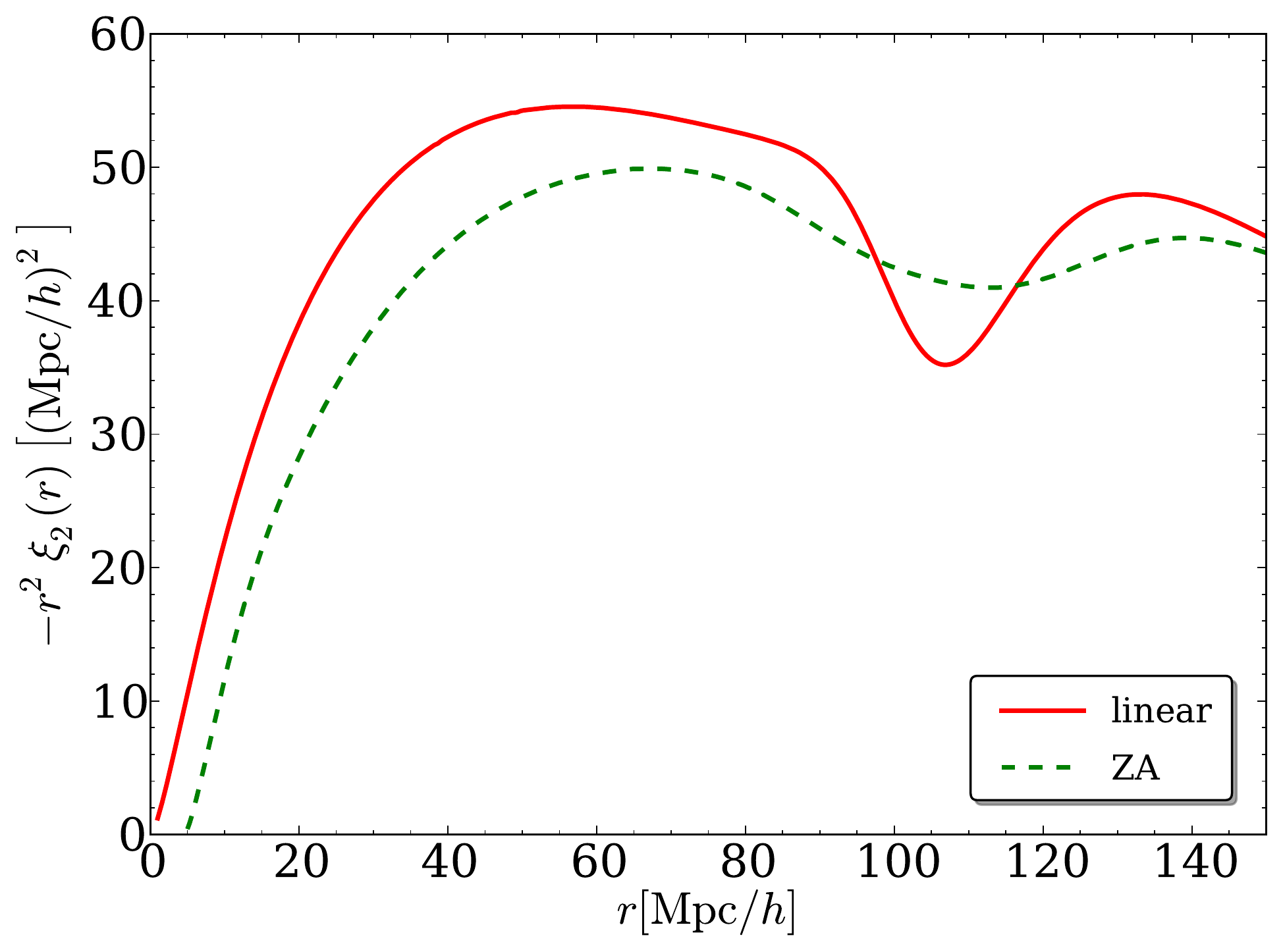}}\hfill
  \subfloat{\includegraphics[width=0.48\textwidth]{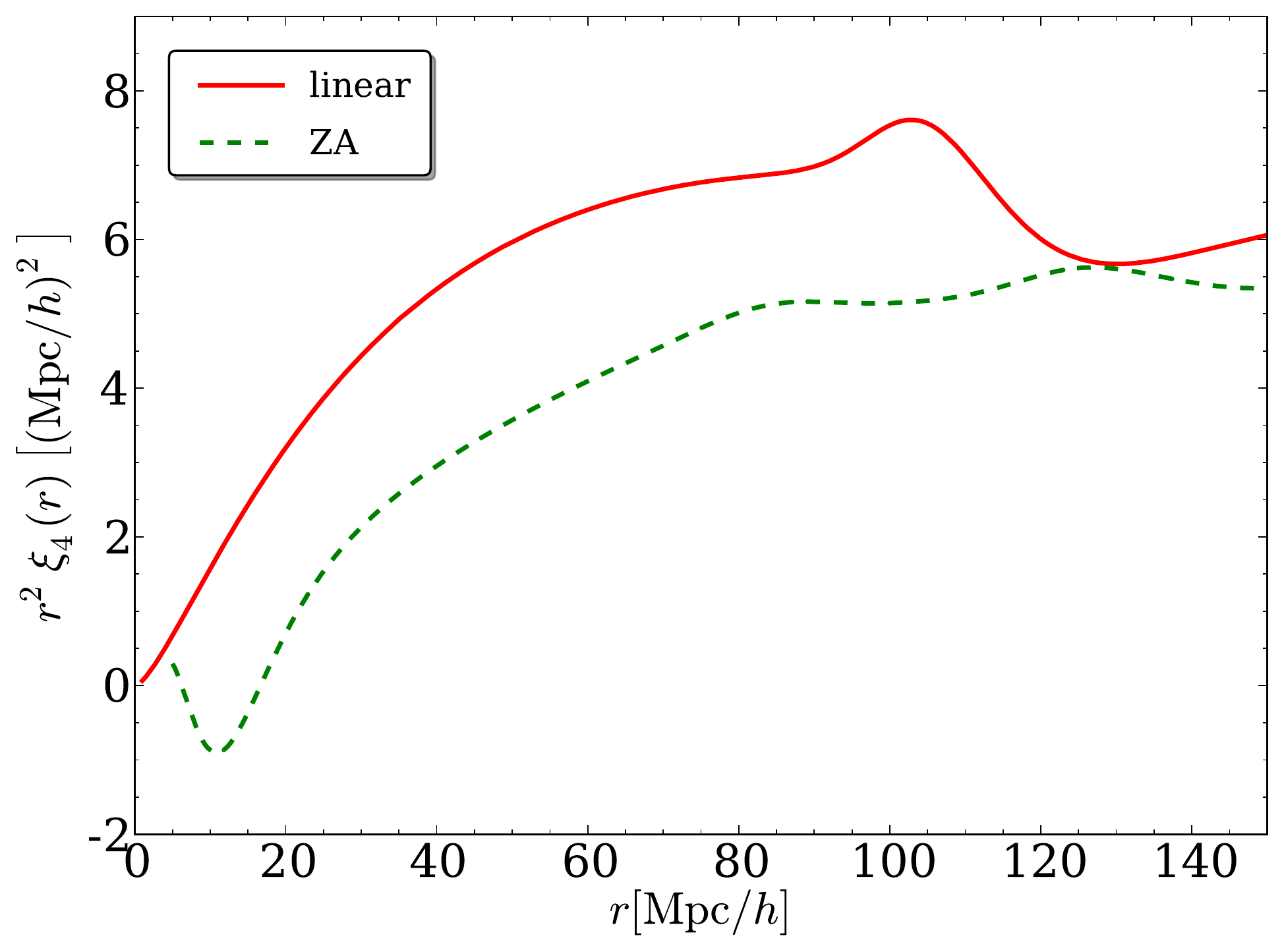}}
\caption{We show the CDM density 2-pt function in real space (upper left), as well as for the redshift space monopole (upper right), quadrupole (lower left) and hexadecapole (lower right) for $z=0.5$ as predicted by Eulerian linear theory and the ZA. Comparing with results in \cite{2013MNRAS.429.1674C}, one can check that the ZA matches N-body results for $r\gtrsim20$Mpc$/h$ for the real space and redshift space monopole and quadrupole; and for $r\gtrsim60$Mpc$/h$ for the redshift space hexadecapole. Eulerian linear theory, however, systematically shows fractional errors of tens of percent especially around the BAO at this redshift.
} \label{fig:2pt}
\end{figure}

Consider Figure~\ref{fig:2pt}, which gives the ZA and linear predictions for the CDM density 2-pt statistics in real space as well as its angle averaged 2-pt monopole, quadrupole and hexadecapole in redshift space. Those results intentionally match the cosmology and redshift ($z=0.5$) of \cite{2013MNRAS.429.1674C} which presents simulation results for those quantities in their Figures 1 and 2. Those authors also show an analytical calculation in a modified ZA perturbation theory, dubbed CLPT \cite{2013MNRAS.429.1674C}, which includes corrections around the full ZA solution. The authors find that their analytical results match the simulation results at separations $r\gtrsim20$Mpc$/h$ for the real space as well as for the redshift space monopole and quadrupole; and match the hexadecapole for  $r\gtrsim60$Mpc$/h$. At the same time linear theory does a surprisingly poor job at matching simulations at those relatively linear scales, especially around the BAO peak, as well as for the whole range of $r$ for the quadrupole. 

One may think their model is so successful because of the higher order corrections to the ZA they included. Yet, overlaying our  results on their plot shows completely negligible difference between our analytical ZA calculation and their higher-order calculation for $r$ where model and N-body match, thus showing those corrections to be irrelevant in explaining the good match with the N-body when compared with CLPT. 

Also note that the fractional difference between linear theory and the ZA (which as described above can be treated as the true result) around the BAO peak is on the order of tens of percent at this redshift. Naively one should have expected a difference of about 0.1 percent as given by the overdensity (as measured by the 2-pt function, $\xi$) at $r\sim100$Mpc$/h$, which is the usual ordering parameter in SPT. This tells us that there is more at play than SPT intuition suggests -- an effect already shown in the simple toy model of \cite{2013arXiv1311.2168P}, and which we will discuss in great detail below.

One may object that the above results are foreshadowed by the  failure of the ZA to get the matter power spectrum right even at relatively low $k$ (see Figure~\ref{fig:cc}). However, in what follows we will see that there is a very good reason why linear theory fails in real space; and that the discrepancy between the ZA power spectrum and the true one is in fact completely within expectations, and that it is simply a matter of chance that linear theory performs better than the ZA in Fourier space. Let us tackle those problems in turn focusing first on real space.



%
%


\section{The matter two-point function as a probability distribution}\label{sec:xi}

As described in the previous section, the fractional difference between the true matter 2-pt function, $\xi$, and its ZA counterpart, $\xi_{\mathrm{ZA}}$, is really small. We need to understand this result better, and more precisely, we need to pinpoint which are the physical parameters which control that difference. Therefore, in this section we proceed to cast the matter 2-pt function in a form that will be suitable for extracting those parameters.

It is impossible to keep the discussion completely devoid of equations, so we have to finally introduce our first one. It describes the trajectories that CDM particles follow. Given a particle's initial position $\q$ (its position in Lagrangian space), its final position $\x$ (its position in Eulerian space) after time $t$ can always be written as:
\be\label{trajMain}
\x(\q,t)=\q+\s(\q,t)
\ee
where $\s$ is the so-called displacement vector field. In the Zel'dovich approximation (see Appendix~\ref{sec:ZA}) the displacement is a Gaussian random variable with zero mean. 

One can show (see for example Appendix~\ref{app:2pt}) that the matter density 2-pt function, $\xi(\Delta \x)$, equals the fractional perturbation to the probability for finding a pair of particles separated by $\Delta \x$. In the notation of (\ref{trajMain}), $\xi$ can also be written (again see Appendix~\ref{app:2pt}) as
\be\label{xiByP}
1+\xi(\Delta \bm{x})=\int \, p(\Delta \bm{x}|\Delta\bm{q})\, d^3\Delta q
\ee
where $p(\Delta\bm{x}|\Delta\bm{q})$ is the conditional probability density for a pair of particles initially separated by a fixed (Lagrangian) distance, $\Delta\bm{q}$, to be separated by $\Delta\bm{x}$ at a later time.  To simplify the notation, from now on we drop the $\Delta$'s from $\x$ and $\q$ unless otherwise specified.

In the ZA, we have (see Appendix~\ref{sec:ZA}) that $p(\x|\q)$ is a simple 3-dimensional Gaussian:
\be\label{pZA}
p_{\mathrm{ZA}}(\x|\q)=\mathcal{N}\left(\q;\langle\Delta s^{\mathrm{ZA}}_i\Delta s^{\mathrm{ZA}}_j\rangle_c\right)
\ee
where $\mathcal{N}(\bm{\mu};\bm{\Sigma})$ is the 3-dimensional normal distribution with a mean vector $\bm{\mu}$ and covariance $\bm{\Sigma}$. 
The covariance in the ZA depends on $q$ (not its direction) and is simply given by the 2-pt function of $\Delta s^{\mathrm{ZA}}_i\equiv s^{\mathrm{ZA}}_i(\q)-s^{\mathrm{ZA}}_i(\bm{0})$. 

\begin{figure}[t!]
\centering
    \subfloat{\includegraphics[width=0.48\textwidth]{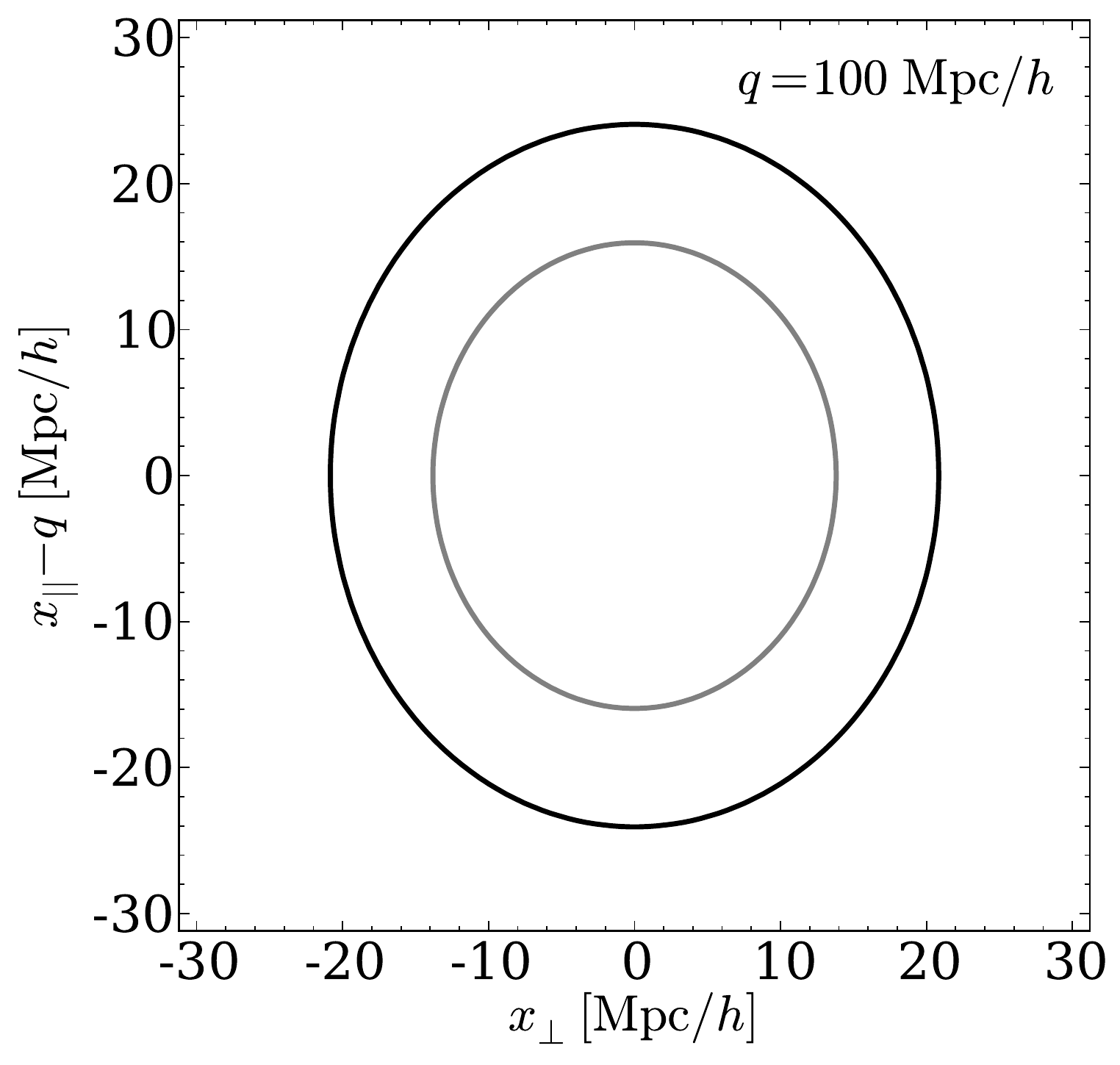}}\hfill
  \subfloat{\includegraphics[width=0.48\textwidth]{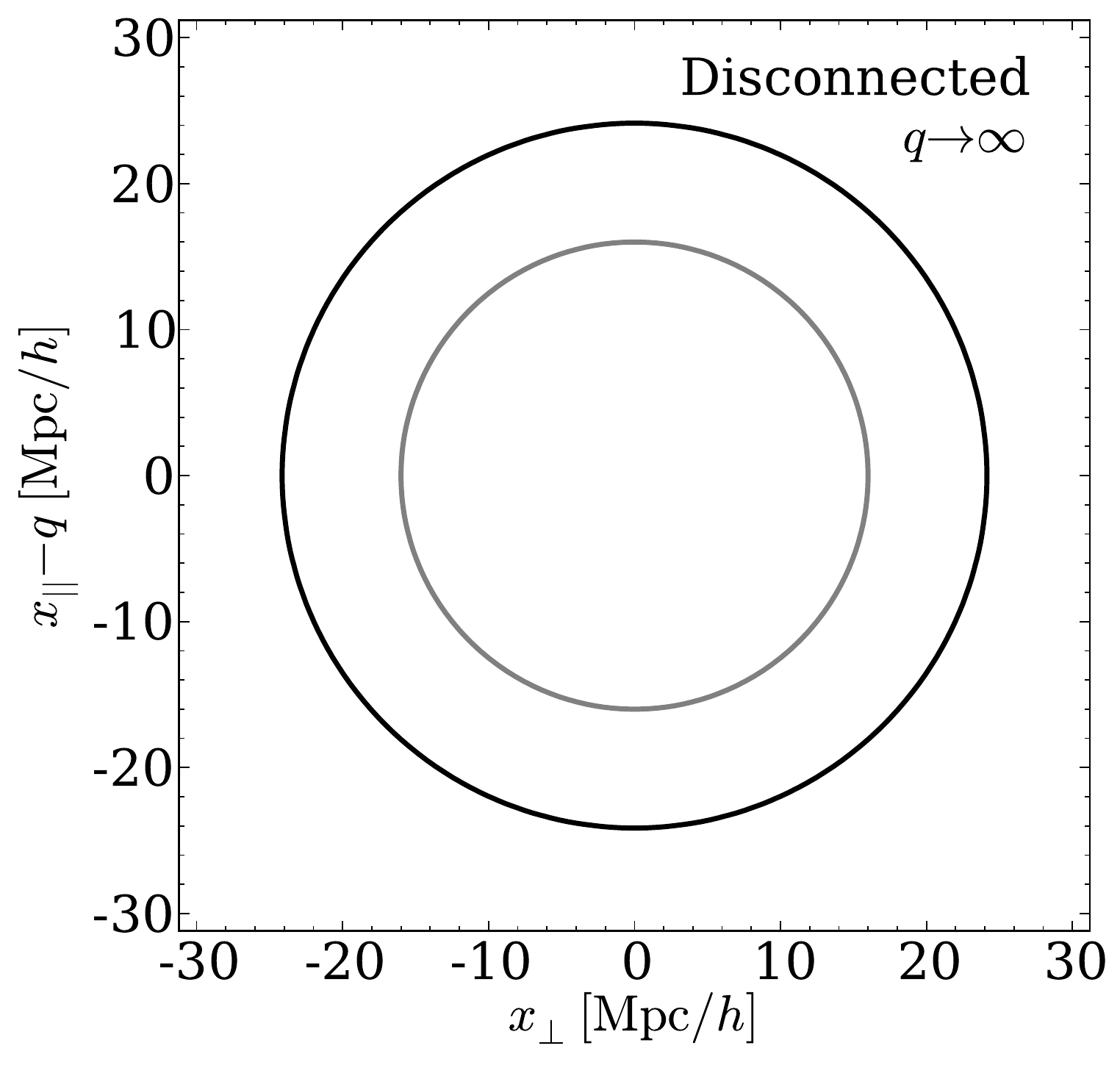}}
\caption{Given an initial particle pair separation $q$ along the $y$-axis, the plots show contours of probability (gray -- 68\%; black -- 95\%) for the final pair separation at $z=0$ both along $\q$ and perpendicular to it.  The  plot is for the ZA. The true probability contours are indistinguishable within the thickness of the lines. }\label{fig:Pqx}
\end{figure}

In Figure~\ref{fig:Pqx} we show contours of equal probability $p_{\mathrm{ZA}}(\bm{x}|\bm{q})$ in $\x$-space for two values of $q$: $q=100$Mpc$/h$ and $q\to\infty$. One can think of the figure as follows: Let us take two particles separated by $q$ and glue the  coordinate system to particle 1, such that particle 1 always sits at coordinates $(0,-q)$. Particle 2 then starts at position $(0,0)$ in Figure~\ref{fig:Pqx} and slowly moves away from the origin with time. At $z=0$, particle 2 will remain inside the gray contour with 68\% probability, while the black contour corresponds to $95\%$ probability. Given that $p_{\mathrm{ZA}}(\bm{x}|\bm{q})$ is Gaussian, the contours are simple ellipses. For an infinitely separated pair (right panel of Figure~\ref{fig:Pqx}), we can see that particle 2 is completely unaware that particle 1 is sitting at $(0,-\infty)$, resulting in isotropic contours. However, once  particle 1 is moved closer, the contours grow anisotropic.

Now let us see what happens to the BAO feature in $\xi$ due to the movement of particle 2. The BAO peak tells us that there is a higher probability that pairs will be separated by $\sim 100$Mpc$/h$ than by nearby distances. From Figure~\ref{fig:Pqx} we see that the pairs that were initially separated by $\sim 100$Mpc$/h$ will diffuse radially by roughly $\pm 10$Mpc$/h$ which results in a smeared BAO feature. Thus, we can write that the piece of $\xi$ containing the BAO feature is given by
\be\label{xiFeature}
\xi_{\mathrm{feature}}(\x)\approx\int d^3q\,  \xi_{\mathrm{feature,linear}}(\q)p(\x|\q)
\ee
where $\xi_{\mathrm{feature,linear}}$ is the linear (i.e. initial) piece of $\xi$ containing the BAO peak, and the true $p$ is being very well approximated by the Gaussian $p_{\mathrm{ZA}}$ (as we discuss below). We derive the above equation in detail in Appendix~\ref{app:length}.

It is important to highlight that (\ref{xiFeature}), as derived in Appendix~\ref{app:length}, explicitly shows that the smoothing of the BAO feature is done by $p$, which depends  on flows which are non-coherent only within the distance scale set by the BAO peak ($\sim100$Mpc$/h$), while flows which coherently move patches of size $\sim100$Mpc$/h$ or larger are completely irrelevant -- a result which is physically very intuitive and which we anticipated at the beginning of Section~\ref{sec:bao}. Yet other work on BAO smoothing has either used an ad hoc smoothing lengthscale (e.g. \cite{2010ApJ...720.1650S}) or has explicitly depended on bulk flows on arbitrarily large scales (e.g. \cite{2009PhRvD..80l3501N}). The latter is unphysical as flows coherent on scales larger than $\sim 100$Mpc$/h$ cannot distort the peak. One should highlight that the above result exactly confirms the intuition of the simple toy model in \cite{2013arXiv1311.2168P}, even though they assume a one dimensional uncorrelated displacement and density fields.

Now let us consider the $p(\bm{x}|\bm{q})$ resulting from the true gravitational dynamics. We can make the same plot as Figure~\ref{fig:Pqx} for the true $p$ obtained from simulations and we will obtain contours which are indistinguishable from the ZA contours, given their line thickness in the plot (based on the numerical results of Appendix~\ref{app:Npt}). So, non-linear evolution changes the ZA result by introducing only small corrections to $p$. Those can affect the covariance of $\Delta s_i$, its 3pt function (i.e. introduce a skewness to $p$),  its 4pt function (i.e. introduce a kurtosis to $p$), as well as higher order $n$-point functions.\footnote{Note that in this case ``$n$-point'' statistics may sound as a misnomer as there are only two points involved and what is changing is which moment of $\Delta s_i$ we evaluate. However, those moments can  be written as $n$-point functions of $\s$ in the limit of distributing the $n$ points within infinitesimal neighborhoods of $\bm{0}$ and $\q$.} It is exactly those corrections we will quantify next.

\section{Expansion parameters I. Analytical considerations}\label{sec:exp1}
In this section we give a schematic derivation of all the parameters controlling the corrections to the ZA, as well as SPT. For those interested beyond the schematics, we give the details in Appendix~\ref{app:Npt}.

\subsection{Expanding around the ZA}\label{sec:expanding}

Motivated by our discussion in the previous section, let us write the true $p(\x|\q)$ schematically as:
\be\label{pSPTexp}
p(\bm{x}|\bm{q})=p_{\mathrm{ZA}}(\bm{x}|\bm{q})\times\bigg[1+l^{-2}\{\mathrm{2pt}_{\mathrm{NL}}\}+l^{-3}\{\mathrm{3pt}\}+l^{-4}\{\mathrm{4pt}\}+ \cdots\bigg]
\ee
where ellipsis correspond to corrections from higher order $n$-pt functions of $\s$.
Above, $\{{\mathrm{2pt}}_{\mathrm{NL}}\}$ corresponds to non-linear corrections to the covariance of $\Delta s_i$; $\{{\mathrm{3pt}}\}$ corresponds to terms due to $\langle(\Delta s)^3\rangle_c$; and  $\{{\mathrm{4pt}}\}$ corresponds to terms due to $\langle(\Delta s)^4\rangle_c$. To make the terms dimensionless we had to introduce a characteristic scale $l$, which we will come back to in Section~\ref{sec:l} where we will see it to be given by the width of the  acoustic peak.

\begin{figure}[t!]
\includegraphics[width=0.8\textwidth]{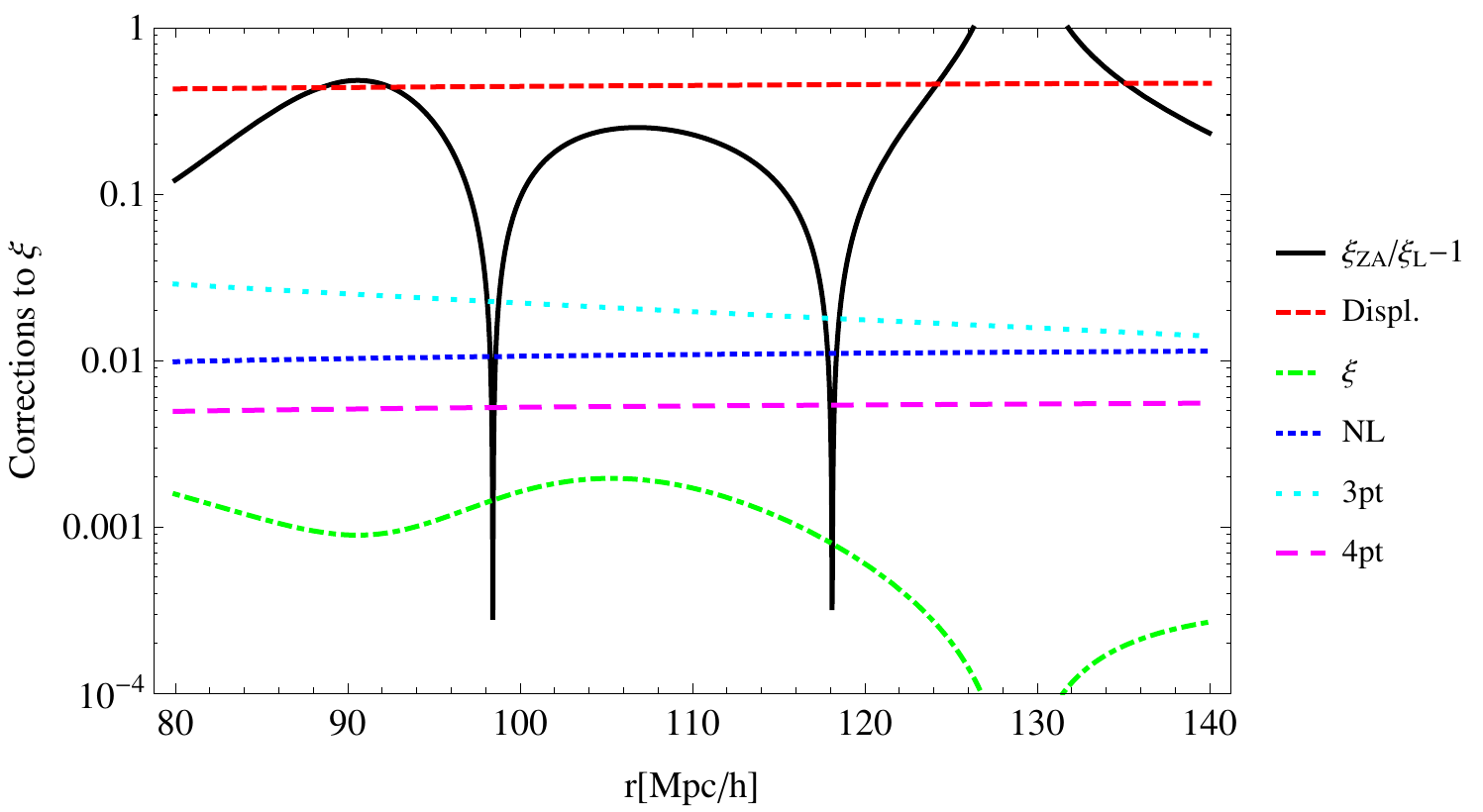}
\caption{Corrections to $\xi$ at $z=0$ from the parameters controling the perturbative expansions in real space around the acoustic scale. Those corrections are practically identical numerically to the expansion parameters themselves. One can see that the naive guess of corrections of order $\xi$ does not materialize. The ZA is controlled by the parameters marked NL, 3pt and 4pt; while SPT features an additional parameter marked by Displ, which matches in size the  fractional difference between the linear theory prediction and the ZA prediction (which one can treat as the truth for these purposes). Note that the ZA parameters are all of order 1\%; while SPT is controled by a parameter of $\mathcal{O}(1)$, rendering the SPT expansion much poorer  at low redshift in real space around  the BAO peak.} \label{fig:correctionsXi}
\end{figure}

 SPT intuition dictates that to lowest order we should have:
\be\label{SPTordering}
l^{-2}\{\mathrm{2pt}_{\mathrm{ZA}}\}\sim \mathcal{O}(\xi)\ &,& \ \ l^{-2}\{{\mathrm{2pt}}_{\mathrm{NL}}\}\sim \mathcal{O}\left(\xi^2\right)\nonumber\\
l^{-3}\{{\mathrm{3pt}}\}\sim \mathcal{O}\left(\xi^2\right)\ &,& \ \ l^{-4}\{{\mathrm{4pt}}\}\sim \mathcal{O}\left(\xi^3\right)
\ee
 with $p_{\mathrm{ZA}}$ controlled by the ZA contribution to the displacement 2-pt function, $l^{-2}\{\mathrm{2pt}_{\mathrm{ZA}}\}$. SPT tells us that all of these terms should be small, the largest being $l^{-2}\{\mathrm{2pt}_{\mathrm{ZA}}\}\sim \xi$, which is $\mathcal{O}(10^{-3})$ at the BAO scale (see the line marked $\xi$ in Figure~\ref{fig:correctionsXi}). Thus, one may complain that (\ref{pSPTexp}) is not a consistent expansion unless we expand $p_{\mathrm{ZA}}$ as well. As we already mentioned in several places, and are still to discuss in great detail, this intuition works well in Fourier space, but is severely violated in real space.

Peaking at the numerical results, (Figure~\ref{fig:correctionsXi}, which we discuss below in detail) around the BAO in real space we see a whole different story than the one that SPT presents. One finds that the term $l^{-2}\{\mathrm{2pt}_{\mathrm{ZA}}\}$ (marked ``Displ.'' in the figure) which enters in  $p_{\mathrm{ZA}}$ is $\mathcal{O}(1)$ at $z=0$ around the BAO peak, which implies that $p_{\mathrm{ZA}}$ cannot be expanded in powers of $l^{-2}\{\mathrm{2pt}_{\mathrm{ZA}}\}$, which is why we kept it unexpanded in (\ref{pSPTexp}). The rest of the terms (the non-ZA terms) turn out to be of similar order, well smaller than $l^{-2}\{\mathrm{2pt}_{\mathrm{ZA}}\}$. Thus, the relevant expansion parameter in real space is not given by $\xi$ (or, equivalently, by the linear power spectrum $P_L$), but there are in fact several expansion parameters\footnote{One may be worried about stream crossing and how it may affect a perturbative 
calculation of the expansion parameters. Note, however, that the the condition for stream crossing $\langle(\Delta \s\cdot \hat q)^2\rangle_c\sim q^2$ is met for $q\sim 2-3$Mpc$/h$ at $z=0$ and is therefore smaller than the NL scale. Thus, any perturbative calculation will stop to be valid long before one reaches scales where stream crossing becomes important. In this paper, however, we circumvent any of these problems by extracting the expansion parameters from simulations.}, given by the terms in the square brackets of (\ref{pSPTexp}). Their leading order corrections to $\xi_\mathrm{ZA}$ are given by their ratios to the lowest order ZA contribution,  $l^{-2}\{\mathrm{2pt}_{\mathrm{ZA}}\}$ (see Appendix~\ref{app:Npt} and eq.~(\ref{EPSILONS}) for the exact definitions):
\be\label{contribsXi}
\mathrm{NL}_{\mathrm{Fig.}\ref{fig:correctionsXi}}&\sim&\frac{\{{\mathrm{2pt}}_{\mathrm{NL}}\}}{\{{\mathrm{2pt}}_{\mathrm{ZA}}\}}\nonumber\\
\mathrm{3pt}_{\mathrm{Fig.}\ref{fig:correctionsXi}}&\sim&l^{-1}\frac{\{\mathrm{3pt}\}}{\{{\mathrm{2pt}}_{\mathrm{ZA}}\}}\nonumber\\
\mathrm{4pt}_{\mathrm{Fig.}\ref{fig:correctionsXi}}&\sim&l^{-2}\frac{\{\mathrm{4pt}\}}{\{{\mathrm{2pt}}_{\mathrm{ZA}}\}}
\ee
The subscripts remind us that those are the quantities shown in Figure~\ref{fig:correctionsXi}. However, we postpone its discussion until Section~\ref{sec:numEps}.
Let us just highlight, however, that since $l^{-2}\{\mathrm{2pt}_{\mathrm{ZA}}\}$ is roughly 1, the expansion parameters themselves are approximately equal to their leading order contributions shown in (\ref{contribsXi}) and Figure~\ref{fig:correctionsXi}.

\subsection{Expanding \`a la SPT}

To reiterate, if we expanded $p(\bm{x}|\bm{q})$ following SPT, we would need to expand $p_{\mathrm{ZA}}$ in powers of $l^{-2}\{\mathrm{2pt}_{\mathrm{ZA}}\}$. Thus SPT features an additional expansion parameter  (see Appendix~\ref{app:Npt} and eq.~(\ref{EPSILONS}) for the exact definition):
\be\label{displEps}
\mathrm{Displ.}_{\mathrm{Fig.}\ref{fig:correctionsXi}}\sim l^{-2}\{\mathrm{2pt}_{\mathrm{ZA}}\}
\ee
with a next-to-leading order correction to $\xi$ of the same magnitude.
This parameter will turn out to be the culprit for the breaking down of SPT around the BAO scale in real space -- exactly confirming the intuition of \cite{2013arXiv1311.2168P}.  But before we see that let us determine the lengthscale $l$. 

\subsection{Features in the 2-pt function}\label{sec:l}

If the matter power spectrum were scale-invariant, there would only be one lengthscale in the problem -- the non-linear scale. Thus, all parameters above would reduce to the ``normal'' SPT result in (\ref{SPTordering}). Then the SPT approach would present an as valid expansion as the ZA, and this paper would not have been written. However, the real world is more complicated. The BAO introduces two additional lengthscales beyond the non-linear scale -- the position and width of the BAO feature. The rms displacements of the particle inside the acoustic scale gives yet another lengthscale. Those lengthscales can be thought of as different $k$ moments of the linear power spectrum, thus encoding its non-scale-invariant shape. Each one of them has the potential to lead to a new ordering parameter, much different from $\xi$. 

Given this plethora of scales, it may not seem straightforward how to choose the relevant $l$. In the full calculation in Appendix~\ref{app:Npt}, from (\ref{xiI}) and (\ref{Iker}), we see that $l^{-1}$ arises in the Fourier transform of $\xi$ and is simply given by $k\sim1/l$. Thus $l$ must be related to the rate of change of $\xi$. And indeed, we know (e.g. Fig.~\ref{fig:2pt}) that all the displacement field really does is smooth  any features in $\xi$ according to (\ref{xiFeature}), while keeping the smooth component of $\xi$ relatively unperturbed. So, around the BAO peak we must have
\be\label{length}
l\sim l_{\hbox{peak width}}\sim20\mathrm{Mpc}/h
\ee

To confirm this, let us concentrate on (\ref{xiFeature}). The presence of $\xi_\mathrm{f}$ in the integrand forces 
$q\sim x_{\hbox{peak position}}\pm l_{\hbox{peak width}}$, 
which in turn is picked up by the nearly Gaussian $p$ to give us the expansion parameter (\ref{displEps}) with $l$ indeed given by (\ref{length}) -- yet again in a nice confirmation of the toy model result of \cite{2013arXiv1311.2168P}.
The above result also nicely confirms our intuition at the beginning of Section~\ref{sec:bao}, where we anticipated that one should compare rms motions within patches of size of the acoustic scale, quantified by $\{\mathrm{2pt}_{\mathrm{ZA}}\}$, with the BAO width. Having established the relevant lengthscale, let us proceed and plot the results.

\section{Expansion parameters II. Numerical results}\label{sec:numEps}

One may want to calculate the expansion parameters above in perturbation theory. Unfortunately, however, they involve expectation values of products of fields evaluated at one and the same point in space (see also Appendix~\ref{app:Npt}). For such short scales perturbation theory breaks down and to proceed, one should renormalize the theory \cite{2013arXiv1311.2168P} by including  corrections, which need to be calibrated using simulations or data. Thus, to simplify our work, after identifying analytically the relevant expansion parameters in the previous section, we proceeded to calculate them numerically as described in detail in Appendix~\ref{app:Npt}. 

To sketch the calculation, we performed N-body simulations (described in that appendix) from which we extracted the quantities $\langle(\Delta \s)^n\rangle_c$ which characterize the deviation of $p(\x|\q)$ from the ZA result. We identified the largest scalar contributions from each of those tensor quantities and used those to make the estimates shown in the rest of this section. For the purposes of these estimates, the uncertainties in those quantities are not important and are omitted from the figures in this section. If one is interested however, one is welcome to look at the  figures in Appendix~\ref{app:Npt} where we do include the errorbars.  Our main results of that appendix are collected in Figures~\ref{fig:correctionsXi} and \ref{fig:correctionsPk}.

\subsection{Real space}

In the previous section, we parametrized the deviation of the matter 2-pt function from the ZA result using a set of parameters listed in (\ref{contribsXi}) (for their exact definitions, see (\ref{EPSILONS}) and (\ref{A2A4})). We calculate those from simulations in Appendix~\ref{app:Npt} and plot them in Figure~\ref{fig:correctionsXi}, using the fact that the Gaussian  $p$ forces us to choose $x\sim q$ for our estimates. We see that all of the corrections to the ZA 2-pt function coming from them are comparable in size and $\mathcal{O}(1\%)$, with the largest being $\lesssim 3\%$. Note that the 4pt function contribution is reassuringly smaller than the 3pt function contribution. This indicates that if one wants to model the BAO to 1\% precision, one may be able to neglect $n$-pt functions above $n\gtrsim 4$.
All of the corrections above are larger than the naive estimate given by $\xi$, but are still small enough that their effects should be perfectly perturbative if one wants to reach that accuracy.

However, one should note that reaching 1\% accuracy would involve calculating at least the NL 2-pt and 3-pt functions. As discussed in the beginning of this section and in Appendix~\ref{app:Npt}, those involve expectation values of products of fields evaluated at zero separation. Given that perturbation theory breaks down at small scales, those quantities receive corrections, which one has to calibrate from simulations, i.e. one must renormalize the theory  \cite{2013arXiv1311.2168P}. Thus, including non-renormalized corrections, such as those in CLPT \cite{2013MNRAS.429.1674C} may not give better predictability than using pure ZA -- a result which will depend on the magnitude of the renormalization corrections.

Going back to Figure~\ref{fig:correctionsXi}, we also see that the  additional parameter given by (\ref{displEps}) controlling SPT (but not the ZA) is $\mathcal{O}(1)$, and is exactly comparable to the fractional difference between linear theory and the ZA (which can be considered the truth for this comparison) -- a nice confirmation of our analysis. Thus, SPT is expected to behave poorly at $z=0$ around the BAO peak. 

One can ask at what redshift does linear theory reach a comparable 3\% accuracy at the BAO, when  SPT results would be perturbative. Given that the  correction (\ref{displEps}) scales as the square of the growth factor, we find that redshift to be $z=4$ -- something which we confirmed by looking at the fractional difference between the linear and ZA $\xi$'s at that redshift. Therefore, even at that redshift, one has to include 1-loop corrections in SPT to achieve a $3\%$ accuracy in the matter 2-pt correlation function.

\subsection{Fourier space}\label{sec:FT}

The results of the previous section are worrying. If there is an $\mathcal{O}(1)$ parameter in real space for SPT, how come we have not seen it in Fourier space -- the space of choice for SPT analysis? We will address this question below. In the process we will confirm that in Fourier space one recovers the usual SPT intuition both for SPT and the ZA, but that comes at a cost of mixing the information about the BAO scale with mildly non-linear modes.

Let us start by writing down the power spectrum in terms of $p(\x|\q)$:
\be\label{PByP}
P(k)=\int \frac{d^3x}{(2\pi)^3}e^{-i\k\cdot\x}\xi(\bm{x})=\int \frac{d^3xd^3q}{(2\pi)^3} e^{-i\k\cdot\x}\, p(\bm{x}|\bm{q})
\ee
with $p$ given by (\ref{pSPTexp}). In order to plot the parameters of the previous section, we need to know the typical sizes of $q$, $r\equiv |\x|$ and $l$ that enter in the above equation ($l$ entering through (\ref{pSPTexp})). In real space, our most complicated task was finding the lengthscale $l$. In Fourier space, our task is simplified tremendously if we actually use the full equations in Appendix~\ref{app:Npt}. Thus, comparing  (\ref{Iker}) with (\ref{pSPTexp}) we immediately see that $l\approx 1/k$. Moreover, the nearly Gaussian kernel $p$ above sets $r\approx q$. So, we are left with estimating $r$.

In other words, our next task is to find the $r$ for which $\xi(r)$ contributes the most to $P(k)$. Simple intuition states that $r\sim 1/k$. And it would be correct. However, let us see this explicitly.

There are two regions in Fourier space that are of interest to us. One is the region around\footnote{The acoustic scale describes a sphere of diameter $2\times 100$Mpc$/h$ (hence a relation $k\sim\pi/(2r)$). One can check numerically that $\xi(r)\approx k^3P(k)/(2\pi^2)$ only if one uses $k\sim\pi/(2r)$. Of course, in the end, factors of 2 such as those are completely irrelevant for our conclusions.} $k\sim\pi/(2\times 100\mathrm{Mpc}/h)$, corresponding to the acoustic scale. That is the scale at which we saw the $\mathcal{O}(1)$ effects in real space. Yet, the BAO oscillations in the power spectrum (see Figure~\ref{fig:cc}) do not complete one cycle until  around  $k\approx0.1h/$Mpc. Thus any study, theoretical or observational, hoping to recover precise information about the peak position needs to go up to at least those $k$ if working in Fourier space. Yet at those $k$, the effects of the width of the BAO peak start to kick in, since $\pi/l_{\hbox{peak width}}\sim 0.1h/$Mpc as well. So, one may wonder whether we will see the same $\mathcal{O}(1)$ parameter as we saw in real space again at those higher $k$.

\begin{figure}[t!]
\centering
  \subfloat{\includegraphics[width=0.48\textwidth]{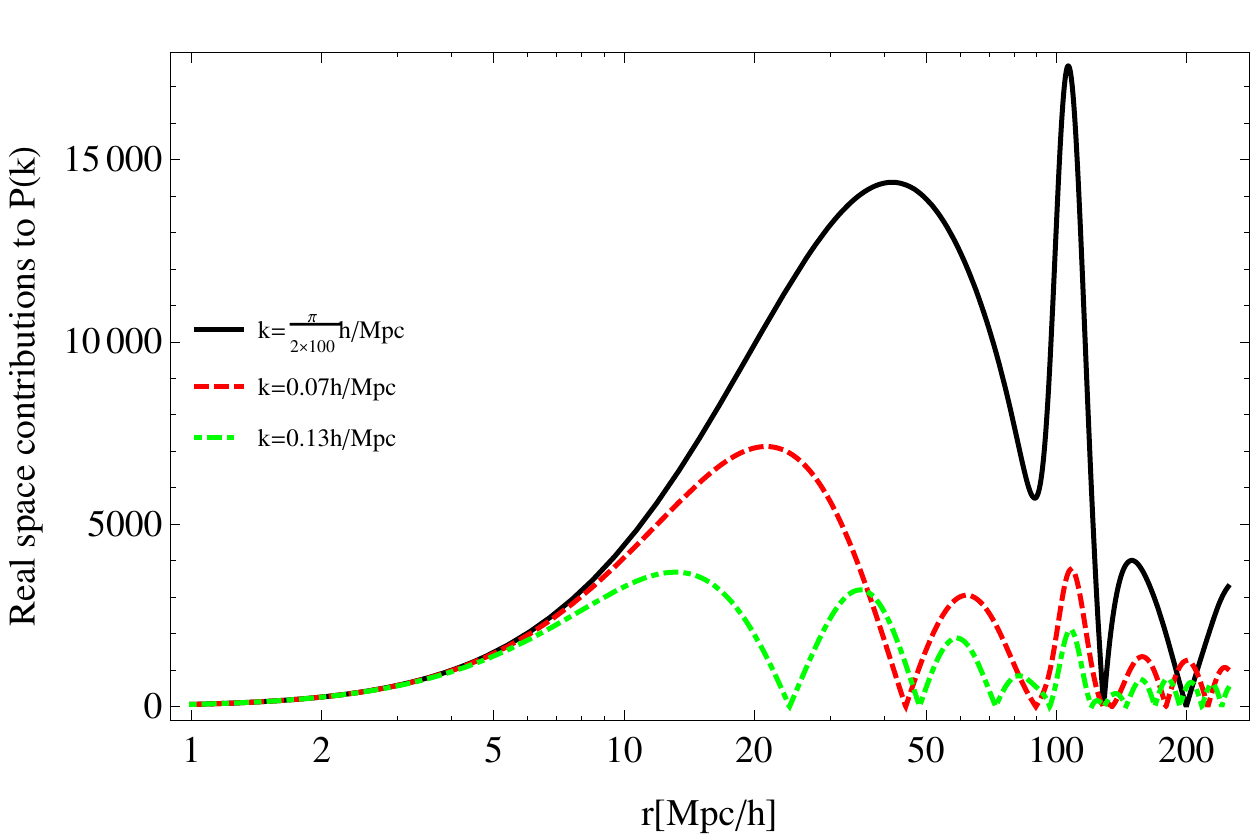}}
\hfill
    \subfloat{\includegraphics[width=0.48\textwidth]{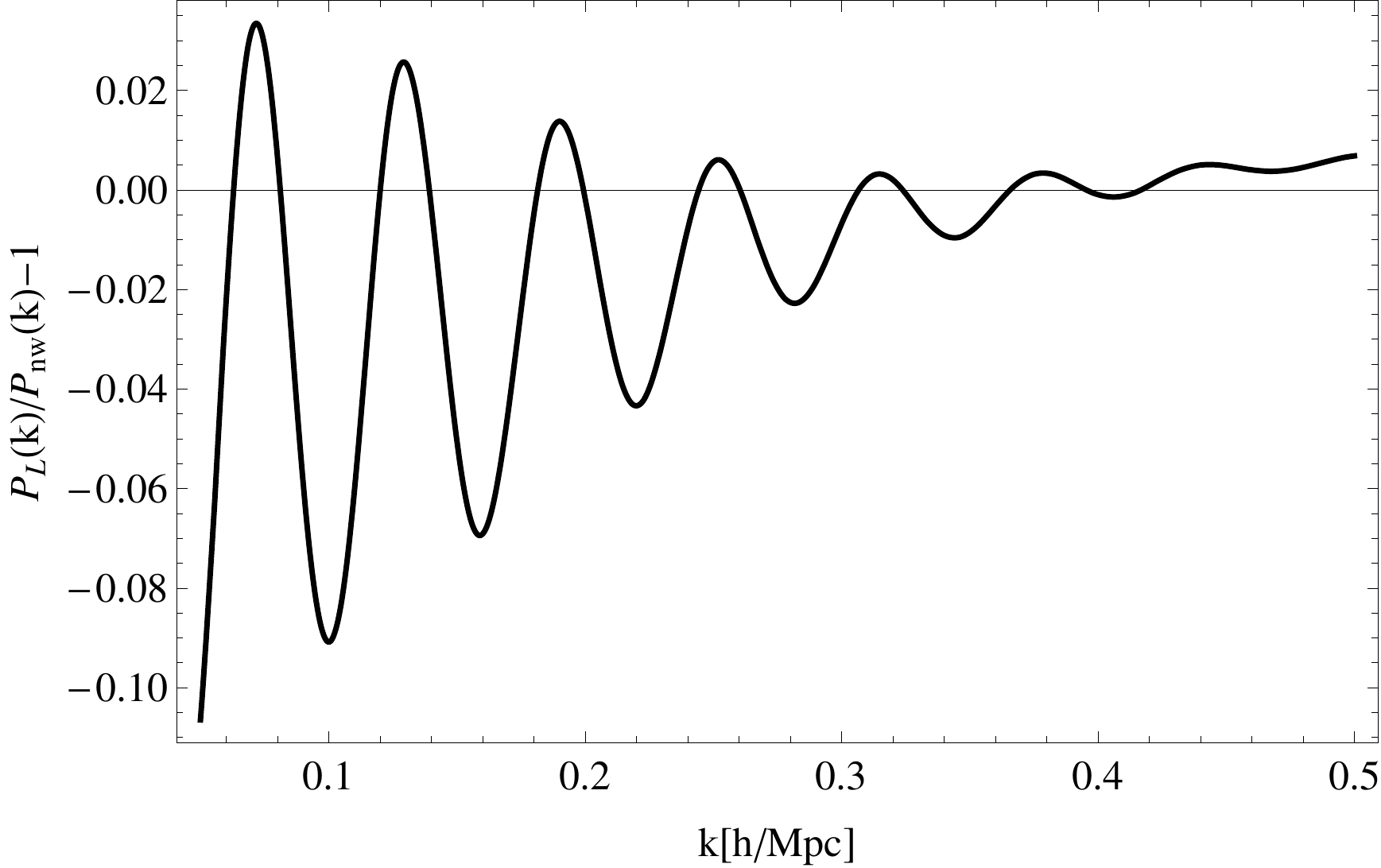}}
\caption{Left: Magnitude of the contributions (per logarithmic $r$-bin) to $P(k)$ at $z=0$ from the different regions of the real space matter 2-pt function. One should note that the acoustic peak (at $r\sim100$Mpc$/h$) is never as important as other scales for all choices of $k$. The values of $k$ are chosen to be: roughly the inverse of the acoustic scale, around the first, and around the second peaks of the acoustic wiggles in $P(k)$ (see right panel of Fig.~\ref{fig:cc}). Right: Fractional change in the linear power spectrum from the presence of the acoustic peak in the matter 2-pt function. One can see that the BAO wiggles contribute only at $\mathcal{O}(5\%)$ of $P(k)$.} \label{fig:contributions_RtoPk}
\end{figure}

In the left panel of Figure~\ref{fig:contributions_RtoPk} we show the magnitude of the real space contributions\footnote{Those are  given by $|4\pi r^3\xi(r)j_0(rk)|$ since $P(k)=\int  \, 4\pi r^3\xi(r)j_0(rk)d\ln r$.}  coming from $\xi$ to the power spectrum, for different values of $k$. Contributions to $P(k)$ from a given logarithmic $r$ range are simply given by the integral under the curve. Thus, we see that for $k\sim\pi/(2\times 100\mathrm{Mpc}/h)$ corresponding to the acoustic scale, the BAO peak brings in only a small contribution. Then, the biggest contribution to $P(k)$ comes from $r\sim1/k$ and, therefore, the BAO width, which was so prominent in our real-space analysis, turns out to be completely irrelevant -- a completely unsurprising result given the low value of $k$.

Now let us turn to the more interesting case of $k\sim 1/l_{\hbox{peak width}}\sim 0.1$Mpc$/h$. To remind the reader,  in real space we found that the peak width was   most important at the acoustic scale $q\sim r\sim 100$Mpc$/h$, which was the combination of $k\sim 1/l$, $q$ and $r$ which gave rise to the order one parameter for SPT in real space. Let us see if that is the case in Fourier space. 

In the left panel of Figure~\ref{fig:contributions_RtoPk} we see that for those higher values of $k$, which probe the width of the peak, the signal from the acoustic peak ($r\sim 100$Mpc$/h$) is in fact swamped\footnote{Note that this did not need to be the case. An example of this would be using a $\xi$, which is zero everywhere except around the BAO peak.} by the signal coming from low $r\sim 1/k$. Still, one may worry about cancellations  occurring due to the oscillations seen in the left panel of Figure~\ref{fig:contributions_RtoPk} (which shows only the magnitude). Therefore, in the right panel of the figure we plot the fractional difference between the linear $P_L(k)$ and a no-wiggle power spectrum, $P_{\mathrm{nw}}(k)$, as calculated using the results of \cite{1998ApJ...496..605E}. We confirm that the BAO signal is swamped by other real-space contributions to $P(k)$, which we already identified to be $r\sim1/k$. These results nicely reproduce the results of the simple toy model of \cite{2013arXiv1311.2168P}.

\begin{figure}[t!]
\includegraphics[width=0.8\textwidth]{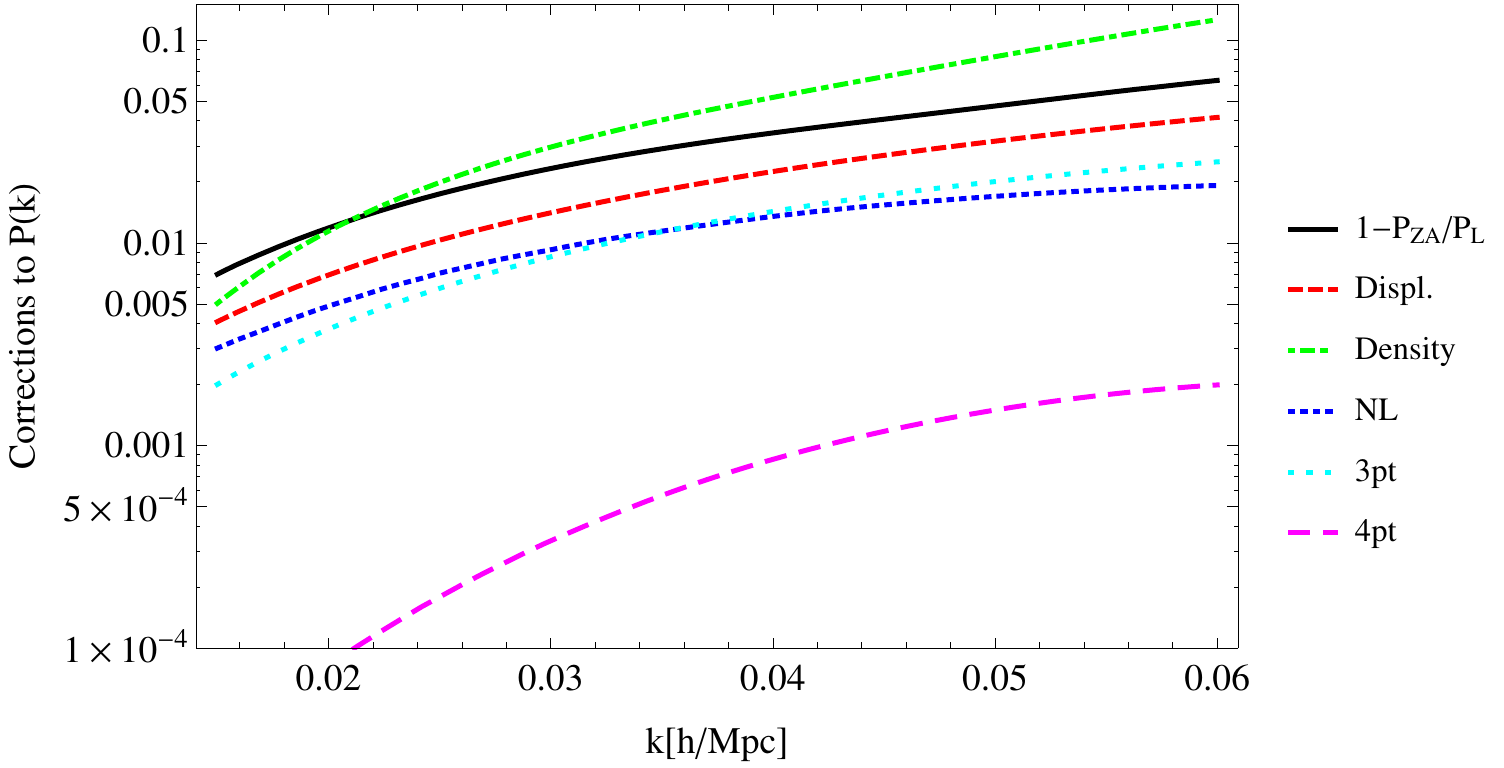}
\caption{
Corrections to $P(k)$ from the parameters controling the perturbative expansions in Fourier space for scales corresponding to 1/acoustic scale. The corrections are of magnitude exactly predicted by ``normal'' SPT intuition both for the ZA and SPT. See the text for further discussion.} \label{fig:correctionsPk}
\end{figure}

So, the prescription for choosing $r$, $q$ and $l$ we end up with in Fourier space  is: $$q\sim r\sim l\sim1/k\ .$$
Using that, in Figure~\ref{fig:correctionsPk} we plot the various  contributions to the power spectrum at acoustic peak scales from the expansion parameters in Section~\ref{sec:expanding}. We also plot the standard overdensity amplitude  per logarithmic $k$-bin $\Delta^2(k)\equiv P(k)k^3/(2\pi^2)$ (denoted ``Density'' in the plot). From the plot, we can see that the SPT intuition given by (\ref{SPTordering}) is  confirmed: the second order fractional contribution from the ZA (``Displ.'') is of the same magnitude as  the lowest order fractional contributions from the NL 2-pt and the 3-pt functions, and in turn those three are comparable with $\Delta^2(k)$; while the 4-pt function (being third order) we see contributes a fractional correction of order $\Delta^4(k)$ to $P(k)$ as expected. Thus, in Fourier space we find no surprises, which explains the long standing usage of SPT in Fourier space to mode the BAO.

Thus, the $\mathcal{O}(1)$ parameter due to BAO we saw in real space fails to materialize in Fourier space.
 But that came at a price. At $k\approx 0.13h/$Mpc, corresponding to the second peak in the power spectrum (Figure~\ref{fig:cc}), the usual ordering parameter, the overdensity amplitude $\Delta^2(k)$, reaches $\mathcal{O}(1)$. That is completely due to small scales physics and has nothing to do with the BAO. So, Fourier space inadvertently mixes the BAO signal with a dominating mildly non-linear component.
 
We still have one issue  to be settled: Why is the ZA so bad at recovering the true power spectrum compared to linear theory as seen in Figure~\ref{fig:cc}. Or is it? Let us look at the fractional difference between $P_{\mathrm{ZA}}(k)$ and $P_{\mathrm{L}}(k)$ (the latter being close to the true result for those $k$) shown in Figure~\ref{fig:correctionsPk}. We see that in fact there is nothing out of the ordinary. The fractional difference between the ZA and linear theory is of the same order as the lowest order corrections between them, which in turn are comparable to $\Delta^2(k)$.    So, the ZA is not behaving unexpectedly (compared to standard SPT intuition) in Fourier space. But is there a sense in which $P_L(k)$ is \textit{too close} to the true result  as seen in Figure~\ref{fig:cc}? The answer is no. The fractional difference between $P_L(k)$ and the truth is still comparable to $\Delta^2(k)$, albeit  with a smaller prefactor. So, we have shown the behavior of the ZA in Fourier space to be a non-issue, captured perfectly with the usual SPT intuition.

\section{Discussion: Why real and Lagrangian?}\label{sec:why}

As we saw in the previous section, Fourier space mixes signal from short scales with the signal from the BAO to the level that non-linear contributions from short scales dominate the BAO signal. These non-linear contributions imply that in Fourier space one has little hope of being certain that their choice of perturbation theory will  be treating the BAO correctly. In contrast, real space cleanly separates the BAO signal from non-linear dynamics at the expense of forcing us to use the ZA, thus working in Lagrangian space, to avoid an $\mathcal{O}(1)$ parameter at $z=0$. 

Despite what the numerics are telling us, these results still may look strange. After all, one can always calculate the fully correct power spectrum in the ZA, and then go to real space and recover the fully correct matter 2-pt function in the ZA. 

However, that is an \textit{inconsistent} procedure for the real world. The reason is that in Fourier space, one should not trust any perturbative results for the real world beyond $k_{\mathrm{NL}}\sim0.25h/$Mpc, as the main contributions there come from mildly non-linear scales, $r\sim 1/k$, as we saw above. So, if one were working  consistently, one should cutoff at that scale in Fourier space. Transforming to real space, one will no longer recover the ZA 2-pt function, but a 2-pt function smoothed by additional $\pi/k_{\mathrm{NL}}\sim 10$Mpc$/h$, which is comparable to the smoothing from the linear bulk motions and the width of the acoustic peak, which in turn destroys all nice properties of the ZA real-space predictions.

But one might then worry that the same exact arguments hold in real space as well. So, for a  model of the real world, one should in fact smooth any real-space predictions at the NL scale of $\pi/k_{\mathrm{NL}}\sim 10$Mpc$h$, \textit{including} the ones from the ZA.
 Indeed, for a generic physical problem this is exactly what one should do. But the gravitational problem is somewhat special. The linear  flows in the universe move particles on the order of $\sim 10$Mpc$/h$ to $z=0$. The non-linear correction to that for a generic physical problem should be $\pi/k_{\mathrm{NL}}\sim 10$Mpc$/h$, i.e. a non-linear correction of the same size as the linear term. Yet, for the gravitational  problem we find that the non-linear contribution to the displacements (our $\{\mathrm{2pt}_{\mathrm{NL}}\}$) is in fact $\sim 1\%$!
 
 Thus, the non-linear contribution to the displacements is \textit{unnaturally} small compared to the linear bulk flows. So, for example, in the ZA center-of-mass positions of halos are therefore easy to recover with a precision much better than the NL scale, which is comparable to the typical linear displacement scale.  But to do that one must take great pains to ensure that one does not mix truly non-linear effects (those coming from the density) with trivially non-linear effects (those coming from the bulk flow). Perturbing  in real space (which is the space in which particles ``stick together'') around the ZA ensures that. Therefore, that should be the  way to go for linear features such as the acoustic peak, in confirmation with our numerical results so far.

\section{Summary}\label{sec:summary}

It is somewhat ironic for the field that the best model (the Zel'dovich approximation) we have so far of the BAO was developed in 1970. In this paper, we made a systematic investigation pinpointing the parameters which regulate its accuracy. We found  the ZA to be accurate to $3\%$ at redshift of $z=0$ in modelling the matter 2-pt function in real space around the acoustic peak. All corrections to the ZA are perfectly perturbative in real space on those scales. However, any attempt to achieve better precision must involve calibration to simulations due to the need to renormalize those corrections.

The Zel'dovich approximation succeeds on the basis of being the lowest order solution for particle trajectories, thus capturing large-scale flows within the acoustic scale, which are dominant in our universe. So, the ZA is inherently a Lagrangian theory. In contrast, theories such as  Eulerian SPT which do not fully preserve the ZA as their solution, receive $\mathcal{O}(1)$  corrections in real space at $z=0$ around the acoustic peak, and are thus performing poorly at low redshift for the BAO. We find that a similar accuracy of $3\%$ for the BAO is achieved by linear SPT only at $z=4$. Thus even when SPT is perturbative, one needs to include loop corrections at $z\lesssim4$.

Real space cleanly separates the BAO signal from non-linear dynamics at the expense of forcing us to use the ZA. In Fourier space, the picture is different. Then ``normal'' SPT intuition is recovered, and SPT and the ZA behave similarly. However, Fourier space mixes signal from short mildly non-linear scales with the linear signal from the BAO to the level that non-linear contributions from short scales dominate the BAO signal. These non-linear contributions imply that in Fourier space one has little hope constructing a systematic theory for the BAO.

Our results give further support to using reconstruction techniques based on the ZA to recover the shape of the acoustic peak \cite{2007ApJ...664..675E,2009PhRvD..79f3523P,2009PhRvD..80l3501N,2010ApJ...720.1650S,2012MNRAS.427.2132P,2012JCAP...10..006T}. Moreover, if our results are confirmed for higher $n$-point functions of the matter density field, this will further  motivate using Gaussian covariance matrices in analysing reconstructed data as done in \cite{2012MNRAS.427.2146X}. 

Given the success of the ZA in capturing physics around the BAO peak, in a companion  paper \cite{ZApaper}  we present an analytic solution for the $n$-point functions in the ZA in real and redshift space, and provide a numerical code to evaluate the matter 2-pt and 3-pt functions.

\appendix
\section{The Zel'dovich approximation}\label{sec:ZA}
This section presents a condensed review of the ZA (see e.g. \cite{ZApaper} for more details). We can always write CDM particle trajectories as
\be\label{traj}
\x(\q,t)=\q+\s(\q,t)
\ee
where  $\q$ is a particle's Lagrangian coordinates (which can be thought of as labeling the particles), $\x$ is its Eulerian coordinates, $t$ is time, and $\s$ is the displacement field. In the Zel'dovich approximation \cite{zeldovich} the displacement field is given by $\s(\q,t)=D(t)\s_0(q)$ ($D$ being the linear growth factor), which is a Gaussian random variable with zero mean. Its variance is given by
\be\label{psi}
\psi_{ij}(\q,t)\equiv\langle s_i(\bm{0},t)s_j(\q,t)\rangle=\int \frac{d^3k}{(2\pi)^3} e^{i\k\cdot(\q)}\frac{k_ik_j}{k^4}P_L(k,t) 
\ee
where $P_L(k,t)$ is the linear power spectrum at time $t$. It is given by $P_L(k,t)=D(t)^2 P_{L}(k,t_0)$, where $t_0$ is defined by $D(t_0)=1$.

After some algebra, the above expression can be written as (we drop the time arguments for brevity)
\be\label{psiEq}
\psi_{ij}(\q)=A^{\mathrm{ZA}}_{01}(q) \delta_{ij}+B^{\mathrm{ZA}}_{01}(q)\hat q_i \hat q_j \ ,
\ee
where
\be\label{ABza}
A^{\mathrm{ZA}}_{01}(q)\equiv \frac{4 \pi}{3}\int \frac{dk}{(2\pi)^3}  P_L(k,t) \left(j_0(q k)+j_2(qk)\right)  , \ \hbox{and} \  
B^{\mathrm{ZA}}_{01}(q)\equiv -4\pi\int \frac{dk}{(2\pi)^3} P_L(k,t) j_2(qk)\nonumber\\
\ee
Note that 
\be\label{psi0}
\psi_{ij}(\q=0)=\delta_{ij}A_{00}^{\mathrm{ZA}}  ,
\ee
 where
\be\label{a00za}
A_{00}^{\mathrm{ZA}}\equiv\frac{4\pi}{3}\int \frac{dk}{(2\pi)^3} P_L(k)
\ee
One should compare the above equations with (\ref{2ptB}).

Given a displacement field (irrespective of whether that is the true $\s$ or $\s$ in the ZA), the density field can be written as
\be\label{density}
1+\delta(x,t)=\int d^3q \d(\x-\q-\s(q,t))
\ee
In the ZA, after some algebra (e.g. \cite{ZApaper}), one can show that 
\be\label{xi}
1+\xi(|\x|)=\langle \delta (\bm{0})\delta (\x)\rangle=\int\frac{d^3q}{(2\pi)^{3/2}}\frac{1}{\sqrt{\det\left[\bm{\Sigma}^{\mathrm{ZA}}_{ij}(q)\right]}}e^{-\frac{1}{2}(x_i-q_i)\left[\bm{\Sigma}^{\mathrm{ZA}}(q)^{-1}\right]_{ij}(x_j-q_j)}
\ee
where 
\be\label{N}
\bm{\Sigma}^{\mathrm{ZA}}_{ij}(\q)\equiv 2\left(A_{00}^{\mathrm{ZA}}\delta_{ij}-\psi_{ij}(\q)\right)
\ee
with $\psi_{ij}$ given by (\ref{psiEq}). Comparing (\ref{xi}) with (\ref{xiByP}) one can read off $p_{\mathrm{ZA}}(\x|\q)$, which is then given in (\ref{pZA}).

\section{Density cross-correlation coefficient between SPT and the real world and halo position accuracy}\label{app:cc}

In this section we estimate the parameter controlling the decay of the density cross-correlation between the ``true'' result and the SPT results. Immediately we encounter a problem, however. We do not have an equation for the true density in order to bracket it with the density in SPT. 

However, remember that the ZA is extremely well correlated with the true result, at least compared to SPT  (\cite{2012JCAP...04..013T}; and Figure~\ref{fig:cc}). So,  for the purposes of this estimate we can write with a good accuracy that
\be
\delta_{\mathrm{true}}(\k,t)\approx R(k,t)\delta_{\mathrm{ZA}}(\k,t)
\ee
with a transfer function $R\sim1$ \cite{2012JCAP...04..013T}. Clearly,  we can expand $\delta_{\mathrm{true}}(\k)$ above in powers of the linear overdensity, $\delta_L$, to obtain a good approximation of the SPT results as well. 

So, following the same calculation as the one in Appendix~\ref{sec:ZA}, we find that the cross-correlation coefficient between the true result and the result up to $n$-th order in SPT ($n$SPT) is given by
\be\label{ccEq}
r_{\mathrm{true,}n \mathrm{SPT}}(k)&\equiv&\frac{\langle\delta_{\mathrm{true}}\delta_{n\mathrm{SPT}}\rangle}{\sqrt{\langle\delta_{\mathrm{true}}\delta_{\mathrm{true}}\rangle\langle\delta_{{n\mathrm{SPT}}}\delta_{n\mathrm{SPT}}\rangle}}\\
&\approx& R^2(k)\frac{e^{-\frac{A_{00}^\mathrm{ZA} k^2}{2}}}{\sqrt{P_\mathrm{true}P_{n\mathrm{SPT}}}}\mathrm{T}_{\lambda,n}
\int\frac{d^3q}{(2\pi)^3}
e^{i\k\cdot\q}
e^{-\lambda^2\frac{A_{00}^\mathrm{ZA} k^2}{2}}
\left[e^{\lambda k_ik_j\psi_{ij}(q)}-1
\right]\nonumber 
\ee
where $P_\mathrm{true}$ and $P_{n\mathrm{SPT}}$ are the power spectra of the true and SPT results. $\mathrm{T}_{\lambda,n}$ in front of the integral reminds us that we have to Taylor expand the integral with respect to $\lambda$ to $n$-th order. 

As an example, to first order we find
\be
r_{\mathrm{true,}\mathrm{1SPT}}(k)\approx R(k)\sqrt{\frac{P_{\mathrm{1SPT}}}{P_{\mathrm{true}}}} e^{-\frac{A_{00}^\mathrm{ZA} k^2}{2}}\sim e^{-\frac{A_{00}^\mathrm{ZA} k^2}{2}}
\ee
In the second approximate equality we used the fact that the dominant decay comes from the exponent. Clearly then, the Taylor expansion removes the possibility of exact cancellation of the bulk flows from the cross-correlation coefficient. From (\ref{ccEq}) we see that the decay in the cross-correlation coefficient between the true and SPT result at any (low) order  will then be dominated by the exponent $A_{00}^{\mathrm{ZA}}k^2/2$, which is entirely due to large-scale coherent flows (see e.g. \cite{2012JCAP...04..013T}; and Appendix~\ref{app:Npt} as well), with the higher orders slowly trying to remedy this decay by resumming $A_{00}^\mathrm{ZA}$. 

So, using low-order SPT one makes an $\mathcal{O}(1)$ mistake at the scale where the exponent is 1, i.e. at 
$k= \sqrt{2/A_{00}^\mathrm{ZA}}\approx0.23h/$Mpc, corresponding to a lengthscale $\pi/k\approx13$Mpc$/h$ at $z=0$. If our universe lived in a box of size $L$, such as in a simulation volume, then 
modes with $k<2\pi/L$ will be set to zero. Moving the lower integration limit in (\ref{a00za}) accordingly, we find that the  lengthscale where low-order SPT will decorrelate with the truth is:
\be\label{SPTfail}
l_{\mathrm{lowSPT}}^2\sim \max\left[\frac{1}{12}\int\limits_{2\pi/L}^\infty dk P_L(k)\ , \ \ \left(\frac{\pi}{k_{\mathrm{NL}}}\right)^2\right]
\ee
So, $l_{\mathrm{lowSPT}}$ is proportional to the particle rms displacements inside the box, unless one hits the non-linear scale, $k_{\mathrm{NL}}$, which we discuss below. When the cross-correlation coefficient is very different from 1, short-scale phases are not captured correctly, and so halo positions cannot be recovered. Thus, $l_{\mathrm{lowSPT}}$ also gives the accuracy with which low-order SPT predicts halo positions. 

The exponential decay in (\ref{ccEq}) can in principle be captured by including higher and higher orders in SPT as the exponential has an infinite radius of convergence. That is true in the simplified model we presented above. Whether it holds for the real universe remains to be seen. Let us be optimistic, however, and see what that implies for high order SPT.

If one wants better halo position accuracy, in Fourier space that implies that one needs to model higher and higher $k$. Eventually, for $k\gtrsim k_{\mathrm{NL}}$ the terms from the exponential decay will be added to truly non-linear terms which perturbation theory will fail to capture. Going back to real space, we are thus always bounded by $\pi/k_{\mathrm{NL}}$. Thus, the NL scale puts an ultimate  limit on the accuracy with which we can recover halo positions for theories which work in Fourier space. To write it out explicitly, we have that even high-order SPT (in the optimistic case that the exponential decay can be captured by it) has a bound on the accuracy with which it predicts halo positions. And that is given by $l_{\mathrm{highSPT}}$:
\be
l_{\mathrm{highSPT}}\sim \frac{\pi}{k_{\mathrm{NL}}}
\ee
For SPT, that scale is $l_{\mathrm{highSPT}}\approx10$Mpc$/h$ at $z=0$, numerically comparable with the rms displacements in the universe.  Thus, SPT halo accuracy can never be better than $\sim10$Mpc$/h$ at $z=0$.

The same analysis above can be repeated for EEFT. Thus, we expect that low-order EEFT will get halo positions with the same accuracy as low-order SPT. Indeed, it is rather unlikely that EEFT can fix problems related to bulk flows, as introducing effective fluid parameters such as sound speed and viscosity cannot possibly compensate for large bulk motions. 
That is unless, those fluid parameters fix the short-scale convergence of the theory, and one is then able to predict the non-bulk-flow contributions at higher $k$, allowing one to separate them from the bulk-flow contributions. 

There are indeed indications that the non-linear scale at $z=0$ in EEFT is higher, maybe as high as $k_{\mathrm{NL}}\approx 0.6h/$Mpc \cite{2013arXiv1310.0464C}. In that case, high-order EEFT may be able to achieve a halo position accuracy of $l_{\mathrm{highEEFT}}\approx5$Mpc$/h$, in the optimistic case that the exponential above can be captured by high order EEFT for the real world. Still, we use that optimistic value in Table~\ref{table}.  

As we see in the main text, the convergence in real space  around the BAO peak of any theory not preserving the ZA (EEFT is such a theory) is  governed by bulk flows -- by the ability to get  halo positions accurate within patches of size of the acoustic scale. More precisely, the parameter controlling the convergence is given by the square of the ratio between that accuracy and the width of the BAO peak. For low-order EEFT,  this ratio will still be  $\mathcal{O}(1)$ as for SPT. If the higher value of $k_{\mathrm{NL}}$ in EEFT is confirmed, however, ultimately higher-order EEFT may achieve a parameter of $\mathcal{O}(0.1)$, which may allow one to recover the acoustic peak shape at high order. Given the uncertainty, however, we assign a ``no'' to it for the  category regarding real-space convergence in Table~\ref{table}.

The ZA does not expand in the bulk flows. But one may wonder how come the ZA is not bounded by $\pi/k_{\mathrm{NL}}$ in its halo position accuracy.  We discuss this in Section~\ref{sec:why}.

\section{Expressing the density 2-pt function using the probability for pair separations}\label{app:2pt}
In this section we remind the reader why the two point function, $\xi$, is given by
\be\label{xiIntQapp}
1+\xi(\bm{x})=\int d^3q\, \tilde p(\bm{x}|\bm{q})
\ee
where $\tilde p(\bm{x}|\bm{q})$ is the conditional probability density for a pair initially separated by a fixed distance, $\bm{q}$, to be separated by $\bm{x}$ at a later time. The tilde is there to remind us of the fact that we still have to show that this is true.

Consider two volume elements (both of volume $d V$) separated by $\bm{x}$ and containing $N_1$ and $N_2$ particles, respectively. Then the number of pairs separated by $\bm{x}$ (with a precision given by the size of $d V$) is $N_1N_2$. The total number of pairs separated by any distance, while keeping volume $1$ pinned to the origin, is $N_1N_{\mathrm{tot}}$, where $N_{\mathrm{tot}}$ is the total (possibly infinite) number of particles. Then the probability, $P_{\bm x}(\bm{x})$, of having two particles separated by $\bm{x}$ is simply 
\be
P_{\bm x}(\bm{x})=\frac{\langle N_1N_2\rangle}{\langle N_1N_{\mathrm{tot}}\rangle}
\ee
We can rewrite the above using the local fractional overdensities $\delta_i$ ($i=1,2$). Then, $N_i=d V \bar n(1+\delta_i)$, where $\bar n$ is the average number density, and we obtain:
\be
P_{\bm x}(\bm{x})=\big(1+\xi(\bm{x})\big)\frac{d V}{V}
\ee
where $V$ is the total volume of the universe (or survey, if one wants to keep it finite). 

Note that the probability density, $p_{\bm x}(\bm{x})$, of having a pair of particles separated by $\bm{x}$ is given by $P_{\bm x}(\bm{x})=p_{\bm x}(\bm{x})dV$, which implies
\be\label{fractionalP}
p_{\bm x}(\bm{x})=\big(1+\xi(\bm{x})\big)\frac{1}{V}=\big(1+\xi(\bm{x})\big)p_{\bm q}
\ee 
where we used the fact that at early times (when $\bm{x}=\bm{q}$) we have $|\xi|\ll 1$, and thus the first equality above gives us the trivial statement 
\be
p_{\bm q}(\bm{q})=1/V=\mathrm{constant \ ,}
\ee
which is the probability density for having a pair initially separated by $\bm{q}$. From (\ref{fractionalP})
we can see that $\xi$ equals the fractional perturbation to the probability for finding a pair of particles separated by $\x$ -- a fact which is used extensively in building estimators of $\xi$ by pair counting (e.g. \cite{1993ApJ...412...64L}).

Using (\ref{xiIntQapp}), we end up with
\be\label{pints}
p_{\bm x}(\bm{x})=\int d^3q\, \tilde p(\bm{x}|\bm{q})p_{\bm{q}}=\int d^3q\, p_{\bm{q},\bm{x}}(\bm{q},\bm{x})
\ee 
where in the last equality we expressed $p_{\bm{x}}$ using the joint probability density $p_{\bm{q},\bm{x}}$. Note that we can equate the two integrands appearing in (\ref{pints}), since we could constrain the whole analysis to include only pairs of certain initial separations when calculating $p_{\bm{x}}$. Thus, $\tilde p(\bm{x}|\bm{q})$ is exactly the conditional probability density $p(\bm{x}|\bm{q})$.

\section{Displacement $n$-point functions and the matter 2-pt function}\label{app:Npt}

Starting with (\ref{traj}) and (\ref{density}), one can show that the true non-linear density 2-pt function can always be written as (e.g. \cite{2013arXiv1311.2168P})
\be\label{xiI}
1+\xi(\bm{x})&=&\int d^3k e^{-i\k\cdot\bm{x}}\int\frac{d^3q}{(2\pi)^3}
e^{i\k\cdot\q}(1+\mathcal{I}(\k,\q))
\ee
with $1+\mathcal{I}(\k,\q)$ being the Fourier transform of $p(\bm{x}|\q)$ with respect to $\x$:
\be\label{Iker}
1&+&\mathcal{I}(\k,\q)\equiv\\
&\equiv&
\exp\left[-\frac{1}{2!}k_ik_j\langle\Delta s_i\Delta s_j\rangle_c-\frac{i}{3!}k_ik_jk_k\langle\Delta s_i\Delta s_j\Delta s_k\rangle_c+\frac{1}{4!}k_ik_jk_kk_l\langle\Delta s_i\Delta s_j\Delta s_k\Delta s_l\rangle_c+\cdots\right]\nonumber
\ee
in the notation of Appendix~\ref{sec:ZA}, with $\Delta \s\equiv \s(\q)-\s(\bm{0})$. Above, ellipses represent higher order $n$-point functions.  Expanding the exponent we get
\be\label{xiContribsExpanded}
\mathcal{I}(\k,\q)&\approx&\left[
-\frac{k_ik_j}{2}\langle\Delta s_i\Delta s_j\rangle_c\right]+\\\nonumber
&+&\left[
-i\frac{k_ik_jk_k}{6}\langle\Delta s_i\Delta s_j\Delta s_k\rangle_c+\frac{k_ik_jk_kk_l}{8}\langle\Delta s_i\Delta s_j\rangle_c\langle\Delta s_k\Delta s_l\rangle_c\right]+
\\\nonumber
&+&\left[
\frac{k_ik_jk_kk_l}{24}\langle\Delta s_i\Delta s_j\Delta s_k\Delta s_l\rangle_c-\frac{k_ik_jk_kk_lk_mk_n}{48}
\langle\Delta s_i\Delta s_j\rangle_c\langle\Delta s_k\Delta s_l\rangle_c\langle\Delta s_m\Delta s_n\rangle_c\right.\\\nonumber
&&\ \ \ + \ \left.
i\frac{k_ik_jk_lk_mk_n}{12}\langle\Delta s_i\Delta s_j\rangle_c\langle\Delta s_k\Delta s_l\Delta s_m\rangle_c
\right]\\\nonumber
&+&\cdots
\ee
Above we grouped the terms by following SPT and thus combined terms of the same order in $P_L$ in square brackets, the first line being first order in $P_L$, the second line being second order, and the third and forth lines being third order in $P_L$. Note that we did not split $\langle\Delta s_i\Delta s_j\rangle_c$ explicitly into a ZA and a non-linear piece above to avoid cluttering the equation. But one should keep in mind that in perturbation theory, to lowest non-zero order the non-linear piece is second order in $P_L$. 

As discussed in the main text, the above ordering is correct in Fourier space, but the relevant expansion parameter in real space is not given by $P_L$. Instead, there are several parameters, given by the non-ZA contributions to $\xi$ (see (\ref{pSPTexp})). We proceed to evaluate them by first looking in more detail at the displacement cumulants.

The displacement cumulants have the following tensorial structure obtained by using the symmetries of the problem: 
\be\label{DeltaCs}
 \langle \Delta s_i  \Delta s_j    \rangle_{\mathrm{c}} &=&A_2 \delta_{ij}   + B_2 {\hat q}_i{\hat q}_j\\
 \langle \Delta s_i  \Delta s_j  \Delta s_k   \rangle_{\mathrm{c}} &=&A_3 \delta_{ij} {\hat q}_k + \mathrm{(2\ perm)} + B_3 {\hat q}_i{\hat q}_j{\hat q}_k\nonumber\\
\langle \Delta s_i  \Delta s_j  \Delta s_k  \Delta s_l  \rangle_{\mathrm{c}} &=& A_4 \delta_{ij}\delta_{kl} + \mathrm{(2\ perm)} +B_4 {\hat q}_i {\hat q}_j\delta_{kl} + \mathrm{(5\ perm)} + C_4 {\hat q}_i{\hat q}_j{\hat q}_k{\hat q}_l\nonumber
\ee
where the coefficients, $A_i,\ B_i,\ C_4$ are $q$-dependent as well as time-dependent. We can of course plug back in $\Delta s_i=s_i(\q)-s_i(\bm{0})$. Then one can see that the above cumulants depend on a new set of $n$-pt expectation values, which for the 2-pt statistics are (compare with (\ref{ABza}) and (\ref{a00za})):
\be
\langle\label{2ptB}
s_i(\bm{0})s_j(\bm{0})\rangle_c&=&
A_{00} \delta_{ij}\\\nonumber
\langle
s_i(\bm{0})s_j(\bm{q})\rangle_c&=&
A_{01} \delta_{ij}+B_{01}\hat q_i\hat q_j\ \ ;
\ee
for the 3-pt statistics:
\be
\langle
s_i(\bm{0})s_j(\bm{0})s_k(\bm{q})\rangle_c&=&
A_{001}\hat q_i\hat q_j\hat q_k+
B_{001}\bigg(\hat q_i\delta_{jk}+\hat q_j\delta_{ik}\bigg)+
C_{001}\hat q_k\delta_{ij}
\nonumber\ \ ;
\ee
and for the 4-pt statistics:
\be
\langle s_i(\bm{0})s_j(\bm{0})s_k(\bm{0})s_l(\bm{0})  \rangle_{\mathrm{c}} &=& A_{0000} \delta_{ij}\delta_{kl} + \mathrm{(2\ perm)}\\
\langle s_i(\bm{0})s_j(\bm{0})s_k(\bm{q})s_l(\bm{q})\rangle_c&=&A_{0011}\delta_{ij}\delta_{kl}+
B_{0011}\bigg(\delta_{il}\delta_{jk}+\delta_{ik}\delta_{jl}\bigg)+
C_{0011}\bigg(\delta_{ij}\hat q_k\hat q_l+\delta_{kl}\hat q_i\hat q_j\bigg)+\nonumber\\
&&D_{0011}\bigg(\delta_{ik}\hat q_j\hat q_l+\delta_{il}\hat q_j\hat q_k+
\delta_{jk}\hat q_i\hat q_l+\delta_{jl}\hat q_i\hat q_k\bigg)+
E_{0011}\bigg(\hat q_i\hat q_j\hat q_k\hat q_l\bigg)\nonumber\\
\langle
s_i(\bm{0})s_j(\bm{q})s_k(\bm{q})s_l(\bm{q})\rangle_c&=&
A_{0111}\bigg(\hat q_i\hat q_j\delta_{kl}+\hat q_i\hat q_k\delta_{jl}+\hat q_i\hat q_l\delta_{jk}\bigg)+
B_{0111}\bigg(\delta_{ij}\delta_{kl}+\delta_{il}\delta_{jk}+\delta_{ik}\delta_{jl}\bigg)+
\nonumber\\
&&C_{0111}\bigg(\hat q_j\hat q_k\delta_{il}+\hat q_j\hat q_l\delta_{ik}+\hat q_i\hat q_j\delta_{kl}\bigg)+
D_{0111}\bigg(\hat q_i\hat q_j\hat q_k\hat q_l\bigg)\nonumber
\ee

Straightforward algebra establishes the relationship between  the different coefficients $A_i$, $B_i$, ... (let us denote them by the generic $\mathcal{C}$):
\be\label{Crels}
A_2&=&2(A_{00}-A_{01})\ , \ \ \ B_2=-2B_{01}\nonumber\\
A_3&=&4B_{001}+2C_{001}\ , \ \ B_3=6A_{001}\nonumber\\
A_4&=&2A_{0000}+2A_{0011}+4B_{0011}-8B_{0111}\nonumber\\
B_4&=&2 C_{0011}+4D_{0011}-4A_{0111}-4C_{0111}\nonumber\\
C_4&=&6 E_{0011}-8D_{0111}
\ee

\begin{figure}[t!]
\centering
\subfloat{\includegraphics[width=0.7\textwidth]{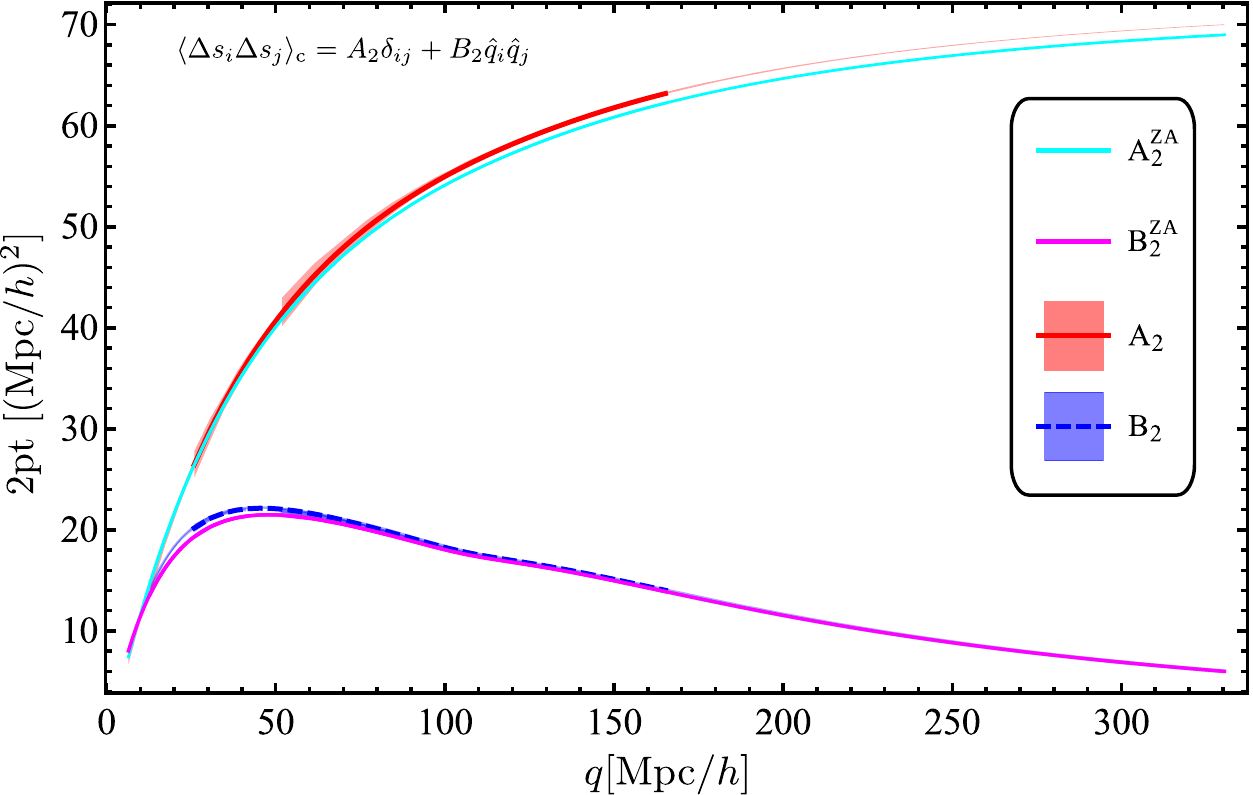}}     
\caption{The components of the displacement 2-pt correlation function as obtained from simulations, and using the ZA. Note how close the ZA predictions are to the truth -- a result extensively used in this paper.} \label{fig:2ptDza}
\end{figure}

One may be tempted to calculate the $\mathcal{C}$'s in perturbation theory. However, as can be seen by inspection, all of the above operators  depend on products of operators evaluated at zero separation (such operators are also known as composite operators). Therefore, they depend on extremely short-scale physics, for which perturbation theory breaks down. Thus, their values receive corrections from short scales which are impossible to obtain within perturbation theory. Those must be obtained by comparing with simulations or data.

\begin{figure}[t!]
\centering
\subfloat{\includegraphics[width=0.48\textwidth]{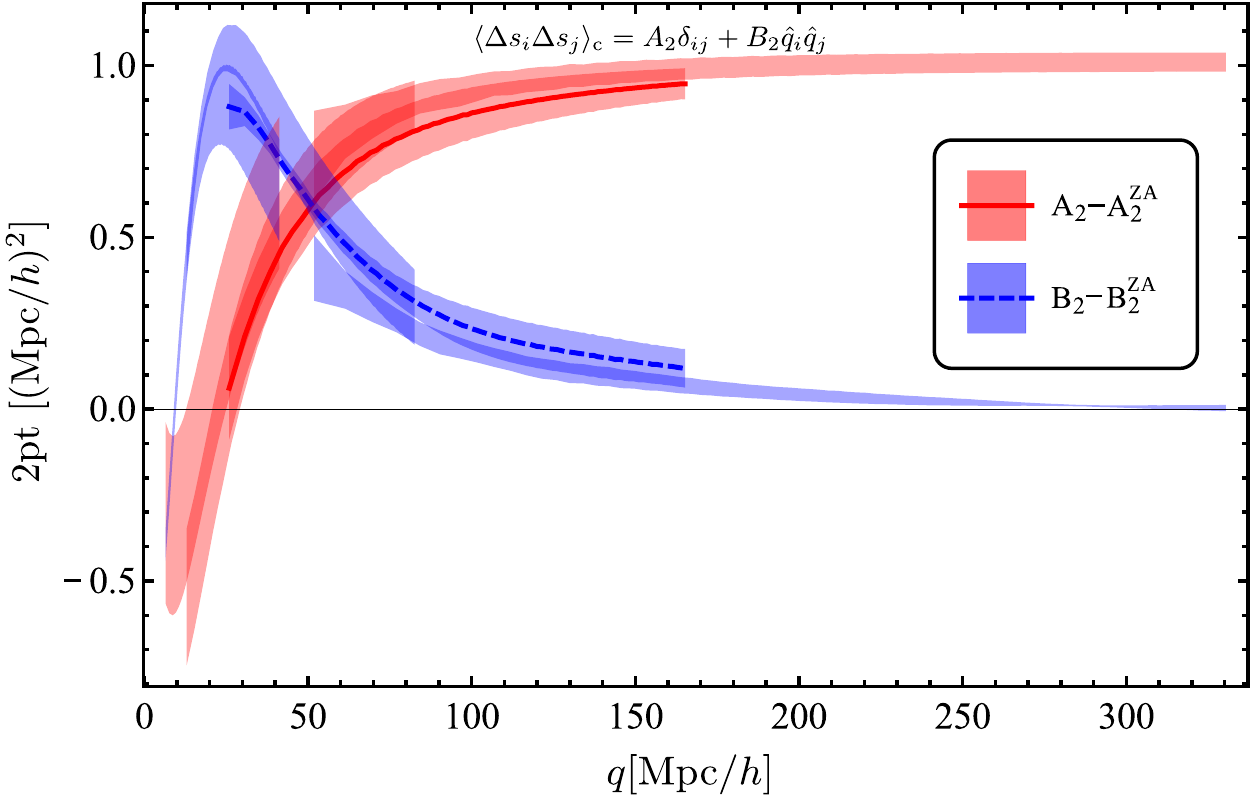}}
\hfill
\subfloat{\includegraphics[width=0.48\textwidth]{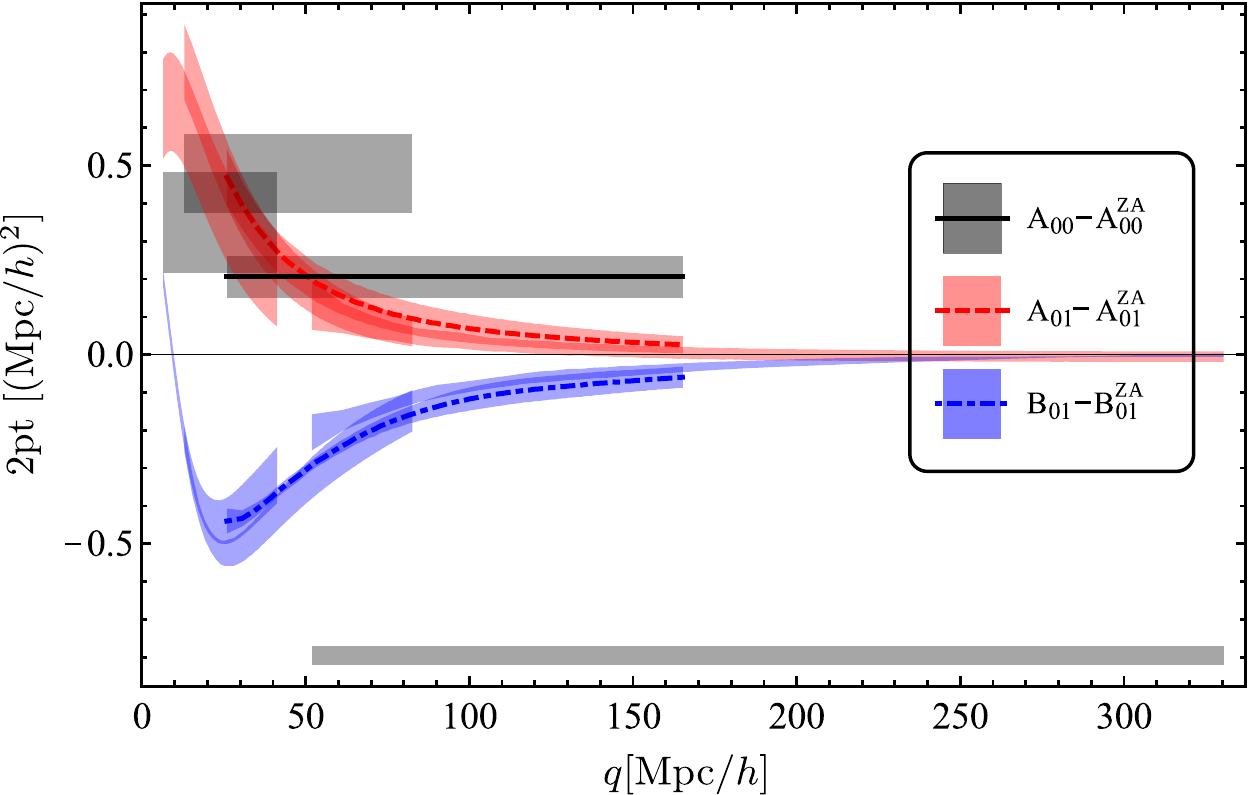}}
\caption{Parameters determining the size of the displacement 2-pt correlation function as obtained from simulations, and using the ZA. See the text for further discussion.} \label{fig:2ptD}
\end{figure}

\begin{figure}[t!]
\centering
\subfloat{\includegraphics[width=0.48\textwidth]{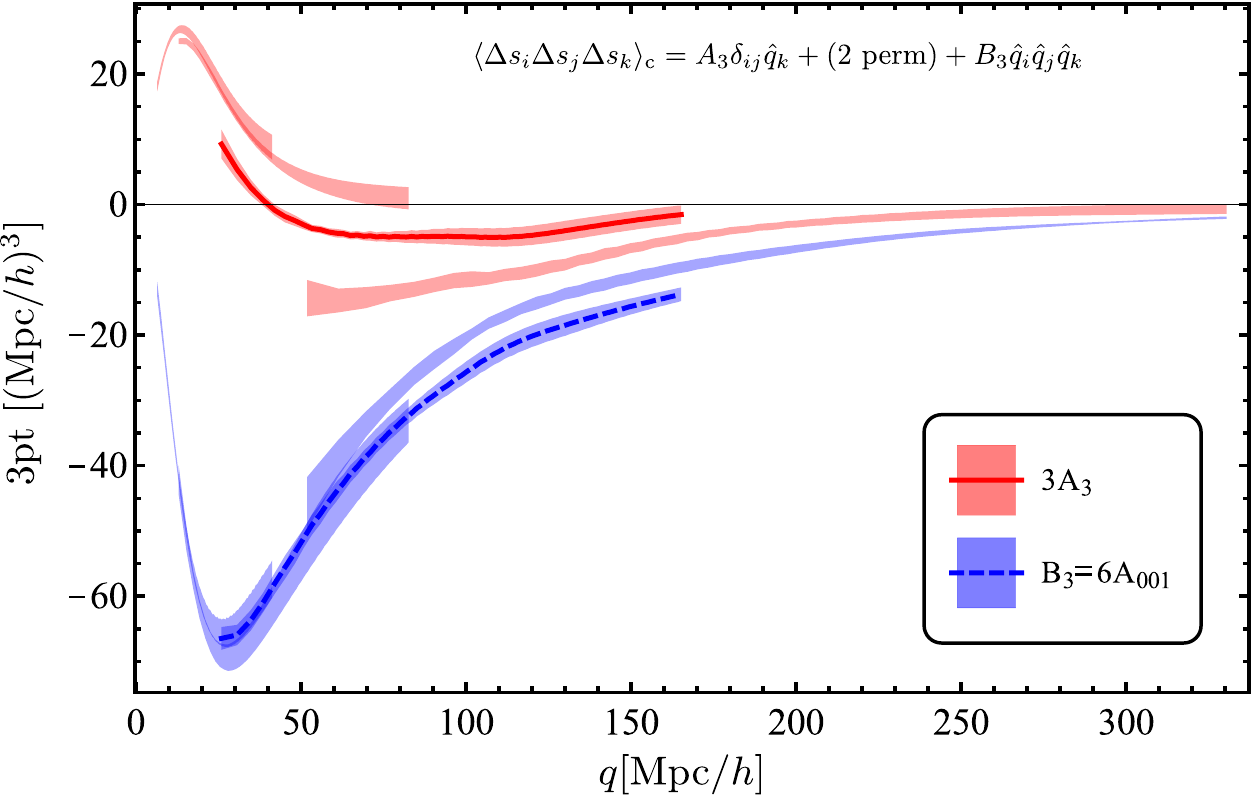}}     
\hfill
\subfloat{\includegraphics[width=0.48\textwidth]{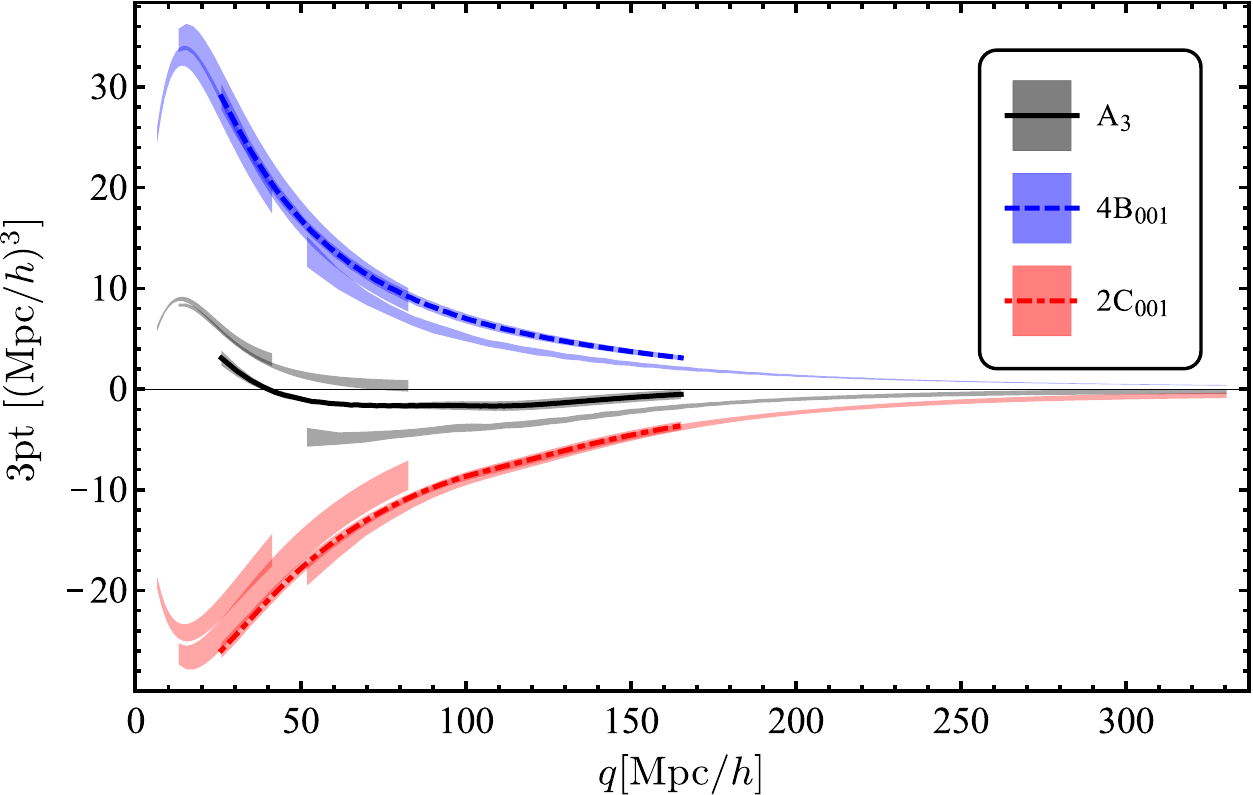}}
\caption{The same as Figure~\ref{fig:2ptD} but for the 3-pt function of the displacement field. The numerical coefficients in the left panel are chosen according to (\ref{orthoParaC}). The numerical coefficients on the right are chosen  according to (\ref{Crels}) so that the black line is a sum of the rest of the lines in that plot. } \label{fig:3ptD}
\end{figure}

\begin{figure}[t!]
\centering
\subfloat{\includegraphics[width=0.48\textwidth]{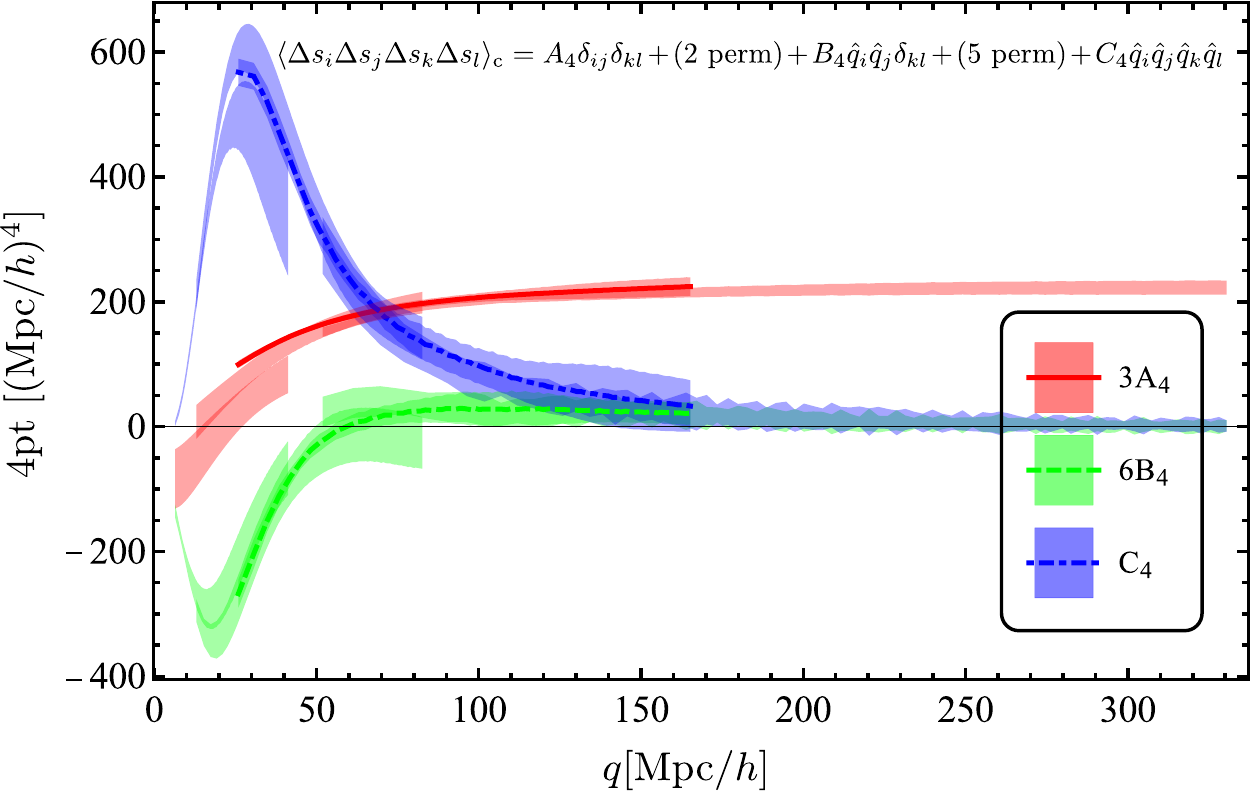}}     
\hfill
\subfloat{\includegraphics[width=0.48\textwidth]{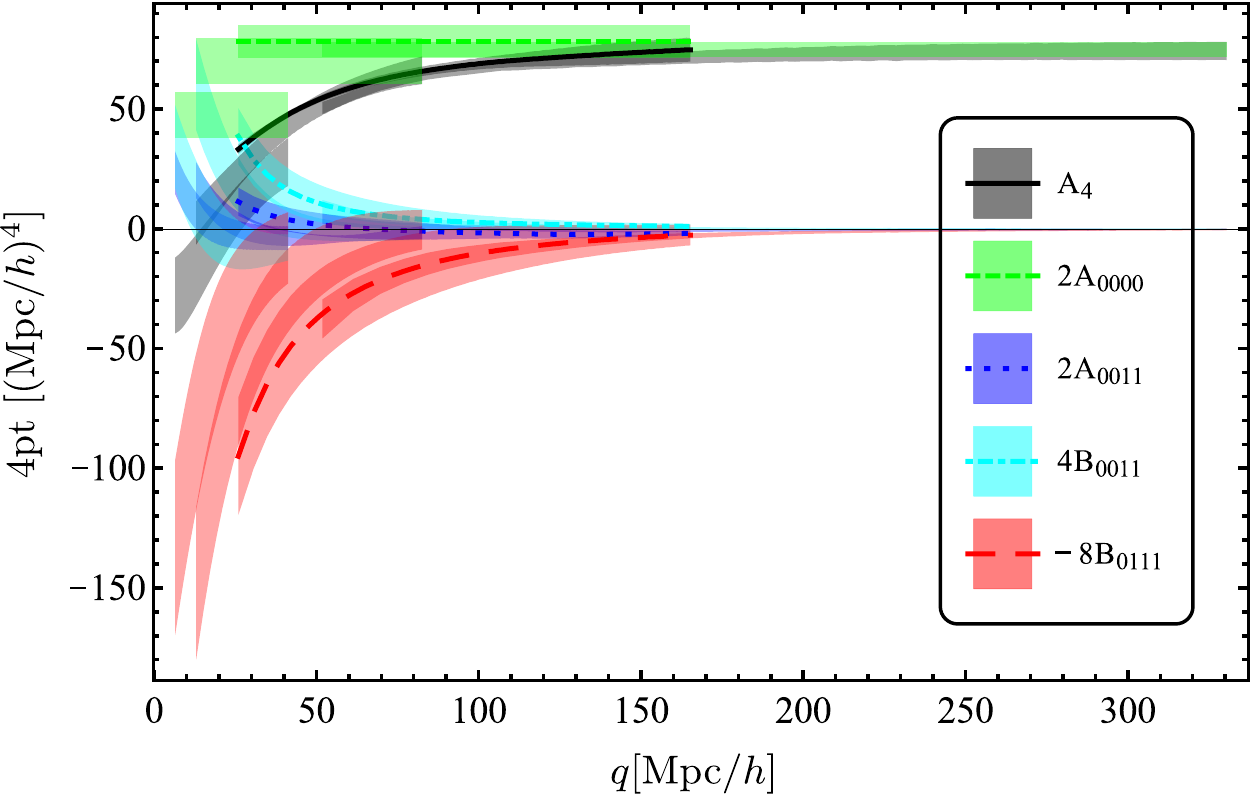}}
\\
\subfloat{\includegraphics[width=0.48\textwidth]{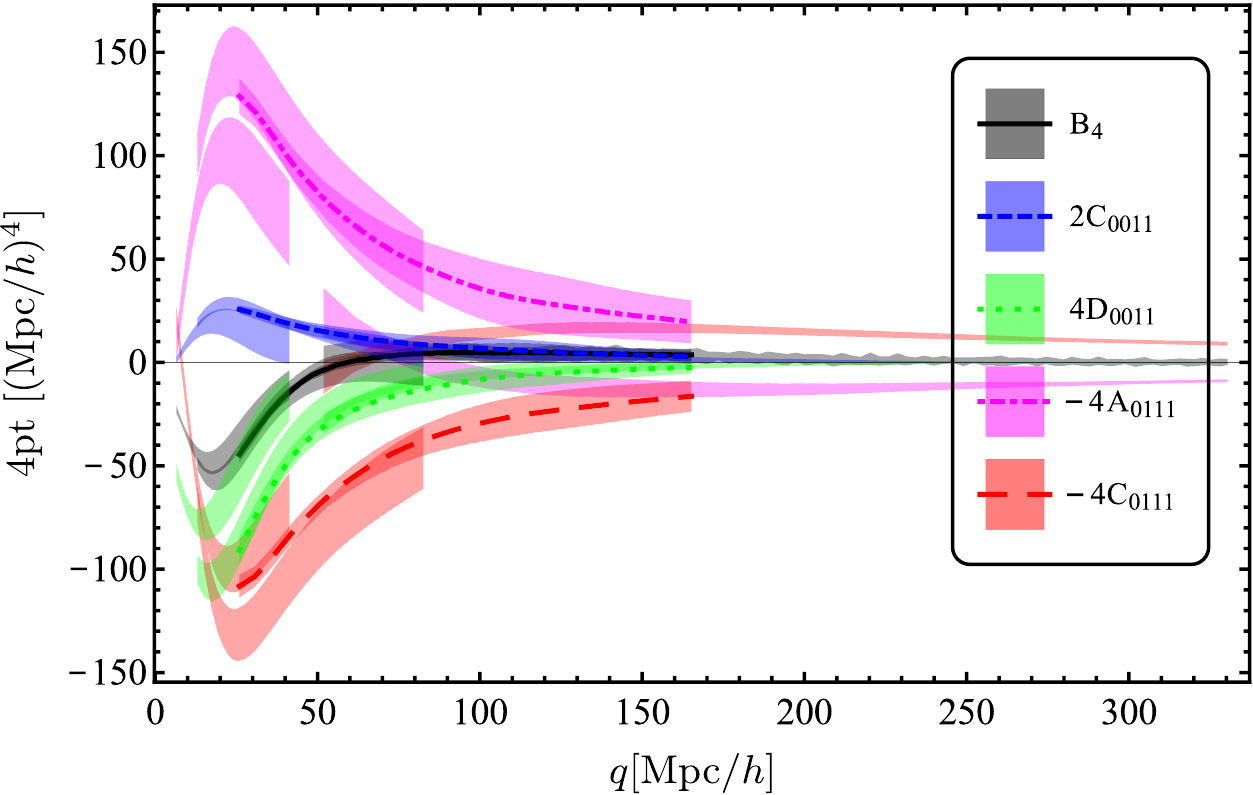}}     
\hfill
\subfloat{\includegraphics[width=0.48\textwidth]{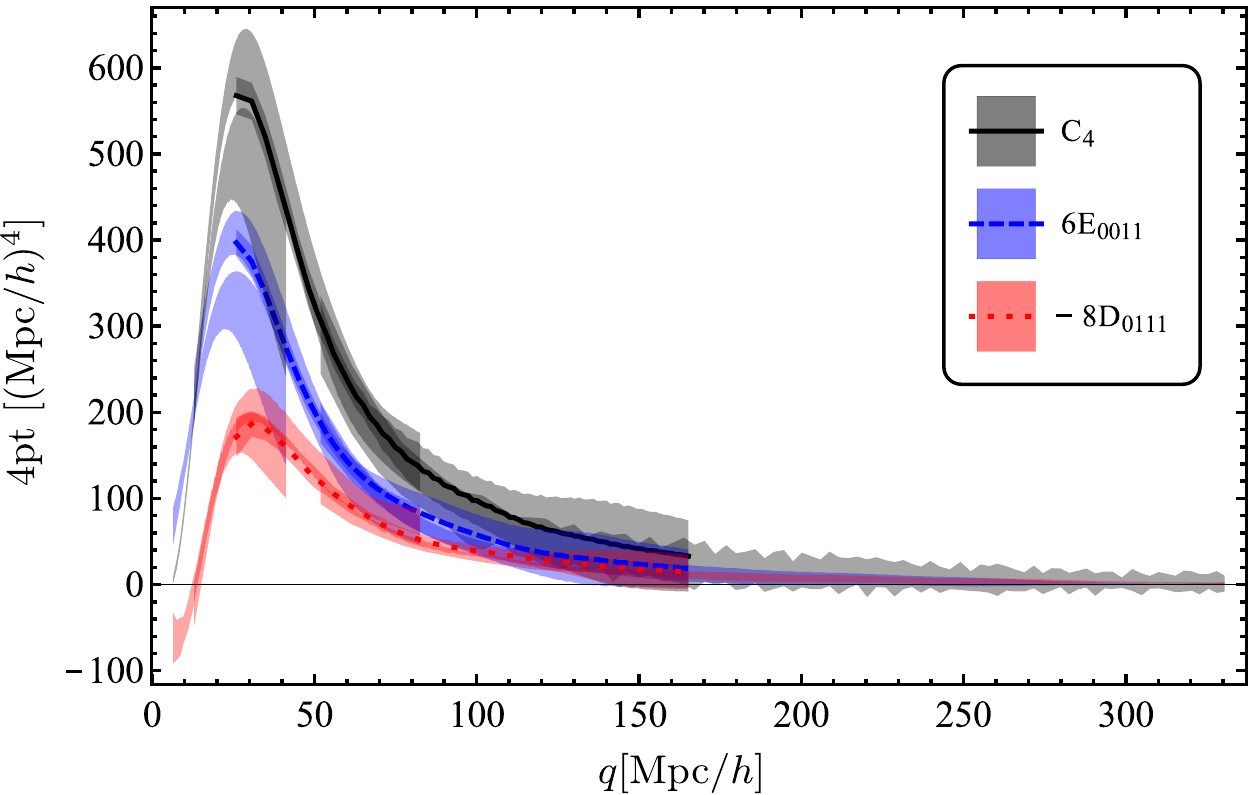}}
\caption{The same as Figure~\ref{fig:2ptD} but for the 4-pt function of the displacement field. The numerical coefficients in the top-left plot are chosen according to (\ref{orthoParaC}). The numerical coefficients in the rest of the panels are chosen  according to  (\ref{Crels}) so that each of the black lines is a sum of the rest of the lines in their respective panel. } \label{fig:4ptD}
\end{figure}

Therefore, to obtain estimates to the different contributions to $\xi$, we abandon perturbation theory completely and use simulations instead\footnote{One may have been worried about the solenoidal component of $\s$. Fortunately, given that our results are extracted from simulations, the solenoidal component is automatically included.}. To obtain the $\mathcal{C}$'s, we performed a total of 24 Gadget-2 \cite{gadget} cheap simulations down to redshift of $z=0$ of the same standard cosmology ($\sigma_8=0.8$, $n_s=0.95$, $\Omega_b=0.046$, $\Omega_{\mathrm{cdm}}=0.228$, $h=0.7$ in the standard notation). All simulations have $256^3$ particles, and the boxes are with sizes $L$ as follows: 8 sims with $L=250$Mpc$/h$; 8 sims with $L=500$Mpc$/h$; 4 sims with $L=1000$Mpc$/h$; 4 sims with $L=2000$Mpc$/h$. The results\footnote{
We calculate these quantities by extracting from the simulations all distinct (as dictated by the symmetries) contractions of $\langle \s^n\rangle_c$ with a suitable number of $\delta_{ij}$ and $\hat q_k$ to get scalar quantities. For example for the two point function we calculate the scalars $I_1=\langle\Delta s_i\Delta s_j\rangle_c\delta_{ij}$ and $I_2=\langle\Delta s_i\Delta s_j\rangle_c\hat q_i\hat q_j$ from the simulations. Calculating the same contractions analytically using the definition of $A_2$ and $B_2$ from (\ref{DeltaCs}), we find $I_1=3A_2+B_2$ and $I_2=A_2+B_2$. We then solve for $A_2$ and $B_2$ in terms of the $I$'s. The same is done for all cumulants in this section.

There are two non-trivial steps in the analysis of the sims which we performed to reduce sample variance, given the small number of cheap boxes. Starting with the same set of 24 initial conditions, we repeated the 24 sims in the ZA as well. In analogy with the estimator constructed for the matter power spectrum in \cite{2012JCAP...04..013T} we write down the following estimator for $\mathcal{C}$:
\be
\tilde{\mathcal{C}}=\mathcal{C}_{\mathrm{Gadget}}-\mathcal{C}_{\mathrm{ZA,\ sims}}+\mathcal{C}_{\mathrm{ZA,\ analytical}}
\ee
where $\mathcal{C}_{\mathrm{ZA,\ analytical}}$ is non-zero only for the 2-pt functions.
Note that the estimators $\tilde{\mathcal{C}}$ are unbiased by construction although not necessarily optimal. As we will see below, the plotted $\mathcal{C}$ are given by $\tilde{\mathcal{C}}$ except for $A_2$. One may wonder whether the raw $\mathcal{C}_{\mathrm{Gadget}}$ agree with $\tilde{\mathcal{C}}$ within their respective errorbars. In theory they should, but in practice these results can be influenced by our initial conditions generator (i.e. how well $\mathcal{C}_{\mathrm{ZA,\ sims}}-\mathcal{C}_{\mathrm{ZA,\ analytical}}$ approaches zero in the infinite ensemble limit). Numerically we find an agreement for all quantities except $A_{0111}$ and $C_{0111}$, which we find to be significantly smaller as given by 
$\tilde{\mathcal{C}}$ than by $\mathcal{C}_{\mathrm{Gadget}}$. However, for our calculations only the sum $A_{0111}+C_{0111}$ is important, and the results for it do match. But if one is interested in extracting renormalization corrections from simulations, one should keep track of such problems in the initial conditions.

The second non-trivial step in the analysis is the following. In the right panel of Figure~\ref{fig:2ptD}, we use $\tilde{\mathcal{C}}$ to plot the shown components of the displacement 2-pt function. There we can see that the quantity $A_{00}-A^{\mathrm{ZA}}_{00}$ (the latter being the analytical ZA result) is not constant across the sets of sims with different $L$. Yet, $A_{00}$ must be constant. That discrepancy is clearly due to the fact that the large boxes do not resolve halos, while the small boxes have too few massive halos.  Thus, we fix $A_{00}$ to a fiducial value of $A_{00}^{\mathrm{fiducial}}=A^{\mathrm{ZA}}_{00}+0.5(\mathrm{Mpc}/h)^2$ for the rest of the plots. Thus, the final estimator we use for the plots is:
\be
\mathcal{C}_{\mathrm{plot}}=\mathcal{C}_{\mathrm{Gadget}}\big[A_{00}^{\mathrm{fiducial}}\big]-\mathcal{C}_{\mathrm{ZA\ sims}}\big[A_{00}^{\mathrm{fiducial}}\big]+\mathcal{C}_{\mathrm{ZA\ analytical}}\big[A_{00}^{\mathrm{fiducial}}\big]
\ee 
where we have indicated that $A_{00}$ must be shifted to $A_{00}^{\mathrm{fiducial}}$ in all disconnected diagrams when calculating the cumulants. We find that only $A_2$ is significantly affected, the rest of the coefficients being shifted only within their errorbars. Thus, for all practical purposes the plots show $\mathcal{C}_{\mathrm{plot}}=\tilde{\mathcal{C}}$ except for $A_2$ (remember, we used $\tilde C$ to plot $A_{00}$). Note that the errors in the mean shown in the plots are calculated assuming a normal distribution between realizations, which combined with the small number of realizations implies that they should really be taken only as an indication of the size of the errors, which is sufficient for the purposes of our estimates.
}
 are shown in Figures~\ref{fig:2ptDza}, \ref{fig:2ptD}, \ref{fig:3ptD} and \ref{fig:4ptD} for the 2pt, 3pt and 4pt displacement statistics.

Those figures show 4 shaded regions for each $\mathcal{C}$ representing the 4 sets of box sizes. The width of the shaded regions gives the 2-sigma errors in the mean. We show the mean value only for $L=1000$Mpc$/h$. The fact that not all shaded regions for a given $\mathcal{C}$ overlap  implies that we have systematics due to finite resolution and finite volume effects. Yet a detailed analysis of convergence is not our goal, as we really aim only at crudely estimating the contributions of these terms to $\xi$. With that modest goal in mind, from the plots we see that we can claim convergence. Therefore, in Figures~\ref{fig:correctionsXi} and \ref{fig:correctionsPk} we use the results from the $L=1000$Mpc$/h$ boxes.

Now we need to establish the relationship between $\mathcal{C}$'s and the contributions to $\xi$.
The quantities above have a complicated tensorial structure. Thus we need to make a simplifying assumption in order to make the estimates. So, let us choose $\bm{k}$ to be either parallel or orthogonal to $\bm{q}$. Using (\ref{DeltaCs}) we find
\be\label{orthoParaC}
\left\langle \big(\Delta s\cdot \hat{k}_{||}\big)^2\right\rangle_c&=&A_2+B_2\nonumber\\
\left\langle \big(\Delta s\cdot \hat{k}_{\bot}\big)^2\right\rangle_c&=&A_2\nonumber\\
\left\langle \big(\Delta s\cdot \hat{k}_{||}\big)^3\right\rangle_c&=&3 A_3+B_3\nonumber\\
\left\langle \big(\Delta s\cdot \hat{k}_{\bot}\big)^3\right\rangle_c&=&0\nonumber\\
\left\langle \big(\Delta s\cdot \hat{k}_{||}\big)^4\right\rangle_c&=&3 A_4+6B_4+C_4\nonumber\\
\left\langle \big(\Delta s\cdot \hat{k}_{\bot}\big)^4\right\rangle_c&=&3 A_4
\ee
Numerically we find that choosing $\k||\q$ gives larger expectation values for the above quantities than $\k\bot\q$ for the 3-pt and 4-pt functions, while for the non-linear piece of the 2-pt function, choosing $\k\bot\q$ and $\k||\q$ are comparable (within less than a factor of 2) and change in relative  importance with scale.  Thus, for consistency, we choose $\k||\q$ for the 2-pt function as well.

Following the discussion after (\ref{xiContribsExpanded}), we use (\ref{xiContribsExpanded}) and (\ref{orthoParaC}) to finally write down the leading order contributions to $\xi$ from each of the expansion parameters (shown in Figure~\ref{fig:correctionsXi}):
\be\label{EPSILONS}
\mathrm{Displ.}_{\mathrm{Fig.}\ref{fig:correctionsXi},\ref{fig:correctionsPk}}&\equiv &\frac{\frac{1}{8}k^4\left\langle \big(\Delta s_{\mathrm{ZA}}\cdot \hat{k}_{||}\big)^2\right\rangle_c^2}{\frac{1}{2}k^2\left\langle \big(\Delta s_{\mathrm{ZA}}\cdot \hat{k}_{||}\big)^2\right\rangle_c}
=
\frac{1}{4}k^2\left(A^{\mathrm{ZA}}_2+B^{\mathrm{ZA}}_2\right)
\nonumber\\
\mathrm{NL}_{\mathrm{guess}}&\equiv &\frac{\frac{1}{2}k^2\left[\left\langle \big((\Delta s)\cdot \hat{k}_{||}\big)^2\right\rangle_c-\left\langle \big((\Delta s_{\mathrm{ZA}})\cdot \hat{k}_{||}\big)^2\right\rangle_c\right]}{\frac{1}{2}k^2\left\langle \big(\Delta s_{\mathrm{ZA}}\cdot \hat{k}_{||}\big)^2\right\rangle_c}
=
\frac{A_2+B_2-A^{\mathrm{ZA}}_2-B^{\mathrm{ZA}}_2}{A^{\mathrm{ZA}}_2+B^{\mathrm{ZA}}_2}\nonumber
\\
\mathrm{3pt}_{\mathrm{Fig.}\ref{fig:correctionsXi},\ref{fig:correctionsPk}}&\equiv &\frac{\frac{1}{6}k^3\left\langle \big(\Delta s\cdot \hat{k}_{||}\big)^3\right\rangle_c}{\frac{1}{2}k^2\left\langle \big(\Delta s_{\mathrm{ZA}}\cdot \hat{k}_{||}\big)^2\right\rangle_c}
=
\frac{k}{3}\frac{3 A_3+B_3}{A^{\mathrm{ZA}}_2+B^{\mathrm{ZA}}_2}
\nonumber\\
\mathrm{4pt}_{\mathrm{guess}}&\equiv&\frac{\frac{1}{24}k^4\left\langle \big(\Delta s\cdot \hat{k}_{||}\big)^4\right\rangle_c}{\frac{1}{2}k^2\left\langle \big(\Delta s_{\mathrm{ZA}}\cdot \hat{k}_{||}\big)^2\right\rangle_c}
=
\frac{k^2}{12}\frac{3 A_4+6B_4+C_4}{A^{\mathrm{ZA}}_2+B^{\mathrm{ZA}}_2}
\ee

There is an important caveat for the NL 2-pt as well as for the 4-pt corrections (thus, the labels ``guess''). The $\mathcal{C}$'s of those two contributions approach constants as $\q\to\infty$ due to non-zero $\Delta A_{00}\equiv A_{00}-A^\mathrm{ZA}_{00}$ and $A_{0000}$. Therefore, at leading order they give exactly zero contributions to $\xi$ and $P(k)$ as can be seen from (\ref{xiI}). To estimate their lowest order non-zero contribution, note that those constants can be taken outside the integral in $q$ and for the power spectrum will lead to $$P(k)\to P(k)\exp\left(-k^2\Delta A_{00}+\frac{k^4}{4}A_{0000}\right)$$
Therefore (still somewhat naively) we may conclude that their lowest order fractional contributions to $P(k)$ are $k^2\Delta A_{00}$ and $k^4A_{0000}/4$ (compare with (\ref{EPSILONS}), where at leading order for the rest of the terms we have the ZA 2-pt function in the denominator). However, as written, these contributions explicitly depend on bulk flows coherent on arbitrarily large scales, which is unphysical. For the purposes of the estimates we are performing, eliminating bulk flows at scales larger than $1/k$ amounts to replacing $\Delta A_{00}$ and $A_{0000}$ with their fully $q$-dependent $A_2$ and $A_4$. Thus, including the combinatorial factors we end up with contributions equal to 
\be\label{A2A4}
\mathrm{NL}_{\mathrm{Fig.}\ref{fig:correctionsXi}}&\equiv&\frac{k^2}{2}(A_2-A_2^{\mathrm{ZA}})\nonumber\\
\mathrm{4pt}_{\mathrm{Fig.}\ref{fig:correctionsXi}}&\equiv& \frac{k^4}{8}A_4\ .
\ee
 Those we find to be dominant in real space (hence we include them in the real space Figure~\ref{fig:correctionsXi})  to the contributions which do decay to zero at infinity: $B_2$, $B_4$, and $C_4$; but we find the terms in (\ref{A2A4}) to be subdominant in Fourier space.\footnote{
One may wonder why we did not split the ZA 2-pt function into two pieces as we did for $\Delta A_{00}$, given that $A_{00}$ itself goes to a constant. The reason is that all the ZA terms are comparable $\sim\mathcal{O}(1)$ and did not warrant making the supposedly simple estimates more cumbersome. Moreover, these splits cannot be made cleanly without a full calculation since $A_{00}$ itself can never be present alone (as it unphysically depends on large-scale bulk flows) but must be in the combination $A_2$ which in turn is no turn longer a constant. In Fourier space we keep the ZA 2-pt function not split for consistency with real space, and from the numerics we see that its magnitude is comparable to the density as it should be, given that what enters in the calculation in Fourier space is the divergence of the displacement field, which is simply the linear density.
 
But then one may ask why we bothered splitting the NL 2-pt and 4-pt pieces in the first place. Indeed if we did not split them, the numerics would not change appreciably for the NL 2-pt contribution in real space, and the 4-pt contributions in real and Fourier space.  However, the NL 2-pt contribution in Fourier space is a different story. In Fourier space, using NL$_\mathrm{guess}$ from (\ref{EPSILONS}), instead of its splitted version (\ref{A2A4}), results in  NL$_\mathrm{guess}$ becoming an unphysical non-negligible constant at low $k$ as both its numerator and denominator becomes constants. Instead it must decay as $k^2$ as is  the case after splitting it (see (\ref{A2A4})). This is the main reason why we proceeded to make these splits explicit.} Those decaying contributions are given as in (\ref{EPSILONS}) except for dropping $\Delta A_2$ and $A_4$, which we already included in (\ref{A2A4}):
\be\label{EPSILONS_Pk}
\mathrm{NL}_{\mathrm{Fig.}\ref{fig:correctionsPk}}&\equiv &
\frac{B_2-B^{\mathrm{ZA}}_2}{A^{\mathrm{ZA}}_2+B^{\mathrm{ZA}}_2}\nonumber
\\
\mathrm{4pt}_{\mathrm{Fig.}\ref{fig:correctionsPk}}&\equiv&
\frac{k^2}{12}\frac{6B_4+C_4}{A^{\mathrm{ZA}}_2+B^{\mathrm{ZA}}_2}
\ee
Using (\ref{length}), in real space we set $k=\pi/(20\mathrm{Mpc}/h)$ for Figures~\ref{fig:correctionsXi}; while in Fourier space we set $q=\pi/(2k)$ for evaluating the $\mathcal{C}$'s for Figure~\ref{fig:correctionsPk}.

\section{Smoothing of the BAO peak}\label{app:length}

In this section we make the simple toy model in Appendix C of \cite{2013arXiv1311.2168P} more realistic  to motivate the choice of $l$.

Let us split the covariance $\bm{\Sigma}_{ij}(q)=\langle \Delta s_i\Delta s_j\rangle_c$ into a smooth part (``sm'') and a part containing the BAO feature (``f'') in analogy with the usual wiggle/no-wiggle split for the power spectrum (e.g. \cite{2010ApJ...720.1650S}). In what follows we use a smooth $\bm{\Sigma}^{\mathrm{sm}}_{ij}$ calculated using the no-wiggle power spectrum of \cite{1998ApJ...496..605E}. At $z=0$, we find that $\bm{\Sigma}^{\mathrm{sm}}_{ii}(100\mathrm{Mpc}/h)=8.5\mathrm{Mpc}/h$, while $\bm{\Sigma}^{\mathrm{f}}_{ii}(100\mathrm{Mpc}/h)=1.9\mathrm{Mpc}/h$. Thus, we can treat $l^{-2}\bm{\Sigma}^{\mathrm{f}}\sim 0.2 l^{-2}\bm{\Sigma}\sim \mathcal{O}(0.1)$ as perturbative and expand to first order. The expansion is easiest done in Fourier space using (\ref{xiI},\ref{Iker}). Going back to real space, we obtain
\be\label{xiAppr}
\xi(\x)\approx\xi_{\mathrm{sm}}(\x)+\frac{1}{2}\partial_i\partial_j\int d^3q\,  \bm{\Sigma}^{\mathrm{f}}_{ij}(q)p_{\mathrm{sm}}(\x|\q)
\ee
Note that since $p(\x|\q)$ is a very close to a Gaussian which is a function of $(\x-\q)$, we can write
\be\label{ddpsm}
\frac{\partial}{\partial x_i}\frac{\partial}{\partial x_j}p_{\mathrm{sm}}(\x|\q)=\frac{\partial}{\partial q_i}\frac{\partial}{\partial q_j}p_{\mathrm{sm}}(\x|\q)+\frac{1}{\Sigma^\mathrm{sm}_{ii}}\left\{\mathcal{O}\left(\frac{\partial_q\Sigma_{ii}^{\mathrm{sm}}}{\sqrt{\Sigma_{ii}^{\mathrm{sm}}}}\right)+\mathcal{O}\left(\partial^2_q\Sigma_{ii}^{\mathrm{sm}}\right)\right\}
\ee
with the last two terms in the figure brackets numerically on the order of a percent around the position of the BAO peak. That is to be expected since the displacement covariance does not change appreciably across the BAO peak. However, we cannot drop them yet as we will see below. We can plug (\ref{ddpsm}) in (\ref{xiAppr}) to get 
\be
\xi(\x)\approx\xi_{\mathrm{sm}}(\x)+\int d^3q\,  \xi_{\mathrm{f,linear}}(\q)p_{\mathrm{sm}}(\x|\q)+\left\{\mathcal{O}\left(\frac{\Sigma^\mathrm{f}_{ii}}{\Sigma^\mathrm{sm}_{ii}}\frac{\partial_q\Sigma_{ii}^{\mathrm{sm}}}{\sqrt{\Sigma_{ii}^{\mathrm{sm}}}}\right)+\mathcal{O}\left(\frac{\Sigma^\mathrm{f}_{ii}}{\Sigma^\mathrm{sm}_{ii}}\partial^2_q\Sigma_{ii}^{\mathrm{sm}}\right)\right\}
\ee
where we used that $\frac{1}{2}\partial_i\partial_j\bm{\Sigma}^{\mathrm{f}}_{ij}=\xi_{\mathrm{f,linear}}$ (since the linear density is given by the divergence of the displacement field), the latter being the linear piece of $\xi$ containing the feature. Now we see that the term containing $\xi_\mathrm{f}$ is at a percent level -- as large as the terms in the figure brackets. However, $\bm{\Sigma}^\mathrm{f}(q)$ is related to $\xi_\mathrm{f}(q)$ by two integrals in $q$, which effectively remove the feature from $\bm{\Sigma}^\mathrm{f}(q)$ -- a fact we checked numerically as well. Thus, the resulting shape of the BAO feature is dominated by the term containing $\xi_{\mathrm{f,linear}}$. Given that we are keeping only terms linear in $\xi_\mathrm{f}$, we can replace $p_{\mathrm{sm}}$ by $p$ to reduce notational overload, which in turn allows us to write (\ref{xiFeature}).

\section*{Acknowledgements}
The author would like to thank Matias Zaldarriaga and Daniel Eisenstein for many stimulating discussions leading up to this work and for commenting on the manuscript. The author would like to especially thank M.~Zaldarriaga for describing his toy model (Appendix C in \cite{2013arXiv1311.2168P}) to the author well before publication.

\bibliography{mildly_NL_v2}

\providecommand{\href}[2]{#2}\begingroup\raggedright\begin{thebibliography}{10}

\bibitem{1992ApJ...396L...1S}
G.~F. {Smoot}, C.~L. {Bennett}, A.~{Kogut}, E.~L. {Wright}, J.~{Aymon}, N.~W.
  {Boggess}, E.~S. {Cheng}, G.~{de Amici}, S.~{Gulkis}, M.~G. {Hauser},
  G.~{Hinshaw}, P.~D. {Jackson}, M.~{Janssen}, E.~{Kaita}, T.~{Kelsall},
  P.~{Keegstra}, C.~{Lineweaver}, K.~{Loewenstein}, P.~{Lubin}, J.~{Mather},
  S.~S. {Meyer}, S.~H. {Moseley}, T.~{Murdock}, L.~{Rokke}, R.~F. {Silverberg},
  L.~{Tenorio}, R.~{Weiss}, and D.~T. {Wilkinson}, {\it {Structure in the COBE
  differential microwave radiometer first-year maps}},  {\em \apjl} {\bf 396}
  (Sept., 1992) L1--L5.

\bibitem{2005ApJ...633..560E}
D.~J. {Eisenstein}, I.~{Zehavi}, D.~W. {Hogg}, R.~{Scoccimarro}, M.~R.
  {Blanton}, R.~C. {Nichol}, R.~{Scranton}, H.-J. {Seo}, M.~{Tegmark},
  Z.~{Zheng}, S.~F. {Anderson}, J.~{Annis}, N.~{Bahcall}, J.~{Brinkmann},
  S.~{Burles}, F.~J. {Castander}, A.~{Connolly}, I.~{Csabai}, M.~{Doi},
  M.~{Fukugita}, J.~A. {Frieman}, K.~{Glazebrook}, J.~E. {Gunn}, J.~S.
  {Hendry}, G.~{Hennessy}, Z.~{Ivezi{\'c}}, S.~{Kent}, G.~R. {Knapp}, H.~{Lin},
  Y.-S. {Loh}, R.~H. {Lupton}, B.~{Margon}, T.~A. {McKay}, A.~{Meiksin}, J.~A.
  {Munn}, A.~{Pope}, M.~W. {Richmond}, D.~{Schlegel}, D.~P. {Schneider},
  K.~{Shimasaku}, C.~{Stoughton}, M.~A. {Strauss}, M.~{SubbaRao}, A.~S.
  {Szalay}, I.~{Szapudi}, D.~L. {Tucker}, B.~{Yanny}, and D.~G. {York}, {\it
  {Detection of the Baryon Acoustic Peak in the Large-Scale Correlation
  Function of SDSS Luminous Red Galaxies}},  {\em \apj} {\bf 633} (Nov., 2005)
  560--574, [\href{http://xxx.lanl.gov/abs/astro-ph/0501171}{{\tt
  astro-ph/0501171}}].

\bibitem{2010arXiv1004.2488B}
D.~{Baumann}, A.~{Nicolis}, L.~{Senatore}, and M.~{Zaldarriaga}, {\it
  {Cosmological non-linearities as an effective fluid}},  {\em \jcap} {\bf 7}
  (July, 2012) 51, [\href{http://xxx.lanl.gov/abs/1004.2488}{{\tt
  arXiv:1004.2488}}].

\bibitem{2012JCAP...04..013T}
S.~{Tassev} and M.~{Zaldarriaga}, {\it {The mildly non-linear regime of
  structure formation}},  {\em \jcap} {\bf 4} (Apr., 2012) 13,
  [\href{http://xxx.lanl.gov/abs/1109.4939}{{\tt arXiv:1109.4939}}].

\bibitem{2013ApJ...769..106S}
N.~S. {Sugiyama} and T.~{Futamase}, {\it {Relation between the Standard
  Perturbation Theory and Regularized Multi-point Propagator Method}},  {\em
  \apj} {\bf 769} (June, 2013) 106,
  [\href{http://xxx.lanl.gov/abs/1303.2748}{{\tt arXiv:1303.2748}}].

\bibitem{2013JCAP...05..031P}
M.~{Peloso} and M.~{Pietroni}, {\it {Galilean invariance and the consistency
  relation for the nonlinear squeezed bispectrum of large scale structure}},
  {\em \jcap} {\bf 5} (May, 2013) 31,
  [\href{http://xxx.lanl.gov/abs/1302.0223}{{\tt arXiv:1302.0223}}].

\bibitem{2013MNRAS.428.1036M}
M.~{Manera}, R.~{Scoccimarro}, W.~J. {Percival}, L.~{Samushia}, C.~K.
  {McBride}, A.~J. {Ross}, R.~K. {Sheth}, M.~{White}, B.~A. {Reid}, A.~G.
  {S{\'a}nchez}, R.~{de Putter}, X.~{Xu}, A.~A. {Berlind}, J.~{Brinkmann},
  C.~{Maraston}, B.~{Nichol}, F.~{Montesano}, N.~{Padmanabhan}, R.~A. {Skibba},
  R.~{Tojeiro}, and B.~A. {Weaver}, {\it {The clustering of galaxies in the
  SDSS-III Baryon Oscillation Spectroscopic Survey: a large sample of mock
  galaxy catalogues}},  {\em \mnras} {\bf 428} (Jan., 2013) 1036--1054,
  [\href{http://xxx.lanl.gov/abs/1203.6609}{{\tt arXiv:1203.6609}}].

\bibitem{2012MNRAS.427.2168M}
K.~T. {Mehta}, A.~J. {Cuesta}, X.~{Xu}, D.~J. {Eisenstein}, and
  N.~{Padmanabhan}, {\it {A 2 per cent distance to z = 0.35 by reconstructing
  baryon acoustic oscillations - III. Cosmological measurements and
  interpretation}},  {\em \mnras} {\bf 427} (Dec., 2012) 2168--2179,
  [\href{http://xxx.lanl.gov/abs/1202.0092}{{\tt arXiv:1202.0092}}].

\bibitem{2012MNRAS.427.2146X}
X.~{Xu}, N.~{Padmanabhan}, D.~J. {Eisenstein}, K.~T. {Mehta}, and A.~J.
  {Cuesta}, {\it {A 2 per cent distance to z = 0.35 by reconstructing baryon
  acoustic oscillations - II. Fitting techniques}},  {\em \mnras} {\bf 427}
  (Dec., 2012) 2146--2167, [\href{http://xxx.lanl.gov/abs/1202.0091}{{\tt
  arXiv:1202.0091}}].

\bibitem{2013arXiv1311.2168P}
R.~A. {Porto}, L.~{Senatore}, and M.~{Zaldarriaga}, {\it {The Lagrangian-space
  Effective Field Theory of Large Scale Structures}},  {\em ArXiv e-prints}
  (Nov., 2013) [\href{http://xxx.lanl.gov/abs/1311.2168}{{\tt
  arXiv:1311.2168}}].

\bibitem{2002PhR...367....1B}
F.~{Bernardeau}, S.~{Colombi}, E.~{Gazta{\~n}aga}, and R.~{Scoccimarro}, {\it
  {Large-scale structure of the Universe and cosmological perturbation
  theory}},  {\em \physrep} {\bf 367} (Sept., 2002) 1--248,
  [\href{http://xxx.lanl.gov/abs/astro-ph/0112551}{{\tt astro-ph/0112551}}].

\bibitem{1995A&A...296..575B}
F.~R. {Bouchet}, S.~{Colombi}, E.~{Hivon}, and R.~{Juszkiewicz}, {\it
  {Perturbative Lagrangian approach to gravitational instability.}},  {\em
  \aap} {\bf 296} (Apr., 1995) 575,
  [\href{http://xxx.lanl.gov/abs/astro-ph/9406013}{{\tt astro-ph/9406013}}].

\bibitem{zeldovich}
Y.~B. {Zel'dovich}, {\it {Gravitational instability: An approximate theory for
  large density perturbations.}},  {\em \aap} {\bf 5} (Mar., 1970) 84--89.

\bibitem{2008PhRvD..78j3521B}
F.~{Bernardeau}, M.~{Crocce}, and R.~{Scoccimarro}, {\it {Multipoint
  propagators in cosmological gravitational instability}},  {\em \prd} {\bf 78}
  (Nov., 2008) 103521, [\href{http://xxx.lanl.gov/abs/0806.2334}{{\tt
  arXiv:0806.2334}}].

\bibitem{2008PhRvD..77b3533C}
M.~{Crocce} and R.~{Scoccimarro}, {\it {Nonlinear evolution of baryon acoustic
  oscillations}},  {\em \prd} {\bf 77} (Jan., 2008) 023533,
  [\href{http://xxx.lanl.gov/abs/0704.2783}{{\tt arXiv:0704.2783}}].

\bibitem{2008PhRvD..77f3530M}
T.~{Matsubara}, {\it {Resumming cosmological perturbations via the Lagrangian
  picture: One-loop results in real space and in redshift space}},  {\em \prd}
  {\bf 77} (Mar., 2008) 063530, [\href{http://xxx.lanl.gov/abs/0711.2521}{{\tt
  arXiv:0711.2521}}].

\bibitem{2009PhRvD..80d3531C}
J.~{Carlson}, M.~{White}, and N.~{Padmanabhan}, {\it {Critical look at
  cosmological perturbation theory techniques}},  {\em \prd} {\bf 80} (Aug.,
  2009) 043531, [\href{http://xxx.lanl.gov/abs/0905.0479}{{\tt
  arXiv:0905.0479}}].

\bibitem{2013MNRAS.429.1674C}
J.~{Carlson}, B.~{Reid}, and M.~{White}, {\it {Convolution Lagrangian
  perturbation theory for biased tracers}},  {\em \mnras} {\bf 429} (Feb.,
  2013) 1674--1685, [\href{http://xxx.lanl.gov/abs/1209.0780}{{\tt
  arXiv:1209.0780}}].

\bibitem{2012JCAP...12..011T}
S.~{Tassev} and M.~{Zaldarriaga}, {\it {Estimating CDM particle trajectories in
  the mildly non-linear regime of structure formation. Implications for the
  density field in real and redshift space}},  {\em \jcap} {\bf 12} (Dec.,
  2012) 11, [\href{http://xxx.lanl.gov/abs/1203.5785}{{\tt arXiv:1203.5785}}].

\bibitem{ZApaper}
S.~{Tassev}, {\it {N-point Statistics of Large-Scale Structure in the
  Zel'dovich Approximation}}, . In prep.

\bibitem{2002PhR...372....1C}
A.~{Cooray} and R.~{Sheth}, {\it {Halo models of large scale structure}},  {\em
  \physrep} {\bf 372} (Dec., 2002) 1--129,
  [\href{http://xxx.lanl.gov/abs/astro-ph/0206508}{{\tt astro-ph/0206508}}].

\bibitem{2013JCAP...06..036T}
S.~{Tassev}, M.~{Zaldarriaga}, and D.~J. {Eisenstein}, {\it {Solving large
  scale structure in ten easy steps with COLA}},  {\em \jcap} {\bf 6} (June,
  2013) 36, [\href{http://xxx.lanl.gov/abs/1301.0322}{{\tt arXiv:1301.0322}}].

\bibitem{2013MNRAS.433.2389M}
P.~{Monaco}, E.~{Sefusatti}, S.~{Borgani}, M.~{Crocce}, P.~{Fosalba}, R.~K.
  {Sheth}, and T.~{Theuns}, {\it {An accurate tool for the fast generation of
  dark matter halo catalogues}},  {\em \mnras} {\bf 433} (Aug., 2013)
  2389--2402, [\href{http://xxx.lanl.gov/abs/1305.1505}{{\tt
  arXiv:1305.1505}}].

\bibitem{1996ApJS..105...37S}
R.~{Scoccimarro} and J.~{Frieman}, {\it {Loop Corrections in Nonlinear
  Cosmological Perturbation Theory}},  {\em \apjs} {\bf 105} (July, 1996) 37,
  [\href{http://xxx.lanl.gov/abs/astro-ph/9509047}{{\tt astro-ph/9509047}}].

\bibitem{1993MNRAS.260..765C}
P.~{Coles}, A.~L. {Melott}, and S.~F. {Shandarin}, {\it {Testing approximations
  for non-linear gravitational clustering}},  {\em \mnras} {\bf 260} (Feb.,
  1993) 765--776.

\bibitem{2010ApJ...720.1650S}
H.-J. {Seo}, J.~{Eckel}, D.~J. {Eisenstein}, K.~{Mehta}, M.~{Metchnik},
  N.~{Padmanabhan}, P.~{Pinto}, R.~{Takahashi}, M.~{White}, and X.~{Xu}, {\it
  {High-precision Predictions for the Acoustic Scale in the Nonlinear Regime}},
   {\em \apj} {\bf 720} (Sept., 2010) 1650--1667,
  [\href{http://xxx.lanl.gov/abs/0910.5005}{{\tt arXiv:0910.5005}}].

\bibitem{2009PhRvD..80l3501N}
Y.~{Noh}, M.~{White}, and N.~{Padmanabhan}, {\it {Reconstructing baryon
  oscillations}},  {\em \prd} {\bf 80} (Dec., 2009) 123501,
  [\href{http://xxx.lanl.gov/abs/0909.1802}{{\tt arXiv:0909.1802}}].

\bibitem{1998ApJ...496..605E}
D.~J. {Eisenstein} and W.~{Hu}, {\it {Baryonic Features in the Matter Transfer
  Function}},  {\em \apj} {\bf 496} (Mar., 1998) 605,
  [\href{http://xxx.lanl.gov/abs/astro-ph/9709112}{{\tt astro-ph/9709112}}].

\bibitem{2007ApJ...664..675E}
D.~J. {Eisenstein}, H.-J. {Seo}, E.~{Sirko}, and D.~N. {Spergel}, {\it
  {Improving Cosmological Distance Measurements by Reconstruction of the Baryon
  Acoustic Peak}},  {\em \apj} {\bf 664} (Aug., 2007) 675--679,
  [\href{http://xxx.lanl.gov/abs/astro-ph/0604362}{{\tt astro-ph/0604362}}].

\bibitem{2009PhRvD..79f3523P}
N.~{Padmanabhan}, M.~{White}, and J.~D. {Cohn}, {\it {Reconstructing baryon
  oscillations: A Lagrangian theory perspective}},  {\em \prd} {\bf 79} (Mar.,
  2009) 063523, [\href{http://xxx.lanl.gov/abs/0812.2905}{{\tt
  arXiv:0812.2905}}].

\bibitem{2012MNRAS.427.2132P}
N.~{Padmanabhan}, X.~{Xu}, D.~J. {Eisenstein}, R.~{Scalzo}, A.~J. {Cuesta},
  K.~T. {Mehta}, and E.~{Kazin}, {\it {A 2 per cent distance to z = 0.35 by
  reconstructing baryon acoustic oscillations - I. Methods and application to
  the Sloan Digital Sky Survey}},  {\em \mnras} {\bf 427} (Dec., 2012)
  2132--2145, [\href{http://xxx.lanl.gov/abs/1202.0090}{{\tt
  arXiv:1202.0090}}].

\bibitem{2012JCAP...10..006T}
S.~{Tassev} and M.~{Zaldarriaga}, {\it {Towards an optimal reconstruction of
  baryon oscillations}},  {\em \jcap} {\bf 10} (Oct., 2012) 6,
  [\href{http://xxx.lanl.gov/abs/1203.6066}{{\tt arXiv:1203.6066}}].

\bibitem{2013arXiv1310.0464C}
J.~J.~M. {Carrasco}, S.~{Foreman}, D.~{Green}, and L.~{Senatore}, {\it {The
  Effective Field Theory of Large Scale Structures at Two Loops}},  {\em ArXiv
  e-prints} (Oct., 2013) [\href{http://xxx.lanl.gov/abs/1310.0464}{{\tt
  arXiv:1310.0464}}].

\bibitem{1993ApJ...412...64L}
S.~D. {Landy} and A.~S. {Szalay}, {\it {Bias and variance of angular
  correlation functions}},  {\em \apj} {\bf 412} (July, 1993) 64--71.

\bibitem{gadget}
V.~{Springel}, {\it {The cosmological simulation code GADGET-2}},  {\em \mnras}
  {\bf 364} (Dec., 2005) 1105--1134,
  [\href{http://xxx.lanl.gov/abs/astro-ph/0505010}{{\tt astro-ph/0505010}}].

\end{thebibliography}\endgroup

\end{document}